\documentclass[3p,authoryear]{elsarticle}

\usepackage[english]{babel}
\addto\captionsenglish{}

\usepackage{graphicx}

\usepackage{amsmath}
\numberwithin{equation}{section}

\graphicspath{{figs/}}

\newcommand{\mymult}{\,}

\newcommand{\cdummy}{\cdot}
\newcommand{\mathd}{\mathrm{d}}
\newcommand{\nobracket}{}
\newcommand{\nocomma}{}
\newcommand{\of}{:}
\newcommand{\tmem}[1]{{\em #1\/}}
\newcommand{\tmmathbf}[1]{\ensuremath{\boldsymbol{#1}}}

\usepackage[usernames,dvipsnames]{xcolor}

\begin{document}

\journal{J.~Mech.~Phys.~Solids}

\begin{frontmatter}

\title{
  Asymptotically exact strain-gradient models\\
  for nonlinear slender elastic structures:\\
  a systematic derivation method}

    \author[cl]{Claire Lestringant}
	\author[ba]{Basile Audoly}

	\address[cl]{Mechanics \& Materials, Department of Mechanical and Process Engineering, ETH Z\"{u}rich, 8092 Z\"{u}rich, Switzerland}

	\address[ba]{Laboratoire de m{\'e}canique des solides, CNRS, Institut
	Polytechnique de Paris, Palaiseau, France}

\begin{abstract}
  We propose a general method for deriving one-dimensional models for
  nonlinear structures.  It captures the contribution to the strain
  energy arising not only from the macroscopic elastic strain as in
  classical structural models, but also from the strain gradient.  As
  an illustration, we derive one-dimensional strain-gradient models
  for a hyper-elastic cylinder that necks, an axisymmetric membrane
  that produces bulges, and a two-dimensional block of elastic
  material subject to bending and stretching.  The method offers three
  key advantages.  First, it is nonlinear and accounts for large
  deformations of the cross-section, which makes it well suited for
  the analysis of localization in slender structures.  Second, it does
  not require any {\tmem{a priori}} assumption on the form of the
  elastic solution in the cross-section, i.e., it is Ansatz-free.
  Thirdly, it produces one-dimensional models that are asymptotically
  exact when the macroscopic strain varies on a much larger length
  scale than the cross-section diameter.
\end{abstract}

	
\end{frontmatter}




\section{Introduction}\label{sec:introduction}

There exists a variety of models for slender structures, going much
beyond the traditional models for the stretching of bars and the
bending of beams.  The applicability of classical models being limited
to materials having linear, homogeneous and isotropic elastic
properties, a number of extensions have been considered to account for
different elastic behaviors such as hyperelastic
materials~{\citep{Cimetiere-Geymonat-EtAl-Asymptotic-theory-and-analysis-1988}}
or more specifically nematic
elastomers~{\citep{Agostiniani-DeSimone-EtAl-Shape-programming-for-narrow-2016}},
for inhomogeneous elastic properties in the cross-section, for the
presence of natural curvature or
twist~{\citep{Freddi-Hornung-EtAl-A-variational-model-for-anisotropic-2016}}
or more generally for the existence of inhomogeneous pre-stress in the
cross-section~{\citep{Lestringant-Audoly-Elastic-rods-with-2016}}.  As
the classical rod models are inapplicable if the cross-section itself
is a slender 2d domain, specific models have been derived, e.g., to
address inextensible
ribbons~{\citep{Sadowsky-Ein-elementarer-Beweis-fur-die-Existenz-1930,Wunderlich-Uber-ein-abwickelbares-Mobiusband-1962}},
as well as thin walled beams having a
flat~{\citep{FreddiMorassiParoni-Thin-Walled-Beams-the-Case-of-the-Rectangular-Cross-Section-2004}}
or
curved~{\citep{Hamdouni-Millet-An-asymptotic-non-linear-model-2006}}
cross-section.  The classical models are inapplicable as well in the
presence of a large contrast of elastic moduli within the
cross-sections, as happens for sandwiched beams: in this case, the
presence of shear is often accounted for using the Timoshenko beam
model.  Specific models are also required to account for physical
effects such as the interaction with a magnetic
field~{\citep{Geymonat-Krasucki-EtAl-Asymptotic-derivation-of-linear-2018}}
or surface tension arising in soft beams immersed in a
fluid~{\citep{Xuan-Biggins-Plateau-Rayleigh-instability-in-solids-2017}}.

One can easily get lost in view of not only the multiplicity of these models
but also their justification (or lack thereof). Rigorous justifications based
on asymptotic expansions have made use of restrictive assumptions: the work in
this direction was initiated in the context of linear
elasticity~{\citep{Bermudez-Viano-Une-justification-des-equations-de-la-thermoelasticite-1984,Sanchez-Hubert-Sanchez-Palencia-Statics-of-curved-rods-1999}},
and extended to finite elasticity under specific assumptions
regarding material
symmetries~{\citep{Cimetiere-Geymonat-EtAl-Asymptotic-theory-and-analysis-1988}}.
The different models, such as Navier-Bernoulli beams, Timoshenko beams, Vlasov
beams, inextensible ribbons, etc., are justified by different arguments each,
and a unified justification method is lacking. There are many phenomena in
slender structures for which no asymptotically justified 1d model is
available, such as the ovalization of tubes subjected to
bending~{\citep{Calladine-Theory-of-shell-structures-1983}} or
pinching~{\citep{MahadevanVaziriDas-Persistence-of-a-pinch-in-a-pipe-2007}} and
the propagative instabilities in shallow
panels~{\citep{Kyriakides-Chang-The-initiation-and-propagation-of-a-localized-1991}}.

In some work, one-dimensional (1d) models have been proposed based on
kinematic hypotheses.  This is the case, for instance, for the
analysis localization in hyperelastic
cylinders~{\citep{Coleman-Newman-On-the-rheology-of-cold-drawing.-1988}}
and tape
springs~{\citep{Picault-Bourgeois-EtAl-A-rod-model-with-thin-walled-2016}}.
Even when these kinematic hypotheses turn out to be valid, their
domain of validity is typically limited and dependent, in a hidden
way, on the simplifying assumptions of the model.  For instance, the
most common assumptions used to derive the classical theory of beams
is that cross-sections remain planar and perpendicular to the center
line, and that the shear in the plane of the cross-sections is zero.
These assumptions are incorrect unless specific material symmetries
are applicable, which is ill-appreciated.  Moreover, they cannot be
used to derive higher-order models, as discussed by
{\citet{Audoly-Hutchinson-Analysis-of-necking-based-2016}}.

This paper proposes a systematic and rigorous dimension reduction method for obtaining 1d models for nonlinear slender elastic structures, which
works under broad assumptions. The main features of the method are as follows.
The reduction method can start from a variety of models, such as a
hyperelastic model for cylinder (e.g., for the stretching of bars), a
nonlinear model for a thin membrane (e.g., for the analysis of bulges in
axisymmetric balloons) or a shell model (e.g., for the tape spring problem).
It can handle arbitrary elastic constitutive laws (including nonlinear and
anisotropic ones), arbitrary pre-stress distributions in the cross-section,
and inhomogeneous material properties in the cross-section. The mechanical and
geometrical properties of the structure are assumed to be invariant in the
longitudinal direction; the extension to slowly variable properties is
straightforward, as discussed in {\textsection}\ref{sec:conclusion}.
Nonlinearity of both the elastic and geometric types are permitted, and large
spatial variations in the deformed shape of the cross-sections are accounted
for. No {\tmem{a priori}} kinematic hypothesis is made, the microscopic
displacement being found by solving the equations of elasticity. Our reduction method is built on a two-scale expansion, assuming slow
variations in the longitudinal directions. As such, it is asymptotically
exact. Its justification is based on a formal expansion, not on a rigorous
proof. We hope that our formal argument can be turned into a rigorous one in
the future.

An important asset of the method is that it captures the gradient
effect, i.e., the dependence of the strain energy on the gradient of
strain and not just on the strain.  This makes it possible to derive
higher-order reduced models offering the following advantages:
({\tmem{i}})~they feature faster convergence towards the solution of
the full (non-reduced) problem and ({\tmem{ii}})~they are well-suited
to the analysis of localization in slender structures.  Localization
is ubiquitous in slender structures, from neck formation in polymer
bars under
traction~{\citep{GSell-Aly-Helal-EtAl-Effect-of-stress-triaxiality-1983}},
to beading in cylinders made up of soft
gels~{\citep{Matsuo-Tanaka-Patterns-in-shrinking-gels-1992,Mora-Phou-EtAl-Capillarity-Driven-Instability-2010}},
to bulges produced by the inflation of cylindrical party
balloons~{\citep{Kyriakides-Chang-On-the-inflation-of-a-long-elastic-1990}},
and to kinks in bent tape
springs~{\citep{Seffen-Pellegrino-Deployment-dynamics-of-tape-1999}}.
Classical reduced models depending on strain only cannot resolve the
sharp interfaces that result from localization, and are mathematically
ill-posed.  By contrast, higher-order models capturing the dependence
on the gradient of strain allows the interfaces to be resolved and are
well-posed in the context of localization.  In prior work, asymptotic
1d strain-gradient models have been obtained as refinements over the
standard theory for linearly elastic
beams~{\citep{Trabucho-Viano-Mathematical-modelling-of-rods-1996,
Cartaud1}}, inextensible
ribbons~{\citep{Sadowsky-Ein-elementarer-Beweis-fur-die-Existenz-1930,Wunderlich-Uber-ein-abwickelbares-Mobiusband-1962}}
and thin-walled
beams~{\citep{FreddiMorassiParoni-Thin-Walled-Beams-the-Case-of-the-Rectangular-Cross-Section-2004}}.
The possibility of using 1d models to analyze localization in slender
structures easily and accurately has emerged recently in the context
of necking in bars and bulging in
balloons~{\citep{Audoly-Hutchinson-Analysis-of-necking-based-2016,Lestringant-Audoly-A-diffuse-interface-model-2018}}.
Several other localization phenomena could be better understood if 1d
models were available.

Our method can be described in general terms as follows.  First, we
introduce the so-called canonical form, which is a unified and
abstract formulation into which the various structural models for
slender structures can be cast.  The canonical form serves as a
starting point for the reduction process.  The set of degrees of
freedom are split between (master) macroscopic degrees of freedom
which are retained in the 1d model, and (slave) microscopic degrees of
freedom which are relaxed; the choice of which degrees of freedom are
retained as master is left to the user.  Next, the asymptotic
expansion is carried out: the equations of elasticity are expanded
about a configuration having finite and inhomogeneous pre-strain,
whereby each cross-section is in a state parameterized by the local
value of the macroscopic degrees of freedom.  This expansion is
implemented as a series of steps (i.e., a mere recipe, albeit a
slightly technical one at places) which ultimately yields a 1d elastic
potential governing the reduced model.  Classical structural model,
such as the Euler-Bernoulli beam model, are recovered at the dominant
order while corrections depending on the strain gradient are obtained
at the subdominant order.  In this first paper, the general method is
presented and illustrated on simple examples for which the 1d
strain-gradient model is already known from the literature; the method
will be applied to original problems in future work.

In section~\ref{sec:mainResults}, we give a general account of the
reduction method: the series of steps needed to carry out the
reduction are listed.  In Sections~\ref{sec:membranes}
to~\ref{sec:hyperCylinder}, three examples of applications are worked
out, by order of increasing complexity: we establish the 1d models for
the bulging of inflated membranes, for a linearly elastic block in 2d,
and for an axisymmetric hyperelastic cylinder.
In Section~\ref{sec:conclusion}, we conclude and make a few general
remarks about the method.  \ref{sec:detailedProofs} presents a
detailed proof of the reduction method.  \ref{app:cylinder} provides
the detailed calculations for the analysis of an axisymmetric
cylinder.

In mathematical formula, we use bold face symbols for vectors and
tensors.  Their components are denoted using plain typeface with a
subscript, as in $\tmmathbf{h}= (h_1, h_2)$.  Functions of a
cross-sectional coordinate are denoted by surrounding their generic
values using curly braces, with their dummy argument in subscript, as
in $f = \{ f (T) \}_T$.  We denote by $S$ the longitudinal coordinate.
In our notation, the primes will be reserved for the derivation with
respect to the longitudinal coordinate $S$,
\[ f' = \frac{\mathd f}{\mathd S} . \]

\section{Main results}\label{sec:mainResults}

We present the dimension reduction method in a generic and abstract
form that will be applied to specific structures in the forthcoming
sections.  We limit attention to a practical description of the
method: the method is justified in full details separately
in~\ref{sec:detailedProofs}.

\subsection{Starting point: full model in canonical form}

The elastic model used as a starting point for the dimension reduction
will be referred to as the {\tmem{full model}}.  One can use a variety
of full models, such as an axisymmetric membrane model (for the
analysis of bulges in balloons), a hyperelastic cylinder (for the
analysis of necking) or a shell model (for the analysis of a tape
spring).  We start by casting the full models into a standardized
form, called the {\tmem{canonical form}}, which exposes their common
properties and hides their specificities.  The conversion of
particular structural models into the canonical form is not discussed
here, and will be demonstrated based on examples in the following
sections.

\begin{figure}
  \centerline{\includegraphics{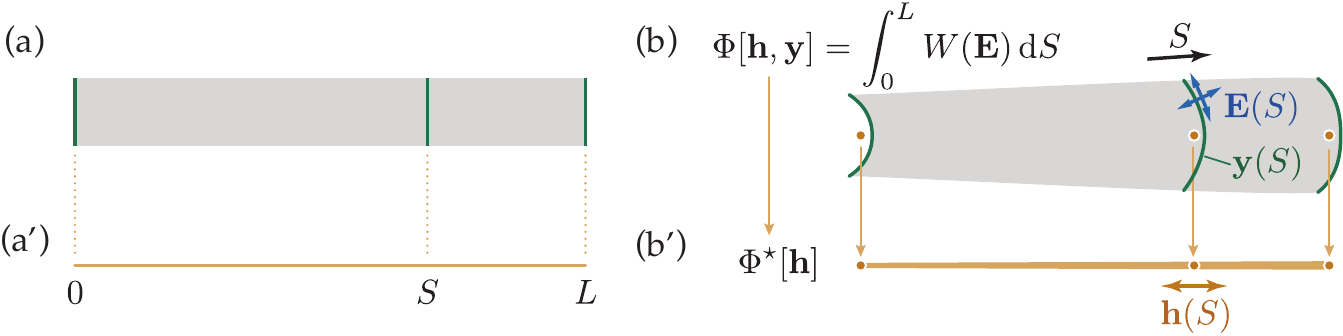}}
  \caption{Dimension reduction for an abstract slender structure. {\tmem{Left
  column}} (a,a'): reference configuration highlighting a particular
  cross-section with coordinate $S$. {\tmem{Right-hand side column}} (b,b'):
  deformed configuration. {\tmem{Top row}} (a,b): full model used as a staring
  point, including a microscopic displacement $\tmmathbf{y} (S)$ and a
  microscopic strain $\tmmathbf{E} (S)$. {\tmem{Bottom row}}
  (a',b'):~equivalent 1d model obtained by dimension reduction,
  in which the details at the scale of cross-section are effectively
  hidden.\label{fig:abstract-geom}}
\end{figure}

We assume that the structure is invariant along its longitudinal
direction in the reference configuration, i.e., it is a block in two
dimensions or a prismatic solid in three dimensions.  The reference
configuration does not need to be stress-free: naturally curved or
twisted elastic rods for instance can be handled.  The extension to
structures whose geometric or mechanical properties are not invariant
but slowly varying in the longitudinal direction is straightforward
and will be discussed at the end of the paper.  We denote by $S$ a
Lagrangian coordinate along the long dimension of the structure.  The
range of variation of $S$ is denoted as $0 \leq S \leq L$, where $L$
typically denotes the natural length of the structure.  The parameter
$S$ is used to label the cross-sections, see
figure~\ref{fig:abstract-geom}(a).

Let $\tmmathbf{E} (S)$ be the strain map over a particular
cross-section $S$ in the current configuration, as sketched in
figure~\ref{fig:abstract-geom}(b).  By strain {\tmem{map}}, we mean
that $\tmmathbf{E} (S)$ is the restriction to a particular
cross-section of the set of strain measures relevant to the particular
structural model.  If we are dealing with a hyperelastic cylinder, for
instance, $\tmmathbf{E} (S)$ collects all the strain components $\{
E_{S \nocomma S} (\cdummy, \cdummy), E_{S \nocomma X} (\cdummy,
\cdummy), E_{X \nocomma X} (\cdummy, \cdummy), E_{S \nocomma Y}
(\cdummy, \cdummy), E_{Y \nocomma Y} (\cdummy, \cdummy), E_{X \nocomma
Y} (\cdummy, \cdummy) \}$, each taking the cross-section coordinates
$(X, Y)$ as arguments.

Next, we introduce two mathematical objects in each cross-section: a
vector of \emph{macroscopic} strain $\tmmathbf{h} (S) = (h_1 (S), h_2
(S), \ldots)$ made up of the strain measures that will survive in the
1d model, and a set of \emph{microscopic} degrees of freedom
$\tmmathbf{y} (S)$ that will be ultimately be eliminated.  Their exact
definitions vary, but typically $\tmmathbf{h}(S)$ is the (apparent) 1d
strain, as calculated from the center line passing through the centers
of all the cross-section, while $\tmmathbf{y}(S)$ parameterizes the
deformation of the cross-section relative to the center line.
Typically, $\tmmathbf{h} (S)$ is a vector of low dimension, while
$\tmmathbf{y} (S)$ is a (collection of) functions defined over the
cross-sections, i.e., an infinite-dimensional vector.  For the
axisymmetric hyperelastic cylinder, for instance, $\tmmathbf{h}$ is
made up of a single entry, the axial stretch, while $\tmmathbf{y} (S)$
is the cross-sectional map of displacement.

Together, the macroscopic strain $\tmmathbf{h}(S)$ and microscopic
degrees of freedom $\tmmathbf{y}(S)$ determine the current
configuration of the structure (up to a rigid-body motion) hence the
microscopic strain $\tmmathbf{E}$.  Therefore, each particular
structural model prescribes a method for calculating the
cross-sectional strain map $\tmmathbf{E} (S)$ in terms of
$\tmmathbf{h}$, $\tmmathbf{y}$ and their longitudinal derivatives,
\begin{equation}
\tmmathbf{E} (S) =\tmmathbf{E} (\tmmathbf{h} (S), \tmmathbf{h}' (S) ;
\tmmathbf{y} (S), \tmmathbf{y}' (S), \tmmathbf{y}'' (S)) .
\label{eq:canonicalFormStrain}
\end{equation}
In the example of the cylinder, the longitudinal strain $E_{S \nocomma
S}$ depends on the longitudinal gradient of the displacement, hence
the dependence of $\tmmathbf{E}$ on $\tmmathbf{y}'$.  

Since $\tmmathbf{y} (S)$ is a function defined on the cross-sections,
the function $\tmmathbf{E}$ in the right-hand side
of~(\ref{eq:canonicalFormStrain}) is a \emph{functional} when the
domain of the cross-section is continuous.  Since $\tmmathbf{E}
(S)$ is a {\tmem{map}} of strain over the cross-section, any
dependence of the strain on the transverse gradients of displacement
is hidden in the definition of $\tmmathbf{E}$ above.  By contrast, we
make sure that the dependence on longitudinal gradients takes place
explicitly through the supplied argument $\tmmathbf{y}'$ (and possibly
$\tmmathbf{y}''$).  The additional dependence on $\tmmathbf{y}''$ will
allow us to handle the bending of plates or shells without change.  It
is easy to take into account an additional dependence of
$\tmmathbf{E}$ on higher-order gradients of $\tmmathbf{h} (S)$ or
$\tmmathbf{y} (S)$; this does not affect any of the results.

In terms of the strain map $\tmmathbf{E} (S)$, the structural model defines a
density of strain energy $W (\tmmathbf{E})$ per unit length $\mathd S$. The
strain energy of the structure therefore writes
\begin{equation}
  \Phi [\tmmathbf{h}, \tmmathbf{y}] = \int_0^L W (\tmmathbf{E} (\tmmathbf{h}
  (S), \tmmathbf{h}' (S) ; \tmmathbf{y} (S), \tmmathbf{y}' (S), \tmmathbf{y}''
  (S))) \mymult \mathd S, \label{eq:generalStrainEnergy}
\end{equation}
where the square brackets emphasize the functional dependence on the
arguments.

Some structural models are conveniently expressed by imposing
kinematic constraints $\tmmathbf{q} (\tmmathbf{y})= \tmmathbf{0}$ on
the microscopic displacement, where $\tmmathbf{q} (\tmmathbf{y}) =
(q_1 (\tmmathbf{y}), q_2 (\tmmathbf{y}), \ldots)$.  For structures
whose cross-section involve infinitely many degrees of freedom,
$\tmmathbf{y} (S)$ is (a set of) functions defined in the
cross-sections, i.e., $\tmmathbf{q}$ is a functional.  We focus
attention on kinematic constraints that are linear and independent of
$S$.  These assumptions can be relaxed easily.  For structures that
are free of kinematic constraints, we set $\tmmathbf{q}$ as the empty
vector, $\tmmathbf{q} (\tmmathbf{y}) = ()$, implying that any term
such as $\tmmathbf{q} (\tmmathbf{y}) \cdot \tmmathbf{x}= 0$ must be
discarded in the following.

We deal with dimension reduction by addressing the following relaxation
problem: the macroscopic strain $\tmmathbf{h} (S)$ is prescribed and we seek
the microscopic variables $\tmmathbf{y} (S)$ making the strain energy $\Phi
[\tmmathbf{h}, \tmmathbf{y}]$ stationary, subject to the kinematic constraint
\begin{equation}
  \forall S \quad \tmmathbf{q} (\tmmathbf{y} (S)) = 0. \label{eq:qConstraint}
\end{equation}
Our goal is to calculate the relaxed strain energy $\Phi^{\star}$ in
terms of the macroscopic strain $\tmmathbf{h} (S)$.  It can be
obtained by inserting the optimal microscopic displacement
$\tmmathbf{y} (S)$ into $\Phi$, as in
\begin{equation}
  \Phi^{\star} [\tmmathbf{h}] = \min_{\tmmathbf{y} \of (\forall S)
  \tmmathbf{q} (\tmmathbf{y} (S)) = 0} \Phi [\tmmathbf{h},
  \tmmathbf{y}] . \label{eq:relaxedEnergy}
\end{equation}
This paper derives an expansion of $\Phi^{\star} [\tmmathbf{h}]$ in
successive derivatives of $\tmmathbf{h}(S)$ using an asymptotic
method.  Note that the relaxed energy $\Phi^{\star} [\tmmathbf{h}]$ is
1d: it no longer makes any reference to the cross-sectional degrees of
freedom.  Once $\Phi^{\star} [\tmmathbf{h}]$ has been obtained, the
equilibrium equations for the 1d model can be derived using standard
variational techniques.

\subsection{Analysis of homogeneous solutions}

The first step in our analysis is to characterize homogeneous
solutions under finite strain.  To do so, we focus attention on the
case where both the macroscopic strain $\tmmathbf{h}= (h_1, h_2,
\ldots)$ and the microscopic displacement $\tmmathbf{y}$ are
independent~\footnote{Here, we are assuming that the splitting of the
microscopic strain $\tmmathbf{E}(S)$ into master ($\tmmathbf{h}(S)$)
and slave ($\tmmathbf{y}(S)$) degrees of freedom has been set up in
such a way that homogeneous solutions correspond to constant
$\tmmathbf{h}$ and constant $\tmmathbf{y}$.  Any reasonable choice of
$\tmmathbf{h}(S)$ and $\tmmathbf{y}(S)$ satisfies this property.} of
$S$. The strain for homogeneous solutions
$\tilde{\tmmathbf{E}} (\tmmathbf{h},\tmmathbf{y})$ is obtained by
setting $\tmmathbf{h}' =\tmmathbf{0}$, $\tmmathbf{y}' =\tmmathbf{0}$
and $\tmmathbf{y}'' =\tmmathbf{0}$ in~(\ref{eq:canonicalFormStrain}) as
\begin{equation}
  \tilde{\tmmathbf{E}} (\tmmathbf{h},\tmmathbf{y}) =\tmmathbf{E}
  (\tmmathbf{h}, \tmmathbf{0}; \tmmathbf{y}, \tmmathbf{0}, \tmmathbf{0}) .
  \label{eq:ETildeH}
\end{equation}
For a given value of the macroscopic strain $\tmmathbf{h}= (h_1, h_2,
\ldots)$, we seek the microscopic displacement
$\tmmathbf{y}=\tmmathbf{y}_{\tmmathbf{h}} =\tmmathbf{y}_{(h_1, h_2, \ldots)}$
such that the cross-sections are in equilibrium. To do so, we seek the
value(s) of $\tmmathbf{y}$ that make stationary the strain energy per unit
length $W (\tilde{\tmmathbf{E}} (\tmmathbf{h},\tmmathbf{y}))$, among those
satisfying the kinematic constraint $\tmmathbf{q} (\tmmathbf{y})$. This yields
the variational problem
\begin{equation}
  \left\{\begin{array}{l}
    \forall \hat{\tmmathbf{y}} \quad - \frac{\mathd W}{\mathd \tmmathbf{E}}(
    \tilde{\tmmathbf{E}} (\tmmathbf{h},\tmmathbf{y}_{\tmmathbf{h}}))\cdot \left(
    \frac{\partial \tilde{\tmmathbf{E}}}{\partial \tmmathbf{y}}
    (\tmmathbf{h},\tmmathbf{y}_{\tmmathbf{h}}) \cdot \hat{\tmmathbf{y}} \right)
    +\tmmathbf{f}_{\tmmathbf{h}} \cdot \tmmathbf{q} (\hat{\tmmathbf{y}}) = 0\\
    \tmmathbf{q} (\tmmathbf{y}_{\tmmathbf{h}}) =\tmmathbf{0},
  \end{array}\right. \label{eq:outlineHomPVW}
\end{equation}
where the unknown $\tmmathbf{f}_{\tmmathbf{h}}$ is a Lagrange
multiplier enforcing the constraint on the second line
($\tmmathbf{f}_h$ can be interpreted as the macroscopic load that is
required for the homogeneous solution to be globally in equilibrium,
such as a transverse external load in the case of a rod subject to a
combination of uniform tension and bending).  For structures whose
cross-sections define a continuous domain in the plane, $W
(\tmmathbf{E})$ is a functional taking on scalar values, and
$\frac{\mathd W}{\mathd \tmmathbf{E}} (\tmmathbf{E}) \cdot \delta
\tmmathbf{E}$ denotes the G{\^a}teaux derivative at $\tmmathbf{E}$ in
the direction $\delta \tmmathbf{E}$.  For structures possessing
discrete cross-sectional degrees of freedom, $\frac{\mathd W}{\mathd
\tmmathbf{E}} (\tmmathbf{E})$ the gradient of the function $W
(\tmmathbf{E})$.

\begin{figure}
  \centerline{\includegraphics{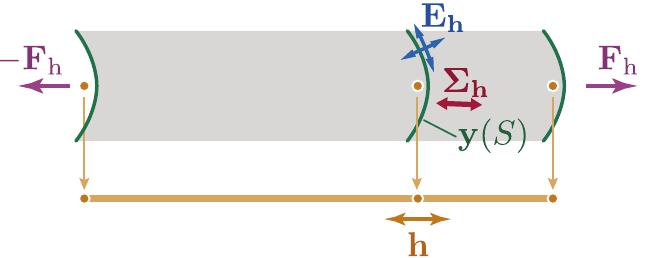}}
  \caption{A homogeneous solution with uniform macroscopic strain
  $\tmmathbf{h}$: microscopic displacement $\tmmathbf{y}_{\tmmathbf{h}}$,
  microscopic strain $\tmmathbf{E}_{\tmmathbf{h}}$ and microscopic stress
  $\tmmathbf{\Sigma}_{\tmmathbf{h}}$. Note that the we are not interested at
  this stage in calculating the external loading $\tmmathbf{F}_{\tmmathbf{h}}$
  that maintains equilibrium with respect to the macroscopic
  variables.\label{fig:homogeneous-geom}}
\end{figure}

Equation~(\ref{eq:outlineHomPVW}) warrants stationarity with respect to
the microscopic displacement, but not with respect to the macroscopic
strain.  For a solution of these equations to represent an actual
equilibrium, one would need to set up macroscopic forces conjugate to
the macroscopic strain, labeled $\tmmathbf{F}_{\tmmathbf{h}}$ in
figure~\ref{fig:homogeneous-geom}.  If the structure is an elastic
cylinder, for instance, equation~(\ref{eq:outlineHomPVW}) imposes the
contraction of cross-sections by Poisson's effect; to maintain the
global equilibrium, a macroscopic tensile load, not discussed here,
would be required.  Macroscopic load do not enter into the dimension
reduction process: they can be introduced directly in the 1d model,
after the dimension reduction.

Equation~(\ref{eq:outlineHomPVW}) is a non-linear elasticity problem
defined on the cross-section: the longitudinal variable has been
removed.  This problem can be solved, most often analytically (see the
examples in the following sections) or in some cases numerically.  By
solving equation~(\ref{eq:outlineHomPVW}) for
$\tmmathbf{y}_{\tmmathbf{h}}$ and $\tmmathbf{f}_{\tmmathbf{h}}$ for
any value of the macroscopic strain $\tmmathbf{h}$, one obtains a
{\tmem{catalog}} of homogeneous solutions, which is at the heart of
the dimension reduction method.  It is derived without any
approximation: the catalog is made up of {\tmem{nonlinear}} solutions.

In terms of the catalog of microscopic displacement
$\tmmathbf{y}_{\tmmathbf{h}}$, we can define the homogeneous strain
$\tmmathbf{E}_{\tmmathbf{h}}$, the homogeneous strain energy density
$W_{\text{hom}} (\tmmathbf{h})$, the homogeneous pre-stress
$\tmmathbf{\Sigma}_{\tmmathbf{h}}$, and the homogeneous tangent stiffness, as
follows,
\begin{equation}
  \begin{array}{rll}
    \tmmathbf{E}_{\tmmathbf{h}} & = & \tilde{\tmmathbf{E}} (\tmmathbf{h},\tmmathbf{y}_{\tmmathbf{h}})\\
    W_{\text{hom}} (\tmmathbf{h}) & = & W (\tmmathbf{E}_{\tmmathbf{h}})\\
    \tmmathbf{\Sigma}_{\tmmathbf{h}} & = & \frac{\mathd W}{\mathd
    \tmmathbf{E}} (\tmmathbf{E}_{\tmmathbf{h}})\\
    \tmmathbf{K}_{\tmmathbf{h}} & = & \frac{\mathd^2 W}{\mathd \tmmathbf{E}^2}
    (\tmmathbf{E}_{\tmmathbf{h}}) .
  \end{array} \label{eq:defineHomogeneousProperties}
\end{equation}
In our notation, the homogeneous quantities are either subscripted with the
letters $\text{`hom'}$, as in $W_{\text{hom}}$, or simply by the vector of
macroscopic strain $\tmmathbf{h}= (h_1, h_2, \ldots)$ on which they depend.

\subsection{Reduced models without gradient
effect}\label{ssec:traditionalModels}

Most 1d models used for slender structures depend on macroscopic
strain variables, but not on their gradients. The Euler-Bernoulli rod model,
for instance, depends on curvature and twist and not on their gradients. These
standard structural models  are governed by the strain energy
\[ \Phi^{\star} [\tmmathbf{h}] \approx \int_0^L W_{\text{hom}} (\tmmathbf{h}
   (S)) \mymult \mathd S \qquad \text{(reduction without strain 
   gradient)}, \]
and can therefore be derived directly from the catalog of homogeneous
solutions.  If we start from an elastic block, for example, and choose
the axial strain and curvature as macroscopic variables, the strain
energy $\int_0^L W_{\text{hom}} (\tmmathbf{h}(S)) \mymult \mathd S$
defines a classical beam model
(see~{\textsection}\ref{ssec:blockDiscussion}).  Note that the 1d
model associated with the energy functional $\Phi^{\star}
[\tmmathbf{h}]$ above might suffer from poorer convexity properties
than the original 3d model; this happens typically when a string model
is derived (\emph{i.e.}, when $\tmmathbf{h}(S)$ is set up to include
just an axial strain variable, so that there is no bending energy in
the resulting 1d model), and in this case an additional relaxation
step is needed to remove the unphysical part of the constitutive law
predicting axial
compression~\citep{Acerbi-Buttazzo-EtAl-A-variational-definition-of-the-strain-1991}.

So far, our method carries out dimension reduction without strain
gradient.  It does so without using any kinematic assumption and works
under very general conditions: no material symmetry has been assumed,
and it can handle inhomogeneous cross-sections and nonlinear elastic
materials.

\subsection{Microscopic correction, energy
expansion\label{ssec:energyExpansionMicroscopicCorrection}}

We return to the main focus of our work, which is on capturing
{\tmem{strain-gradient}} effects.  Given a distribution of macroscopic
strain $\tmmathbf{h} (S)$ with $0 \leq S \leq L$, we aim at
calculating the optimal microscopic displacement $\tmmathbf{y} (S)$
and, thus, the relaxed energy $\Phi^{\star}$ appearing in
equation~(\ref{eq:relaxedEnergy}).  We do so by assuming slow
variations in the longitudinal direction: a proper stretched variable
is introduced in the detailed proof of \ref{sec:detailedProofs}, but
it will suffice here to assume that the successive longitudinal
derivatives of quantities such as $\tmmathbf{h} (S)$ scale like
$\tmmathbf{h}=\mathcal{O} (1)$, $\tmmathbf{h}' =\mathcal{O} (\gamma)$,
$\tmmathbf{h}'' =\mathcal{O} (\gamma^2)$, etc., where $\gamma \ll 1$
is a slenderness parameter.


We seek the microscopic displacement $\tmmathbf{y} (S)$ that achieves
the optimum in equation~(\ref{eq:relaxedEnergy}) in the form:
\begin{equation}
\tmmathbf{y} (S) =\tmmathbf{y}_{\tmmathbf{h} (S)} +\tmmathbf{z} (S) . \label{eq:guess}
\end{equation}
In words, we use the leading order microscopic displacement
$\tmmathbf{y}_{\tmmathbf{h} (S)}$ obtained by looking up our catalog
of homogeneous solutions $\tmmathbf{h} \mapsto
\tmmathbf{y}_{\tmmathbf{h}}$ as a first approximation; this look-up is
performed with the parameter $\tmmathbf{h}$ set to the local
prescribed value of the macroscopic strain, \
$\tmmathbf{h}=\tmmathbf{h} (S)$.  We refine this approximation by a
correction $\tmmathbf{z} (S)$ proportional to the gradient term
$\tmmathbf{h}'$, which we calculate next.

To reflect the change of unknown from $\tmmathbf{y} (S)$ to
$\tmmathbf{z} (S)$ in~(\ref{eq:guess}), let us first define the
function $\tmmathbf{e}_{\tmmathbf{h}}$ that yields the
strain as in~(\ref{eq:canonicalFormStrain}):
\begin{equation}
  \tmmathbf{e}_{\tmmathbf{h}} (\tmmathbf{h}^{\dag}, \tmmathbf{h}^{\ddag} ;
  \tmmathbf{z}, \tmmathbf{z}^{\dag}, \tmmathbf{z}^{\ddag}) =\tmmathbf{E}
  (\tmmathbf{h}, \tmmathbf{h}^{\dag} ; \tmmathbf{y}_{\tmmathbf{h}}
  +\tmmathbf{z}, \tmmathbf{h}^{\dag} \cdot \nabla \tmmathbf{y}_{\tmmathbf{h}}
  +\tmmathbf{z}^{\dag}, \tmmathbf{h}^{\ddag} \cdot \nabla
  \tmmathbf{y}_{\tmmathbf{h}} +\tmmathbf{h}^{\dag} \cdot \nabla^2
  \tmmathbf{y}_{\tmmathbf{h}} \cdot \tmmathbf{h}^{\dag} +\tmmathbf{z}^{\ddag})
  . \label{eq:strainInTermsOfZ}
\end{equation}
The variables bearing a dag ($\dag$) or a double dag $(\ddag)$ are
those that will be set later to the local value of the first or second
gradients.  The quantity $\tmmathbf{h}^{\dag}$, for instance, is a
dummy variable that will be set later to $\tmmathbf{h}^{\dag} =\tmmathbf{h}'
(S)$.  Besides, the $\nabla$ in
equation~(\ref{eq:strainInTermsOfZ}) stands for gradients with respect
to the macroscopic strain $\tmmathbf{h}$,
\begin{equation}
  \nabla^k \tmmathbf{y}_{\tmmathbf{h}} = \frac{\mathd^k
  \tmmathbf{y}_{\tmmathbf{h}}}{\mathd \tmmathbf{h}^k} .
  \label{eq:nablaDefinition}
\end{equation}

Anticipating on the fact that we will need to expand the strain
in~(\ref{eq:strainInTermsOfZ}), we define the {\tmem{structure
coefficients}} $\tmmathbf{e}^{i \nocomma j}_{k \nocomma l \nocomma m}
(\tmmathbf{h})$ as the gradients of $\tmmathbf{e}_{\tmmathbf{h}}$,
evaluated at a homogeneous solution: for any set of integers $(i, j,
k, l, m)$,
\begin{equation}
  \tmmathbf{e}^{i \nocomma j}_{k \nocomma l \nocomma m} (\tmmathbf{h}) =
  \frac{\partial^{(i + j + k + l + m)} \tmmathbf{e}_{\tmmathbf{h}}}{[\partial
  \tmmathbf{h}^{\dag}]^i \mymult [\partial \tmmathbf{h}^{\ddag}]^j \mymult
  [\partial \tmmathbf{z}]^k \mymult [\partial \tmmathbf{z}^{\dag}]^l \mymult
  [\partial \tmmathbf{z}^{\ddag}]^m} (\tmmathbf{0}, \tmmathbf{0};
  \tmmathbf{0}, \tmmathbf{0}, \tmmathbf{0}) . \label{eq:structureCoefs}
\end{equation}
Note that the upper set of indices correspond to gradients with
respect to the gradients of the macroscopic strain parameters
$\tmmathbf{h}^{\dag}$ and $\tmmathbf{h}^{\ddag}$ while the lower set
of indices correspond to gradients with respect to the microscopic
variable $\tmmathbf{z}$ and its gradients $\tmmathbf{z}^{\dag}$ and
$\tmmathbf{z}^{\ddag}$.  The quantities $\tmmathbf{e}^{i \nocomma
j}_{k \nocomma l \nocomma m} (\tmmathbf{h})$ are either tensors, or
operators (if the cross-sectional degrees of freedom are continuous
and at least one integer among $k$, $l$, $m$ are non-zero): they will
always appear contracted $i$ times with $\tmmathbf{h}^{\dag}$, $j$
times with $\tmmathbf{h}^{\ddag}$, $k$ times with $\tmmathbf{z}$, etc.
Each one of these contractions will be denoted by a dot, representing
either the standard contraction of tensors or the application of the
operator. 

The Taylor expansion of the strain~(\ref{eq:strainInTermsOfZ}) near a
homogeneous solution can be written in terms of the structure
coefficients as
\begin{equation*}
\tmmathbf{e}_{\tmmathbf{h}} (\tmmathbf{h}^{\dag}, \tmmathbf{h}^{\ddag} ;
\tmmathbf{z}, \tmmathbf{z}^{\dag}, \tmmathbf{z}^{\ddag}) =   \tmmathbf{e}_{\tmmathbf{h}} (\tmmathbf{0}, \tmmathbf{0};
\tmmathbf{0}, \tmmathbf{0}, \tmmathbf{0})+
\tmmathbf{e}^{1 \nocomma 0}_{0
\nocomma 0 \nocomma 0} (\tmmathbf{h}) \cdot \tmmathbf{h}^{\dag}
+\tmmathbf{e}^{0 \nocomma 0}_{1 \nocomma 0 \nocomma 0} (\tmmathbf{h})
\cdot \tmmathbf{z} + 
\frac{1}{2}\,\left(
2\,\tmmathbf{h}^{\dag} \cdot \tmmathbf{e}^{1 \nocomma 0}_{1
	\nocomma 0 \nocomma 0} (\tmmathbf{h}) \cdot 
	\tmmathbf{z}+\cdots\right)+\cdots
\end{equation*}
Structure coefficients will be calculated explicitly in the second
part of the paper, when explicit structures are considered.

In terms of the structure coefficients, we further introduce the following
operators,
\begin{equation}
  \begin{array}{rcl}
    \tmmathbf{A}_{\tmmathbf{h}} \cdot \tmmathbf{h}^{\dag} & = &
    \tmmathbf{\Sigma}_{\tmmathbf{h}} \cdot (\tmmathbf{e}^{1 \nocomma 0}_{0
    \nocomma 0 \nocomma 0} (\tmmathbf{h}) \cdot \tmmathbf{h}^{\dag})\\
    \tmmathbf{C}_{\tmmathbf{h}}^{(0)} \cdot \tmmathbf{h}^{\ddag} & = &
    \tmmathbf{\Sigma}_{\tmmathbf{h}} \cdot (\tmmathbf{e}^{0 \nocomma 1}_{0
    \nocomma 0 \nocomma 0} (\tmmathbf{h}) \cdot \tmmathbf{h}^{\ddag})\\
    \tmmathbf{C}_{\tmmathbf{h}}^{(1)} \cdot \tmmathbf{z}^{\dag} & = &
    \tmmathbf{\Sigma}_{\tmmathbf{h}} \cdot (\tmmathbf{e}^{0 \nocomma 0}_{0
    \nocomma 1 \nocomma 0} (\tmmathbf{h}) \cdot \tmmathbf{z}^{\dag})\\
    \frac{1}{2} \mymult \tmmathbf{h}^{\dag} \cdot
    \tmmathbf{B}_{\tmmathbf{h}}^{(0)} \cdot \tmmathbf{h}^{\dag} & = &
    \frac{1}{2} \mymult (\tmmathbf{e}^{1 \nocomma 0}_{0 \nocomma 0 \nocomma 0}
    (\tmmathbf{h}) \cdot \tmmathbf{h}^{\dag}) \cdot
    \tmmathbf{K}_{\tmmathbf{h}} \cdot (\tmmathbf{e}^{1 \nocomma 0}_{0 \nocomma
    0 \nocomma 0} (\tmmathbf{h}) \cdot \tmmathbf{h}^{\dag}) + \frac{1}{2}
    \mymult \tmmathbf{\Sigma}_{\tmmathbf{h}} \cdummy (\tmmathbf{h}^{\dag} \cdot
    \tmmathbf{e}^{2 \nocomma 0}_{0 \nocomma 0 \nocomma 0} (\tmmathbf{h}) \cdot
    \tmmathbf{h}^{\dag}) -\tmmathbf{h}^{\dag} \cdot \nabla
    \tmmathbf{C}_{\tmmathbf{h}}^{(0)} \cdot \tmmathbf{h}^{\dag}\\
    \tmmathbf{h}^{\dag} \cdot \tmmathbf{B}_{\tmmathbf{h}}^{(1)} \cdot
    \tmmathbf{z} & = & (\tmmathbf{e}^{1 \nocomma 0}_{0 \nocomma 0 \nocomma 0}
    (\tmmathbf{h}) \cdot \tmmathbf{h}^{\dag}) \cdot
    \tmmathbf{K}_{\tmmathbf{h}} \cdot (\tmmathbf{e}^{0 \nocomma 0}_{1 \nocomma
    0 \nocomma 0} (\tmmathbf{h}) \cdot \tmmathbf{z}_{})
    +\tmmathbf{\Sigma}_{\tmmathbf{h}} \cdot (\tmmathbf{h}^{\dag} \cdot
    \tmmathbf{e}^{1 \nocomma 0}_{1 \nocomma 0 \nocomma 0} (\tmmathbf{h}) \cdot
    \tmmathbf{z}) -\tmmathbf{h}^{\dag} \cdot \nabla
    \tmmathbf{C}_{\tmmathbf{h}}^{(1)} \cdot \tmmathbf{z}\\
    \frac{1}{2} \mymult \tmmathbf{z} \cdot \tmmathbf{B}_{\tmmathbf{h}}^{\text{}
    (2)} \cdot \tmmathbf{z} & = & \frac{1}{2} \mymult (\tmmathbf{e}^{0 \nocomma
    0}_{1 \nocomma 0 \nocomma 0} (\tmmathbf{h}) \cdot \tmmathbf{z}) \cdot
    \tmmathbf{K}_{\tmmathbf{h}} \cdot (\tmmathbf{e}^{0 \nocomma 0}_{1 \nocomma
    0 \nocomma 0} (\tmmathbf{h}) \cdot \tmmathbf{z}) + \frac{1}{2} \mymult
    \tmmathbf{\Sigma}_{\tmmathbf{h}} \cdummy (\tmmathbf{z} \cdot
    \tmmathbf{e}^{0 \nocomma 0}_{2 \nocomma 0 \nocomma 0} (\tmmathbf{h}) \cdot
    \tmmathbf{z}).
  \end{array} \label{eq:preliminaryOperatorsDef}
\end{equation}
They depend on the (local) macroscopic strain $\tmmathbf{h}$.  They
operate on the cross-sectional degrees of freedom $\tmmathbf{z}$ and
$\tmmathbf{z}^{\dag}$ (but not $\tmmathbf{z}^{\ddag}$) and on the
local values of the derivatives $\tmmathbf{h}^{\dag}$ and
$\tmmathbf{h}^{\ddag}$ of the macroscopic strain.

As shown in \ref{sec:detailedProofs}, the expansion of the energy
$\Phi [\tmmathbf{h}, \tmmathbf{y}_{\tmmathbf{h}} +\tmmathbf{z}]$ in
powers of the successive gradients of macroscopic strain can be
expressed in terms of these operators as
\begin{multline}
    \Phi [\tmmathbf{h}, \tmmathbf{y}_{\tmmathbf{h}} +\tmmathbf{z}] = \int_0^L
    W_{\text{hom}} (\tmmathbf{h} (S)) \mymult \mathd S + \int_0^L
    \tmmathbf{A}_{\tmmathbf{h} (S)} \cdot \tmmathbf{h}' (S) \mymult \mathd S
    \cdots\\
    {} + [\tmmathbf{C}_{\tmmathbf{h}}^{(0)} \cdot
    \tmmathbf{h}' +\tmmathbf{C}_{\tmmathbf{h}}^{(1)} \cdot \tmmathbf{z}]_{S =
    0}^L + \int_0^L \left( \frac{1}{2} \mymult \tmmathbf{h}' \cdot
    \tmmathbf{B}_{\tmmathbf{h}}^{(0)} \cdot \tmmathbf{h}' +\tmmathbf{h}' \cdot
    \tmmathbf{B}_{\tmmathbf{h}}^{(1)} \cdot \tmmathbf{z}+ \frac{1}{2} \mymult
    \tmmathbf{z} \cdot \tmmathbf{B}_{\tmmathbf{h}}^{\left( \text{} 2 \right)}
    \cdot \tmmathbf{z} \right)_S \mymult \mathd S \cdots
	\\
	{}
	+\mathcal{O} (| \tmmathbf{h}'
    |^3, | \tmmathbf{h}'' | | \tmmathbf{h}' |^2, | \tmmathbf{h}''' |) .
	\label{eq:PhiExpansionAnnounce} 
\end{multline}
In the boundary term in square brackets, both the arguments
$\tmmathbf{h}$ in subscript of the operators and the operands
$\tmmathbf{h}'$ and $\tmmathbf{z}$ must be evaluated at $S = 0$ and $S
= L$, respectively.  Likewise in the integrand on the second line, the
quantities $\tmmathbf{h}$, $\tmmathbf{h}'$ and $\tmmathbf{z}$ must be
evaluated at the current point $S$. 

The form of the strain gradient model above is similar to that derived
in different contexts, see for example
in~\cite{Bardenhagen-Triantafyllidis-Derivation-of-higher-order-1994};
our main contribution is a method for calculating the coefficients
$\tmmathbf{A}_{\tmmathbf{h} (S)}$,
$\tmmathbf{B}_{\tmmathbf{h}}^{(1)}$, etc.\ explicitly.

\subsection{Optimal correction}\label{ssec:generalMethodOptimalCorrection}

The last step in the reduction process is to determine the correction
$\tmmathbf{z} (S)$ such that the microscopic
displacement~(\ref{eq:guess}) satisfies the optimality
condition~(\ref{eq:relaxedEnergy}).  All derivatives of the unknown
$\tmmathbf{z} (S)$ can be eliminated from
equation~(\ref{eq:PhiExpansionAnnounce}), thanks to an integration by
parts, as shown in \ref{ssec:eliminateZDot}.  The benefit is that the
relaxation of the unknown $\tmmathbf{z}$ leads to a {\tmem{local}}
problem in the cross-sections: as established in
\ref{sec:detailedProofs}, the optimal correction $\tmmathbf{z} (S)$
$S$ is
\[ \tmmathbf{z} (S) =\tmmathbf{z}_{\text{opt}} (S) +\mathcal{O} (|
   \tmmathbf{h}' |^2), \]
where the dominant contribution $\tmmathbf{z}_{\text{opt}}
=\mathcal{O} (| \tmmathbf{h}' |)$ is the one that minimizes the local
elastic potential $\tmmathbf{z} \mapsto \left( \frac{1}{2} \mymult
\tmmathbf{h}' (S) \cdot \tmmathbf{B}_{\tmmathbf{h} (S)}^{(0)} \cdot
\tmmathbf{h}' (S) +\tmmathbf{h}' (S) \cdot \tmmathbf{B}_{\tmmathbf{h}
(S)}^{(1)} \cdot \tmmathbf{z}+ \frac{1}{2} \mymult \tmmathbf{z} \cdot
\tmmathbf{B}_{\tmmathbf{h} (S)}^{\left( \text{} 2 \right)} \cdot
\tmmathbf{z} \right)$, subject to the constraint $\tmmathbf{q}
(\tmmathbf{z}) = 0$.  The correction $\tmmathbf{z}_{\text{opt}} (S)$
is therefore the solution to the following variational problem,
\begin{equation}
  \left\{\begin{array}{l}
    \forall \hat{\tmmathbf{z}} \quad \tmmathbf{h}' (S) \cdot
    \tmmathbf{B}_{\tmmathbf{h} (S)}^{(1)} \cdot \hat{\tmmathbf{z}}
    +\tmmathbf{z}_{\text{opt}} (S) \cdot \tmmathbf{B}_{\tmmathbf{h}
    (S)}^{\text{} (2)} \cdot \hat{\tmmathbf{z}} -\tmmathbf{f}_{\text{opt}} (S)
    \cdot \tmmathbf{q} (\hat{\tmmathbf{z}}) = 0\\
    \tmmathbf{q} \left( \tmmathbf{z}_{\text{opt}} (S) \right) =\tmmathbf{0},
  \end{array}\right. \label{eq:localOptimizationProblem}
\end{equation}
where $\tmmathbf{f}_{\text{opt}} (S)$ is a Lagrange multiplier, to be
determined as part of the solution process.

This variational problem is linear with respect to the local value of
the strain gradient $\tmmathbf{h}' (S)$.  This implies that its
solution $\tmmathbf{z}_{\text{opt}} (S)$ is proportional to
$\tmmathbf{h}' (S)$, i.e., there exists a catalog of corrections
$\tmmathbf{Z}_{\text{opt}}^{\tmmathbf{h} (S)}$ such that
\begin{equation}
  \tmmathbf{z}_{\text{opt}} (S) =\tmmathbf{Z}_{\text{opt}}^{\tmmathbf{h} (S)}
  \cdot \tmmathbf{h}' (S) . \label{eq:correctiveDisplacementFactorOutHPrime}
\end{equation}
The catalog $\tmmathbf{Z}_{\text{opt}}^{\tmmathbf{h}}$ is found by
solving~(\ref{eq:localOptimizationProblem}).  It can be determined
once for all in terms of the geometric and mechanical properties of a
reference cross-section and in terms of the macroscopic strain 
$\tmmathbf{h}$, as we show in the examples.

Equation~(\ref{eq:localOptimizationProblem}) is a problem of linear
elasticity in the cross-section.  The first term $\tmmathbf{h}' (S)
\cdot \tmmathbf{B}_{\tmmathbf{h} (S)}^{(1)} \cdot \hat{\tmmathbf{z}}$
can be interpreted as a pre-stress arising from the presence a
gradient (an interpretation of this pre-stress term will be obtained
based on the analysis of specific structures, see~\S\ref{ssec:localOptimizationNecking} in particular).  The
second term $\tmmathbf{z}_{\text{opt}} (S) \cdot
\tmmathbf{B}_{\tmmathbf{h} (S)}^{\text{} (2)} \cdot
\hat{\tmmathbf{z}}$ is an elastic stiffness term which, in view of the
definition of $\tmmathbf{B}_{\tmmathbf{h} (S)}^{\text{} (2)}$ in
equation~(\ref{eq:preliminaryOperatorsDef}) has two contributions: a
tangent elastic stiffness $\tmmathbf{K}_{\tmmathbf{h} (S)}$, and a
geometric stiffness arising from the pre-stress
$\tmmathbf{\Sigma}_{\tmmathbf{h} (S)}$ associated with the local state
of stress.

\subsection{Relaxed energy}

The relaxed energy $\Phi^{\star} [\tmmathbf{h}]$ is finally obtained by
inserting the optimal displacement $\tmmathbf{y} (S)
=\tmmathbf{y}_{\tmmathbf{h} (S)} +\tmmathbf{z}_{\text{opt}} (S) + \cdots$ into
the energy expansion in equation~(\ref{eq:PhiExpansionAnnounce}). The result
is
\begin{equation}
  \Phi^{\star} [\tmmathbf{h}] = \int_0^L W_{\text{hom}} (\tmmathbf{h} (S))
  \mymult \mathd S + \int_0^L \tmmathbf{A}_{\tmmathbf{h} (S)} \cdot
  \tmmathbf{h}' (S) \mymult \mathd S + [\tmmathbf{C}_{\tmmathbf{h} (S)} \cdot
  \tmmathbf{h}' (S)]_0^L + \frac{1}{2} \mymult \int_0^L \tmmathbf{h}' (S) \cdot
  \tmmathbf{B}_{\tmmathbf{h} (S)} \cdot \tmmathbf{h}' (S) \mymult \mathd S +
  \ldots \label{eq:RelaxedEnergy}
\end{equation}
Here, the operator $\tmmathbf{A}_{\tmmathbf{h}}$ has been introduced in
equation~(\ref{eq:preliminaryOperatorsDef}) and the additional elastic moduli
$\tmmathbf{B}_{\tmmathbf{h}}$ and $\tmmathbf{C}_{\tmmathbf{h}}$ are defined by

\begin{equation}
  \begin{array}{rcl}
    \tmmathbf{B}_{\tmmathbf{h}} & = & \tmmathbf{B}_{\tmmathbf{h}}^{(0)} -
    \left( \tmmathbf{Z}^{\tmmathbf{h}}_{\text{opt}} \right)^T \cdot
    \tmmathbf{B}_{\tmmathbf{h}}^{\text{} (2)} \cdot
    \tmmathbf{Z}^{\tmmathbf{h}}_{\text{opt}},\\
    \tmmathbf{C}_{\tmmathbf{h}} & = & \tmmathbf{C}_{\tmmathbf{h}}^{(0)}
    +\tmmathbf{C}_{\tmmathbf{h}}^{(1)} \cdot
    \tmmathbf{Z}_{\text{opt}}^{\tmmathbf{h}}.
  \end{array} \label{eq:finalOperatorsDef}
\end{equation}

The energy functional in equation~(\ref{eq:RelaxedEnergy}) and the
explicit expression for the strain-gradient modulus
$\tmmathbf{B}_{\tmmathbf{h}}$ are the main results of this paper.

In equation~(\ref{eq:RelaxedEnergy}), the leading order term in the
expansion depends $W_{\text{hom}}$, and defines structural models
without the gradient effect, see \S~\ref{ssec:traditionalModels}.  The
second term depending on $\tmmathbf{A}_{\tmmathbf{h}}$ yields an
energy contribution that is linear with respect to the gradient
$\tmmathbf{h}'$: it is zero in most cases due to symmetry reasons, as
shown in the forthcoming examples.  The terms depending on
$\tmmathbf{C}_{\tmmathbf{h}}$ is a boundary term arising from a
gradient effect, while the last term is the bulk strain-gradient term.

For further reference, we note that the strain gradient term is available in
alternative form as
\[ \frac{1}{2} \mymult \int_0^L \tmmathbf{h}' (S) \cdot
   \tmmathbf{B}_{\tmmathbf{h} (S)} \cdot \tmmathbf{h}' (S) \mymult \mathd S =
   \frac{1}{2} \mymult \int_0^L \tmmathbf{h}' (S) \cdot
   \tmmathbf{B}_{\tmmathbf{h} (S)}^{(0)} \cdot \tmmathbf{h}' (S) \mymult \mathd
   S - \frac{1}{2} \mymult \int_0^L \tmmathbf{z}_{\text{opt}} \cdot
   \tmmathbf{B}_{\tmmathbf{h}}^{\text{} (2)} \cdot \tmmathbf{z}_{\text{opt}}
   \mymult \mathd S. \]
\subsection{A necessary stability condition at the microscopic
scale}\label{ssec:generalStability}

A necessary condition for the microscopic correction derived in
section~\ref{ssec:generalMethodOptimalCorrection} to be stable (and, hence,
for the relaxed energy $\Phi^{\star}$ to be meaningful) is that the stiffness
operator $\tmmathbf{B}_{\tmmathbf{h}}^{\text{} (2)}$ appearing in the
microscopic problem in equation~(\ref{eq:localOptimizationProblem}) is
non-negative,
\begin{equation}
  \left( \forall \tmmathbf{z} \text{ such that $\tmmathbf{q} (\tmmathbf{z})
  =\tmmathbf{0}$} \right) \quad \tmmathbf{z} \cdot
  \tmmathbf{B}_{\tmmathbf{h}}^{\text{} (2)} \cdot \tmmathbf{z} \geq 0.
  \label{eq:stabilityCondition}
\end{equation}
Note that this condition does not warrant that the matrix
$\tmmathbf{B}_{\tmmathbf{h}}$ of strain-gradient moduli is
non-negative, see equation~(\ref{eq:finalOperatorsDef}) (a matrix
$\tmmathbf{B}_{\tmmathbf{h}}$ having negative eigenvalues is indeed
obtained for the elastic block,
see~{\textsection}\ref{ssec:blockDiscussion}).  However,
equation~(\ref{eq:stabilityCondition}) does warrant
\[ \frac{1}{2} \mymult \tmmathbf{h}' (S) \cdot \tmmathbf{B}_{\tmmathbf{h} (S)}
   \cdot \tmmathbf{h}' (S) \leq \frac{1}{2} \mymult \tmmathbf{h}' (S) \cdot
   \tmmathbf{B}_{\tmmathbf{h}}^{(0)} \cdot \tmmathbf{h}' (S),
   \label{eq:ColemanNewmanInequality} \]
which, as discussed in section~\ref{sec:conclusion}, shows that our 1d
model relaxes the elastic energy better than strain gradient models
derived from the \emph{ad hoc} kinematic assumption $\tmmathbf{z}
(S)=\tmmathbf{0}$: this benefit is a consequence of the fact that our
1d model is asymptotically exact.

\section{Application to an axisymmetric membrane}\label{sec:membranes}

Upon inflation, axisymmetric rubber membranes feature localized
deformations in the form of propagating
bulges~{\citep{Kyriakides-Chang-The-initiation-and-propagation-of-a-localized-1991}}.
Standard dimension reduction without gradient terms yields a
non-convex elastic potential $W_{\text{hom}}$ and thus fails at
describing the details of localization.  Localized solutions can be
analyzed using the full membrane
model~{\citep{Fu-Pearce-EtAl-Post-bifurcation-analysis-of-a-thin-walled-2008,Pearce-Fu-Characterization-and-stability-of-localized-2010}},
but are more easily and very accurately described based on a 1d
strain-gradient model, as recently shown by the authors, starting from
the theory of axisymmetric elastic membranes and using a typical
constitutive law for
rubber~{\citep{Lestringant-Audoly-A-diffuse-interface-model-2018}}.
This 1d model is rederived here as a first illustration of the general
reduction method presented in section~\ref{sec:mainResults}.

\subsection{Full axisymmetric membrane model}\label{eq:membraneFullModel}

\begin{figure}
  \centerline{\includegraphics{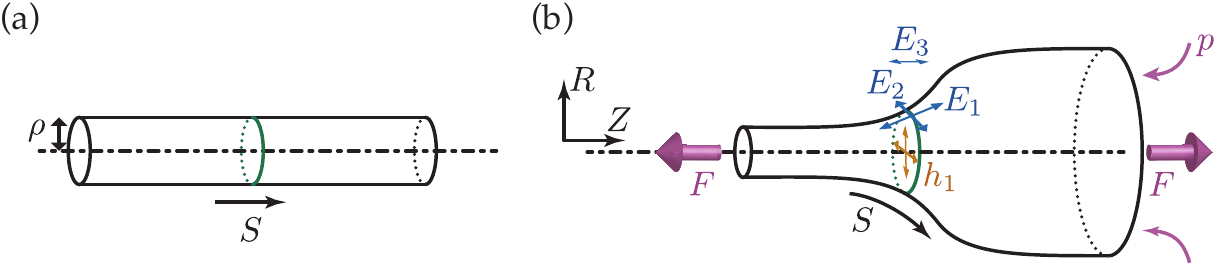}}
  \caption{An axisymmetric membrane: \ (a)~reference and~(b)~current
  configurations.\label{fig:balloon-geom}}
\end{figure}

The reference configuration is chosen as the natural, cylindrical
configuration of the membrane, and the natural radius of the circular
membrane is denoted by $\rho$.  In the current configuration, the
membrane is deformed under the action of an inflating pressure $p$,
and a pulling force $F$ equally distributed over the terminal
cross-sections, see figure~\ref{fig:balloon-geom}.  Natural boundary
conditions are used, i.e., there is no restraint on the terminal
cross-sections.

An axisymmetric configuration of the membrane is parameterized by two
functions $Z (S)$ and $R (S)$, such that the cross-section with
arc-length coordinate $S$ in the reference configuration is
transformed into a circle perpendicular to the axis of the shell, with
axial coordinate $Z (S)$ and radius $R (S)$, see
figure~\ref{fig:balloon-geom}(b).  We consider a standard set of
strain measures from the theory of finite-strain axisymmetric elasticity,
$\tmmathbf{E}= \left(\begin{array}{c|c|c} \sqrt{Z^{\prime 2} +
R^{\prime 2}} & \frac{R}{\rho} & Z' \end{array}\right)$: $E_1 =
\sqrt{Z^{\prime 2} + R^{\prime 2}}$ and $E_2 = \frac{R}{\rho}$,
usually denoted as $(E_1, E_2) = (\lambda_S, \lambda_{\Theta})$, are
the membrane stretches in the (principal) longitudinal and
circumferential directions, respectively.  The additional `strain'
$E_3$ has been included for convenience, as it allows us to write the
potential energy of the pulling force $F$ as $- F \mymult [Z (S)]_0^L
= - F \int_0^L E_3 \mymult \mathd S$.

The sum of the membrane strain energy, and the potential energy of the
loads $p$ and $F$ is captured by an effective potential $W
(\tmmathbf{E})$ per unit length $\mathd S$,
\[ W (\tmmathbf{E}) = \overline{W} (E_1, E_2) - p \mymult \pi \mymult \rho^2
   \mymult E_2^2 \mymult E_3 - F \mymult E_3, \]
where $\overline{W} (E_1, E_2) = \overline{W} (\lambda_S,
\lambda_{\Theta})$ is the strain energy of the hyperelastic membrane
model (we use bars generally for quantities relating to the full
model).  Upon integration with respect to $S$, the second term yields
$(- p)$ times the volume enclosed by the membrane, which is the
potential energy of the pressure force.
Note that we have chosen to include the potential energy of the loads
$p$ and $F$ into the potential $\Phi = \int_0^L W \mymult \mathd S$
which normally captures the strain energy only; in line with this, the
loading parameters $p$ and $F$ are considered constant.

\subsection{Macroscopic and microscopic variables}

A natural choice of macroscopic strain parameter is the apparent axial
stretch $Z'(S)$: this is the stretch of a virtual bar obtained by
collapsing all the circular cross-sections to a point located at their
center.  However, this choice has the drawback that, for typical
constitutive laws for rubber, there can be several homogeneous
solutions corresponding to a given value of the apparent stretch.  To
work around this difficulty, it is preferable to define instead the
macroscopic strain parameter as the hoop stretch $h_1 (S) = E_2 (S) =
\frac{R (S)}{\rho}$.  As we will see, it is possible to reconstruct
the apparent axial stretch $Z'(S)$ in terms of this $h_1 (S)$.  We
thus apply the general formalism using a single macroscopic strain and
a single microscopic degree of freedom, defined as
\[ \tmmathbf{h} (S) = (h_1 (S)) = \left( \frac{R (S)}{\rho} \right), \quad \tmmathbf{y} (S) = (y_1 (S)) = (Z' (S)).\]
With this choice of macroscopic and microscopic variables, it is
possible to reconstruct the configuration using $R (S) = \rho \mymult
h_1 (S)$ and $Z (S) = Z (0) + \int_0^S y_1 (\overline{S}) \mymult
\mathd \overline{S}$, where $Z (0)$ is an unimportant rigid-body
translation.  As we do not need any constraint for this particular
structural model, we set $\tmmathbf{q} (\tmmathbf{y}) = ()$ and drop
all the terms containing $\tmmathbf{q} (\tmmathbf{y})$ in the general
formalism.

The strain vector $\tmmathbf{E}$ for the axisymmetric membrane given in
section~\ref{eq:membraneFullModel} can be cast in the canonical form from
equation~(\ref{eq:canonicalFormStrain}) by choosing the strain function as
\[ \tmmathbf{E} (\tmmathbf{h}, \tmmathbf{h}^{\dag} ; \tmmathbf{y},
   \tmmathbf{y}^{\dag}, \tmmathbf{y}^{\ddag}) = \left(\begin{array}{c|c|c}
     \sqrt{\rho^2 \mymult h_1^{\dag 2} + y_1^2} & h_1 & y_1
   \end{array}\right), \]
where the arguments are vectors whose length matches that of the macroscopic
strain $\tmmathbf{h}$ and microscopic variable $\tmmathbf{y}$ proposed above,
i.e., $\tmmathbf{h}= (h_1)$, $\tmmathbf{h}^{\dag} = (h_1^{\dag})$,
$\tmmathbf{y}= (y_1)$, $\tmmathbf{y}^{\dag} = (y_1^{\dag})$ and
$\tmmathbf{y}^{\ddag} = (y_1^{\ddag}) .$

\subsection{Homogeneous solutions}

Homogeneous solutions are first analyzed, by setting to zero the
derivative terms in the definition of the strain, see
equation~(\ref{eq:ETildeH}).  This yields the homogeneous strain as

\[ \tilde{\tmmathbf{E}} (\tmmathbf{h}, \tmmathbf{y}) =\left(\begin{array}{c|c|c}
y_1^{} & h_1 & y_1
\end{array}\right).
\]
The generalized stress in the homogeneous solution is given by the
gradient of the potential $\tmmathbf{E} (S)$,
\[ \frac{\mathd W}{\mathd E} (\tilde{\tmmathbf{E}} (\tmmathbf{h}, \tmmathbf{y})) = \left(\begin{array}{c|c|c}
     \overline{\Sigma}_S (y_1, h_1)&
     \overline{\Sigma}_{\Theta} (y_1, h_1) - 2 \mymult p \mymult \pi \mymult
     \rho^2 \mymult h_1 \mymult y_1&
     - F - p \mymult \pi \mymult \rho^2 \mymult h_1^2
   \end{array}\right), \]
where $\overline{\Sigma}_S (\lambda_S, \lambda_{\Theta}) =
\frac{\partial \overline{W}}{\partial \lambda_S} (\lambda_S,
\lambda_{\Theta})$ and $\overline{\Sigma}_{\Theta} (\lambda_S,
\lambda_{\Theta}) = \frac{\partial \overline{W}}{\partial
\lambda_{\Theta}} (\lambda_S, \lambda_{\Theta})$ are the components of
the Piola-Kirchhoff stress in the longitudinal and circumferential
directions, respectively, as predicted by the elastic strain potential
$\overline{W} (\lambda_S, \lambda_{\Theta})$ characterizing the
elastic properties of the membrane.

Next, we proceed to write and solve the principle of virtual
work~(\ref{eq:outlineHomPVW}) for homogeneous solutions. Noting that
$\frac{\partial \tilde{\tmmathbf{E}}}{\partial \tmmathbf{y}}
(\tmmathbf{h},\tmmathbf{y}) = \left(\begin{array}{c|c|c}
  1 & 0 & 1
\end{array}\right)$, it writes
\begin{equation}
  - \overline{\Sigma}_S (y_1^{(h_1)}, h_1) + F + p \mymult \pi \mymult \rho^2
  \mymult h_1^2 = 0. \label{eq:bulgingHomogeneousEquil}
\end{equation}
This is an equation for the longitudinal stretch $y_1^{(h_1)}$ in a
homogeneous solution, in terms of the hoop stretch $h_1 =
\frac{R}{\rho}$.  The load parameters $F$ and $\rho$ are considered
fixed, i.e., the dependence on $F$ and $\rho$ will always be silent.
Equation~(\ref{eq:bulgingHomogeneousEquil}) expresses the equilibrium
of a homogeneous solution in the longitudinal direction.  For typical
constitutive laws, equation~(\ref{eq:bulgingHomogeneousEquil}) cannot
be solved explicitly for $y_1^{(h_1)}$ in terms of $h_1$, and will be
viewed as an implicit equation.

In terms of the homogeneous solution $y_1^{(h_1)}$, we obtain the
properties of homogeneous solution from
equation~(\ref{eq:defineHomogeneousProperties}) as
\begin{equation}
  \begin{array}{lll}
    \tmmathbf{E}_{(h_1)} & = &\left(\begin{array}{c|c|c}
   y_1^{(h_1)} & h_1 & y_1^{(h_1)}
    \end{array}\right)\\
    W_{\text{hom}} (h_1) & = & \overline{W} (y_1^{(h_1)}, h_1) -
    \overline{\Sigma}_S (y_1^{(h_1)}, h_1) \mymult y_1^{(h_1)}\\
    \tmmathbf{\Sigma}_{(h_1)} & = & \left(\begin{array}{c|c|c}
      \overline{\Sigma}_S (y_1^{(h_1)}, h_1) & \ast & - \overline{\Sigma}_S
      (y_1^{(h_1)}, h_1)
    \end{array}\right)\\
    \tmmathbf{K}_{(h_1)} & = & \left(\begin{array}{ccc}
      \overline{K}_{S \nocomma S} (y_1^{(h_1)}, h_1) & \ast & 0\\
      \ast & \ast & \ast\\
      0 & \ast & 0
    \end{array}\right) .
  \end{array} \label{eq:bulgingHomogeneousQties}
\end{equation}
Here, the star symbol ($\ast$) denote quantities that play no role and do not
need to be calculated. The quantity $\overline{K}_{S \nocomma S} (\lambda_S,
\lambda_{\Theta}) = \frac{\partial^2 \overline{W}}{(\partial \lambda_S)^2}
(\lambda_S, \lambda_{\Theta})$ is the tangent elastic modulus, as calculated
from the membrane model.

\subsection{Change of microscopic variable}

According to the general method, we introduce a correction $\tmmathbf{z} (S) =
(z_1 (S))$ to the microscopic variable by $\tmmathbf{y} (S)
=\tmmathbf{y}_{\tmmathbf{h}} (S) +\tmmathbf{z} (S)$, i.e., $y_1 (S) =
y_1^{(h_1 (S))} + z_1 (S)$. In terms of the new unknown, the strain function
reads, see equation~(\ref{eq:strainInTermsOfZ}),
\[ \tmmathbf{e}_{\tmmathbf{h}} (\tmmathbf{h}^{\dag}, \tmmathbf{h}^{\ddag} ;
   \tmmathbf{z}, \tmmathbf{z}^{\dag}, \tmmathbf{z}^{\ddag}) =
   \left(\begin{array}{c|c|c}
     \sqrt{\rho^2 \mymult h_1^{\dag 2} + (y_1^{(h_1)} + z_1)^2} & h_1 &
     y_1^{(h_1)} + z_1
   \end{array}\right), \]
where again the arguments are vectors whose dimension is imposed by
the macroscopic strain and microscopic variable as
$\tmmathbf{h}^{\dag} = (h_1^{\dag})$, $\tmmathbf{h}^{\ddag} =
(h_1^{\ddag})$, $\tmmathbf{z}= (z_1)$, $\tmmathbf{z}^{\dag} =
(z_1^{\dag})$ and $\tmmathbf{z}^{\ddag} = (z_1^{\ddag})$.  Note that
the strain function $\tmmathbf{e}_{\tmmathbf{h}}
=\tmmathbf{e}_{(h_1)}$ depends on the macroscopic strain $h_1$ which
appears in subscript, and that we have made use of the catalog of
homogeneous solutions $y_1^{(h_1)}$ in the right-hand side.

The structure coefficients $\tmmathbf{e}^{i \nocomma j}_{k \nocomma l \nocomma
m} (\tmmathbf{h})$ are the successive partial derivatives of the right-hand
side above, see equation~(\ref{eq:structureCoefs}). These partial derivatives
are most easily found by identifying
$\tmmathbf{e}_{\tmmathbf{h}} (\tmmathbf{h}^{\dag}, \tmmathbf{h}^{\ddag} ;
\tmmathbf{z}, \tmmathbf{z}^{\dag}, \tmmathbf{z}^{\ddag})$ with its Taylor expansion $\left(\begin{array}{c|c|c}
y_1^{(h_1)} + z_1 + \frac{\rho^2 \mymult h_1^{\dag 2}}{2 \mymult y_1^{(h_1)}} &
h_1 & y_1^{(h_1)} + z_1
\end{array}\right)$. The result is
\begin{equation}
  \begin{array}{c}
    \begin{array}{llll}
      \tmmathbf{e}^{1 \nocomma 0}_{0 \nocomma 0 \nocomma 0} (\tmmathbf{h})
      =\tmmathbf{0} & \tmmathbf{e}^{2 \nocomma 0}_{0 \nocomma 0 \nocomma 0}
      (\tmmathbf{h}) = \left(\begin{array}{c|c|c}
        \frac{\rho^2}{y_1^{(h_1)}} & 0 & 0
      \end{array}\right) & \tmmathbf{e}^{0 \nocomma 1}_{0 \nocomma 0 \nocomma
      0} (\tmmathbf{h}) =\tmmathbf{0} & \tmmathbf{e}^{1 \nocomma 0}_{1
      \nocomma 0 \nocomma 0} (\tmmathbf{h}) =\tmmathbf{0}
    \end{array}\\
    \begin{array}{lll}
      \tmmathbf{e}^{0 \nocomma 0}_{1 \nocomma 0 \nocomma 0} (\tmmathbf{h}) =
      \left(\begin{array}{c|c|c}
        1 & 0 & 1
      \end{array}\right) & \tmmathbf{e}^{0 \nocomma 0}_{2 \nocomma 0 \nocomma
      0} (\tmmathbf{h}) =\tmmathbf{0} & \tmmathbf{e}^{0 \nocomma 0}_{0
      \nocomma 1 \nocomma 0} (\tmmathbf{h}) =\tmmathbf{0}.
    \end{array}
  \end{array} \label{eq:membraneStructureCoefficients}
\end{equation}
These are the only structure coefficients that are required in the following.
Recall that the dimension of both $\tmmathbf{h}$ and $\tmmathbf{y}$ is one for
an axisymmetric membrane: the tensors $\tmmathbf{e}^{i \nocomma j}_{k \nocomma
l \nocomma m}$, which are of dimensions $3 \times 1 \times \cdots \times 1$
according to the general rule, where the one's are repeated $i + j + k + l +
m$ times, have been identified with vectors of dimension 3.

When these expressions are combined with those for homogeneous quantities
obtained in~(\ref{eq:bulgingHomogeneousQties}), one can calculate the first
batch of operators from equation~(\ref{eq:preliminaryOperatorsDef}) as
\[ \begin{array}{rcl}
     \tmmathbf{A}_{\tmmathbf{h}} \cdot \tmmathbf{h}^{\dag} & = &
     \tmmathbf{\Sigma}_{\tmmathbf{h}} \cdot (\tmmathbf{e}^{1 \nocomma 0}_{0
     \nocomma 0 \nocomma 0} (\tmmathbf{h}) \cdot \tmmathbf{h}^{\dag})
     =\tmmathbf{\Sigma}_{\tmmathbf{h}} \cdot \tmmathbf{0}= 0\\
     \tmmathbf{C}_{\tmmathbf{h}}^{(0)} \cdot \tmmathbf{h}^{\ddag} & = &
     \tmmathbf{\Sigma}_{\tmmathbf{h}} \cdot (\tmmathbf{e}^{0 \nocomma 1}_{0
     \nocomma 0 \nocomma 0} (\tmmathbf{h}) \cdot \tmmathbf{h}^{\ddag})
     =\tmmathbf{\Sigma}_{\tmmathbf{h}} \cdot \tmmathbf{0}= 0\\
     \tmmathbf{C}_{\tmmathbf{h}}^{(1)} \cdot \tmmathbf{z}^{\dag} & = &
     \tmmathbf{\Sigma}_{\tmmathbf{h}} \cdot (\tmmathbf{e}^{0 \nocomma 0}_{0
     \nocomma 1 \nocomma 0} (\tmmathbf{h}) \cdot \tmmathbf{z}^{\dag})
     =\tmmathbf{\Sigma}_{\tmmathbf{h}} \cdot \tmmathbf{0}= 0\\
     \frac{1}{2} \mymult \tmmathbf{h}^{\dag} \cdot
     \tmmathbf{B}_{\tmmathbf{h}}^{(0)} \cdot \tmmathbf{h}^{\dag} & = &
     \frac{1}{2} \mymult \tmmathbf{0} \cdot \tmmathbf{K}_{\tmmathbf{h}} \cdot
     \tmmathbf{0}+ \frac{1}{2} \mymult \tmmathbf{\Sigma}_{\tmmathbf{h}} \cdummy
     \left(\begin{array}{c|c|c}
       \frac{\rho^2}{y_1^{(h_1)}} \mymult (h_1^{\dag})^2 & 0 & 0
     \end{array}\right) -\tmmathbf{h}^{\dag} \cdot \tmmathbf{0} \cdot
     \tmmathbf{h}^{\dag} = \frac{1}{2} \mymult \overline{\Sigma}_S
     (y_1^{(h_1)}, h_1) \mymult \frac{\rho^2}{y_1^{(h_1)}} \mymult
     (h_1^{\dag})^2\\
     \tmmathbf{h}^{\dag} \cdot \tmmathbf{B}_{\tmmathbf{h}}^{(1)} \cdot
     \tmmathbf{z} & = & \tmmathbf{0} \cdot \tmmathbf{K}_{\tmmathbf{h}} \cdot
     (\ast) +\tmmathbf{\Sigma}_{\tmmathbf{h}} \cdot
     \tmmathbf{0}-\tmmathbf{h}^{\dag} \cdot \tmmathbf{0} \cdot \tmmathbf{z}=
     0\\
     \frac{1}{2} \mymult \tmmathbf{z} \cdot
     \tmmathbf{B}_{\tmmathbf{h}}^{\text{} (2)} \cdot \tmmathbf{z} & = &
     \frac{1}{2} \mymult \left(\begin{array}{c|c|c}
       1 & 0 & 1
     \end{array}\right) \cdot \tmmathbf{K}_{\tmmathbf{h}} \cdot
     \left(\begin{array}{c|c|c}
       1 & 0 & 1
     \end{array}\right) + \frac{1}{2} \mymult \tmmathbf{\Sigma}_{\tmmathbf{h}}
     \cdummy \tmmathbf{0}= \frac{1}{2} \mymult \overline{K}_{S \nocomma S}
     (y_1^{(h_1)}, h_1) \mymult (z_1)^2.
   \end{array} \]

\subsection{Local optimization problem}

The local optimization problem~(\ref{eq:localOptimizationProblem}) is
particularly simple, because it has no source term
($\tmmathbf{B}_{\tmmathbf{h}}^{(1)} =\tmmathbf{0}$) and no constraint term
($\tmmathbf{q} (\tmmathbf{y}) = ()$). In view of the operators just derived,
it reads
\[ \forall \hat{z}_1 \quad \hat{z}_1 \mymult \overline{K}_{S \nocomma S}
   (y_1^{(h_1)}, h_1) \mymult z_1^{\text{opt}} = 0. \]
We rule out the possibility of a material instability in the membrane
model, i.e., $\overline{K}_{S \nocomma S} (y_1^{(h_1)}, h_1) > 0$
(note that with this assumption of material stability at the
`microscopic' level, the stability condition from
section~\ref{ssec:generalStability} is automatically satisfied).  The
variational problem above can then be solved for
$\tmmathbf{z}_{\text{opt}} = \left( z_1^{\text{opt}} \right)$ as
$\tmmathbf{z}_{\text{opt}} (\tilde{S}) = \tmmathbf{0}$.  The
correction to the microscopic variable arising from the gradient
effect is zero for this particular structure.  To comply with the
general form of
equation~(\ref{eq:correctiveDisplacementFactorOutHPrime}), we set
accordingly $\tmmathbf{Z}_{\text{opt}}^{\tmmathbf{h}} =
\left(\begin{array}{c} 0 \end{array}\right)$.

\subsection{Regularized model}

In view of equation~(\ref{eq:finalOperatorsDef}), we obtain the operators
entering into the strain-gradient model as $\tmmathbf{A}_{\tmmathbf{h}}
=\tmmathbf{0}$, $\tmmathbf{B}_{\tmmathbf{h}} = (B_{1 \nocomma 1}^{(h_1)})$
where $B_{1 \nocomma 1}^{(h_1)} = \frac{\rho^2 \mymult \overline{\Sigma}_S
(y_1^{(h_1)}, h_1)}{y_1^{(h_1)}}$ and $\tmmathbf{C}_{\tmmathbf{h}}
=\tmmathbf{0}$.

We switch to the more standard notation $\lambda_{\Theta} = h_1 =
\frac{R}{\rho}$ for the hoop stretch and $\lambda_S = y_1$ for the apparent
axial stretch, and recapitulate the main results for the axisymmetric membrane
as follows. We must first solve the implicit
equation~(\ref{eq:bulgingHomogeneousEquil}) for the apparent axial stretch
$y_1^{(h_1)} = \lambda_S^{\text{hom}} (\lambda_{\Theta})$, which reads
$\overline{\Sigma}_S \left( \lambda_S^{\text{hom}} (\lambda_{\Theta}),
\lambda_{\Theta} \right) = F + p \mymult \pi \mymult \rho^2 \mymult
\lambda_{\Theta}^2$ and yields the homogeneous equilibria of the balloon. In
this equation, $\overline{\Sigma}_{\Theta} (\lambda_S, \lambda_{\Theta}) =
\frac{\partial \overline{W}}{\partial \lambda_{\Theta}} (\lambda_S,
\lambda_{\Theta})$ is the hoop stress in the homogeneous solution. In terms of
this catalog of homogeneous solutions, we can calculate $W_{\text{hom}}
(\lambda_{\Theta})$ by~(\ref{eq:bulgingHomogeneousQties}). The balloon is
governed by the strain-gradient bar model, see
equation~(\ref{eq:RelaxedEnergy}),
\begin{subequations}
	\label{eq:balloon1dModelFinal}
\begin{equation}
  \Phi^{\star} [\lambda_{\Theta}] \approx \int_0^L \left[ W_{\text{hom}}
  (\lambda_{\Theta} (S)) + \frac{1}{2} \mymult B (\lambda_{\Theta} (S)) \mymult
  \lambda_{\Theta}^{\prime 2} (S) \right] \mymult \mathd S,
\end{equation}
where the strain-gradient modulus reads
\begin{equation}
	B (\lambda_{\Theta}) = \frac{\rho^2 \mymult \overline{\Sigma}_S \left(
   \lambda_S^{\text{hom}} (\lambda_{\Theta}), \lambda_{\Theta}
   \right)}{\lambda_S^{\text{hom}}},
   \end{equation}
\end{subequations}
and where $\overline{\Sigma}_S (\lambda_S, \lambda_{\Theta}) = \frac{\partial
\overline{W}}{\partial \lambda_S} (\lambda_S, \lambda_{\Theta})$ is the
longitudinal stress in the homogeneous solution.

\subsection{Comments}\label{ssec:balloonComents}

\begin{figure}
  \centerline{\includegraphics{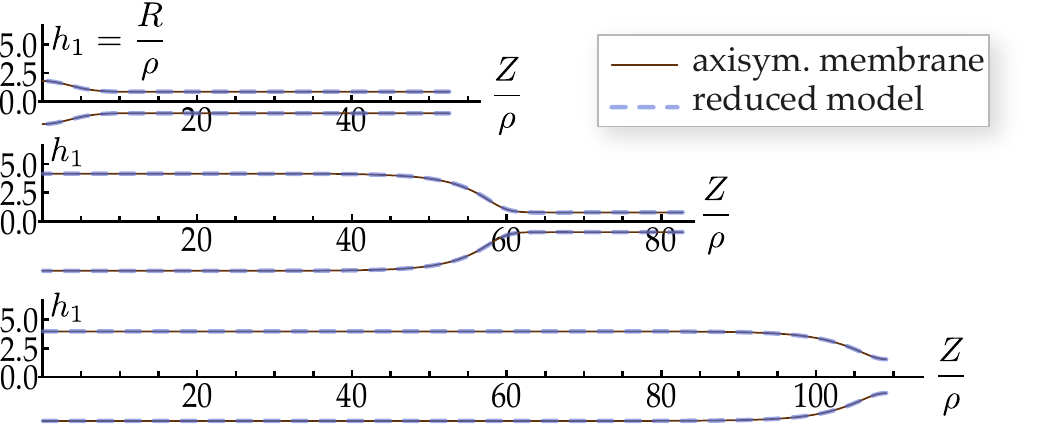}}
  \caption{Solutions for a propagating bulge in an axisymmetric membrane with
  initial aspect ratio $L / \rho = 30$: comparison of the predictions of the
  full axisymmetric membrane model ({\textsection}\ref{eq:membraneFullModel})
  and of the reduced model in equation~(\ref{eq:balloon1dModelFinal}),
  from~{\citet{Lestringant-Audoly-A-diffuse-interface-model-2018}}. The
  material model for rubber proposed
  by~{\citet{Ogden-Large-deformation-isotropic-1972}} is used, with the same
  set of material parameters as used in the previous experimental work
  of~{\citet{Kyriakides-Chang-The-initiation-and-propagation-of-a-localized-1991}},
  see also {\textsection}2 in~{\citet{Lestringant-Audoly-A-diffuse-interface-model-2018}}.\label{fig:balloonComparison}}
\end{figure}

We have recovered the model established
by~{\citet{Lestringant-Audoly-A-diffuse-interface-model-2018}}.
Typical solutions predicted by the 1d model are compared to those of
the full axisymmetric model in figure~\ref{fig:balloonComparison}: the
1d models appears to be highly accurate, even in the regime where the
bulges are fully localized.

The axisymmetric membrane model, which we used as a starting point was
already a 1d model: it does not make use of any transverse variable,
and has discrete degrees of freedom $(Z, R)$ in each cross-section.
The reduction method led us to another 1d model and it therefore is
improper to speak of dimension reduction in this case.  The reduction
method is still useful, as reduced model is simpler and, more
importantly, much more standard: it is the well-known
diffuse-interface model introduced by van der Walls in the context of
liquid-vapor phase transition, as discussed by
by~{\citet{Lestringant-Audoly-A-diffuse-interface-model-2018}}.

Even when bulges are fully formed, the typical length of the interface
between the bulged and unbulged regions never gets much less than
$\sim \rho$, i.e., remains always much larger than than the membrane's
thickness $t$ (assuming the membrane is thin in a first place $\rho
\gg t$).  This warrants that the assumptions underlying the membrane
model remain valid.  To address the case of thick membranes, i.e.,
when the ratio $t/\rho$ is not small, one could apply our reduction
method to a theory of thick membranes, or to a finite-strain model for
a hyperelastic cylinder in 3d.

\section{Application to a linearly elastic block}

Our next example is a homogeneous block of linearly elastic material
in 2d, having length $L$ and thickness $a$, as sketched in
figure~\ref{fig:block-geom}.  We account for both stretching and
bending of the block.  In the first step of the dimension reduction,
we will recover the classical beam model.  Its energy is convex,
implying that this particular structure does not tend to localize.
The strain-gradient model obtained at the next step is still of
interest as its solutions generally~\footnote{It is known, however,
that boundary conditions can prevent strain-gradient models from
converging faster.  This happens when the imposed boundary conditions
are incompatible with the kinematics $\tmmathbf{y} (S)
=\tmmathbf{y}_{\text{hom}} (\tmmathbf{h} (S))
+\tmmathbf{z}_{\text{opt}} + \cdots$ of the strain-gradient model at
the microscopic scale.  We do not address this question in this paper,
and limit attention to natural boundary
conditions.%
} converge faster towards those of the full (2d) elasticity model than
those of the classical beam model.

There is a large amount of work on higher-order asymptotic expansions
for prismatic solids in the specific context of linear elasticity with the
aim to derive \emph{linear} higher-order beam theories, see
for instance the work of
\citet{Trabucho-Viano-Mathematical-modelling-of-rods-1996}.  The
forthcoming analysis shows that these results can be easily recovered
with our method.  It also reveals that the assumption of linear
elasticity brings in severe, somewhat hidden limitations.

The elastic block is our first example where a cross-section possesses
infinitely many degrees of freedom.

\subsection{Full model: a linearly elastic block in
2d}\label{ssec:blockFullModel}

\begin{figure}
  \centerline{\includegraphics{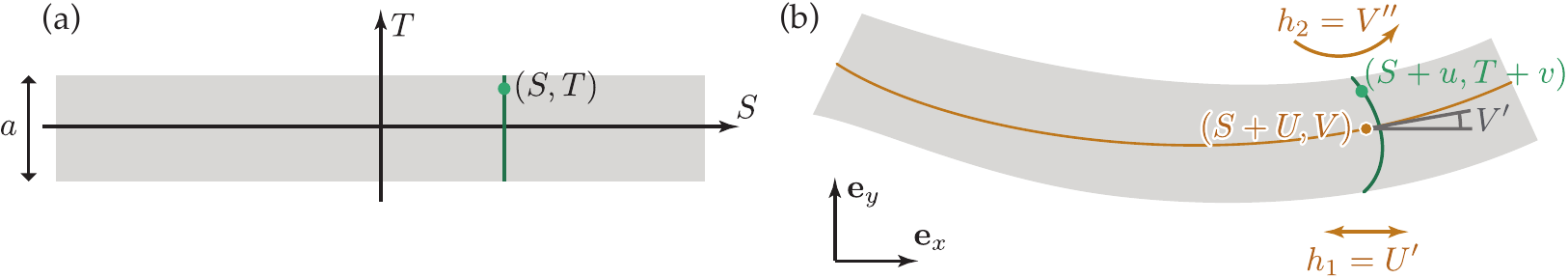}}
  \caption{A block of a linearly elastic material in (a)~reference
  and~(b)~current configuration. The 1d model makes use of the
  center line (brown curve), defined as the curve passing through the centers
  of mass (brown dots) of the cross-section. The macroscopic strain are the
  apparent stretch $h_1 (S) = U' (S)$ and the apparent curvature $h_2 (S) =
  V'' (S)$ of the center line.\label{fig:block-geom}}
\end{figure}

We consider an elastic block in reference configuration. The axial and
transverse coordinates in reference configuration are denoted as $S$ and $T$,
respectively, and are used as Lagrangian coordinates. Their domains are $0
\leq S \leq L$ and $- a / 2 \leq T \leq a / 2$.

We introduce the displacement $(u, v)$ in a Cartesian frame $(\tmmathbf{e}_x,
\tmmathbf{e}_y)$ aligned with axes of the undeformed block: a point with
position $(S, T)$ in reference configuration gets mapped to $\tmmathbf{x} (S,
T) = (S + u (S, T), T + v (S, T))$ in the current configuration, see
figure~\ref{fig:block-geom}(b). The linear strain is presented in vector form
as
\begin{equation}
  \tmmathbf{E}= \left(\begin{array}{c|c|c}
\frac{\partial u}{\partial S} & \frac{\partial
	v}{\partial T} & \frac{1}{2} \mymult \left( \frac{\partial u}{\partial T} +
\frac{\partial v}{\partial S}\right)
  \end{array}\right)
  \label{eq:slabLinearStrainDef}
\end{equation}
where $E_1$, $E_2$
and $E_3$ are respectively the $S \nocomma S$, $T \nocomma T$ and $S \nocomma T$ components of the 2-d strain
tensor from linear elastic theory.
We use a linear isotropic and uniform constitutive in 2d (Hookean elasticity),
corresponding to an elastic potential per unit length $\mathd S$
\begin{equation}
  W (\tmmathbf{E}) = \frac{1}{2} \mymult \int_{- a + 2}^{+ a / 2} (2 \mymult \mu
  \mymult (E_1^2 + E_2^2 + 2 \mymult E_3^2) + \lambda \mymult (E_1 + E_2)^2)
  \mymult \mathd T, \label{eq:slabWofE}
\end{equation}
where the elastic constants $\mu$ and $\lambda$ are known as the Lam{\'e}
parameters.

\subsection{Macroscopic and microscopic variables}

We choose to define the center line as the curve passing through the centers
of mass of the cross-sections. The components of the center line displacement
are therefore
\begin{equation}
  U (S) = \langle u \rangle (S) \quad V (S) = \langle v \rangle (S),
  \label{eq:slabUVisuvAverage}
\end{equation}
where $\langle f \rangle (S) = \frac{1}{a} \mymult \int_{- a / 2}^{+ a / 2} f
(S, T) \mymult \mathd T$ denotes the cross-section average of a function $f (S,
T)$.

The deformed center line is parametrized as $(S + U (S)) \mymult \tmmathbf{e}_x
+ V (S) \mymult \tmmathbf{e}_y$. In the theory of linear elasticity, it is
associated with an apparent longitudinal strain $U' (S)$, deflection angle $V'
(S)$, and curvature $V'' (S)$, where by `apparent' we emphasize the fact
that the center line is non-material. In our reduction of the elastic
block to a 1d model, we use as macroscopic strain measures these
apparent axial strain and curvature,
\[ \tmmathbf{h} (S) = (U' (S), V'' (S)) . \]

Let $\tilde{\tmmathbf{x}} (S, T)$ the final position of the point initially at
position $S \mymult \tmmathbf{e}_x + T \mymult \tmmathbf{e}_y$ if the cross-section $S$ were to undergo a rigid body motion following the center line,
namely the combination of a rigid-body translation $(U (S) \mymult
\tmmathbf{e}_x + V (S) \mymult \tmmathbf{e}_y)$ and a rigid-body rotation with
angle $V' (S)$. Since $V' (S)$ is infinitesimal, the unit normal to the center
line writes $- V' (S) \mymult \tmmathbf{e}_x +\tmmathbf{e}_y$ and so
$\tilde{\tmmathbf{x}} (S, T) = [(S + U (S)) \mymult \tmmathbf{e}_x + V (S)
\mymult \tmmathbf{e}_y] + T \mymult [- V' (S) \mymult \tmmathbf{e}_x
+\tmmathbf{e}_y]$.

We choose to define the microscopic displacement $\tmmathbf{y} (S, T)$ as the
difference between the actual position $\tmmathbf{x} (S, T) = (S + u (S, T))
\mymult \tmmathbf{e}_x + (T + v (S, T)) \mymult \tmmathbf{e}_y$ and 
$\tilde{\tmmathbf{x}} (S, T)$:
\[ \tmmathbf{y} (S, T) =\tmmathbf{x} (S, T) - \tilde{\tmmathbf{x}} (S, T) = (u
   (S, T) - U (S) + V' (S) \mymult T) \mymult \tmmathbf{e}_x - (v (S, T) - V
   (S)) \mymult \tmmathbf{e}_y . \]
The Cartesian components are found as $y_1 (S, T) = u (S, T) - U (S) +
V' (S) \mymult T$ and $y_2 (S, T) = v (S, T) - V (S)$.  This
definition of $\tmmathbf{y}$ warrants $\tmmathbf{y} (S, T)
=\tmmathbf{0}$ automatically whenever the block is moved rigidly, since
$\tmmathbf{x} (S, T) = \tilde{\tmmathbf{x}} (S, T)$ in this case.

In our general presentation of the method in section~\ref{sec:mainResults},
$\tmmathbf{y} (S)$ (with a single argument) was defined as the
{\tmem{collection}} of the microscopic degrees of freedom on a given
cross-section $S$. To comply with this convention, we define $\tmmathbf{y} (S)
= (y_1 (S), y_2 (S))$ as a pair of {\tmem{functions defined on the
cross-section}} taking the
transverse coordinate $T$ as an argument,
\[ \begin{array}{lll}
     y_1 (S) & = & \{ (u (S, T) - U (S) + V' (S) \mymult T) \}_T\\
     y_2 (S) & = & \{ v (S, T) - V (S) \}_T.
   \end{array} \]
We recall that $\{ g (T) \}_T$ is a notation for the function $g$ that
maps $T$ to $g (T)$, the index $T$ appearing in subscript after a
curly brace being a dummy variable.

In view of equation~(\ref{eq:slabUVisuvAverage}), the microscopic displacement
must satisfy the condition $\langle u \rangle (S) = U (S)$ and $\langle v \rangle
(S) = V (S)$. Upon elimination of $(u, v)$ in favor of $(y_1, y_2)$, this
yields $\langle y_1 (S) \rangle = \langle y_2 (S) \rangle = 0$ for all $S$. We
handle these constraints by setting
\[ \tmmathbf{q} (\tmmathbf{y}) =\frac{1}{a} \left( \int_{- a / 2}^{a / 2} y_1 (T) \mymult
   \mathd T, \int_{- a / 2}^{a / 2} y_2 (T) \mymult \mathd T \right) \]
in the general formalism of section~\ref{sec:mainResults}.

The displacement in the Cartesian basis is $u (S, T) = U (S) + [y_1 (S)] (T) -
V' (S) \mymult T$ and $v (S, T) = [y_2 (S)] (T) + V (S)$, and therefore the
strain in equation~(\ref{eq:slabLinearStrainDef}) can be expressed as
\[\tmmathbf{E}(S)=
 \left(\begin{array}{c|c|c}
 h_1 (S) + [y_1' (S)] (T) - h_2 (S) \mymult T & \partial_T
 [y_2 (S)] (T) & \frac{1}{2} \mymult (\partial_T [y_1 (S)] (T) + [y_2' (S)] (T))
 \end{array}\right).\]
Since primes are reserved for derivatives with respect to the
longitudinal variable $S$, we use the symbol $\partial_T$ for
transverse derivatives.

In the above expression, $y_1' (S)$ denotes the function $\{ y_1' (S, T)
\}_T$, and similarly $y_2' (S) = \{ y_2' (S, T) \}_T$. For consistency with
the discrete case, we define the strain function $\tmmathbf{E} (\ldots)$ as an
operator that takes as arguments the pair of {\tmem{functions}}
$\tmmathbf{y}=\tmmathbf{y} (S) = (y_1 (S), y_2 (S))$, and their derivatives
$\tmmathbf{y}^{\dag} =\tmmathbf{y}' (S)$, as well as the pair of scalars
$\tmmathbf{h} (S) = (h_1 (S), h_2 (S))$, and returns the strain {\tmem{map}}
in the cross-section,
\begin{equation}
\tmmathbf{E} (\tmmathbf{h}, \tmmathbf{h}^{\dag}, \tmmathbf{y},
   \tmmathbf{y}^{\dag}, \tmmathbf{y}^{\ddag}) =
 \left(\begin{array}{c|c|c}
 \{ h_1 + y_1^{\dag} (T)
 - h_2 \mymult T \}_T & \{ \partial_T y_2 (T) \}_T & \left\{ \frac{1}{2} \mymult
 (\partial_T y_1 (T) + y_2^{\dag} (T)) \right\}_T
 \end{array}\right).
\label{eq:StrainCanonicalBlock}
\end{equation} 
We use the same ordering conventions for the strain components as in
equation~(\ref{eq:slabLinearStrainDef}), i.e.~the longitudinal,
transverse and shear strain appear in this order.  We continue to use
the same notation as earlier whereby variables bearing a
dagger, such as $\tmmathbf{h}^{\dag} = (h_1^{\dag}, h_2^{\dag})$ are
dummy variables that are intended to hold the local value of the
derivative, here $\tmmathbf{h}' (S)$.

\subsection{Homogeneous solutions}

In the homogeneous case, the arguments of $\tmmathbf{E}$ corresponding to
axial gradients in~(\ref{eq:StrainCanonicalBlock}) (thus, bearing a single or a double dagger) are all set to
zero, see equation~(\ref{eq:ETildeH}). Doing so, we are left with the map of
homogeneous strain,
\[ \tilde{\tmmathbf{E}}(\tmmathbf{h},\tmmathbf{y}) =
 \left(\begin{array}{c|c|c}
 \{ h_1 - h_2 \mymult T \}_T & \{
 \partial_T y_2 (T) \}_T & \left\{ \frac{1}{2} \mymult \partial_T y_1 (T)
 \right\}_T
 \end{array}\right). \]
The first variation of the strain energy~(\ref{eq:slabWofE}) is calculated as
\begin{eqnarray}
     \delta W &= &\frac{\mathd W}{\mathd \tmmathbf{E}}
     (\tilde{\tmmathbf{E}}_{\tmmathbf{h}}) \cdot \delta \tmmathbf{E}= \int_{-
     a + 2}^{+ a / 2} \left(\begin{array}{c|c|c}
       \ast & \lambda \mymult E_1 + (2 \mymult \mu + \lambda) \mymult E_2 & 4
       \mymult \mu \mymult E_3
     \end{array}\right) \cdot \delta \tmmathbf{E} (T) \mymult \mathd T
	 \nonumber\\
     \nobracket \nobracket & =&\int_{- a + 2}^{+ a / 2}
     \left(\begin{array}{c|c|c}
       \ast & \lambda \mymult (h_1 - h_2 \mymult T) + (2 \mymult \mu + \lambda)
       \mymult \partial_T y_2 & 2 \mymult \mu \mymult \partial_T y_1
     \end{array}\right) \cdot \delta \tmmathbf{E} (T) \mymult \mathd T,
	 \nonumber
 \end{eqnarray}
where again stars denote values that play no role in the following.

By setting the variation of strain as $\delta \tmmathbf{E}= \frac{\partial
\tilde{\tmmathbf{E}}(\tmmathbf{h},\tmmathbf{y})}{\partial \tmmathbf{y}} \cdot
\hat{\tmmathbf{y}} = \left(\begin{array}{c|c|c}
\{ 0 \}_T & \{ \partial_T \hat{y}_2 \}_T & \left\{
\frac{1}{2} \mymult \partial_T \hat{y}_1 \right\}_T
\end{array}\right)$ as in
equation~(\ref{eq:outlineHomPVW}), we obtain the principle of virtual work as
\[ \forall (\hat{y}_1, \hat{y}_2) \quad \int_{- a / 2}^{+ a / 2} \left( - (2
   \mymult \mu \mymult \partial_T y_1) \mymult \frac{1}{2} \mymult \partial_T
   \hat{y}_1 + f^{\tmmathbf{h}}_1 \mymult \hat{y}_1 \right) \mymult \mathd T +
   \int_{- a / 2}^{+ a / 2} (- (\lambda \mymult (h_1 - h_2 \mymult T) + (2
   \mymult \mu + \lambda) \mymult \partial_T y_2) \mymult \partial_T \hat{y}_2 +
   f^{\tmmathbf{h}}_2 \mymult \hat{y}_2) \mymult \mathd T = 0 \]
where $\tmmathbf{f}^{\tmmathbf{h}} = (f_1^{\tmmathbf{h}}, f_2^{\tmmathbf{h}})$
is a Lagrange multiplier.

The solution satisfying the constraint $\tmmathbf{q} (\tmmathbf{y})
=\tmmathbf{0}$ is found as
\[ \begin{array}{rll}
     y_1^{\tmmathbf{h}} & = &\left\{0\right\}_T\\
     y_2^{\tmmathbf{h}} & = & \left\{- \nu \mymult h_1 \mymult T + \nu \mymult h_2
     \mymult \left( \frac{T^2}{2} - \frac{a^2}{24} \right)\right\}_T\\
     \tmmathbf{f}^{\tmmathbf{h}} & = & \tmmathbf{0}
   \end{array} \]
where we have defined the 2-d Poisson's ratio $\nu = \frac{\lambda}{2
\mymult \mu + \lambda}$.  We also define the Young's modulus in 2d as
$ Y = \frac{4 \mymult \mu \mymult (\lambda + \mu)}{2 \mymult \mu +
\lambda}$ (note that these expressions of $\nu$ and $Y$ are valid for
2d elasticity but not for 3d elasticity).

We can then calculate the quantities characterizing the homogeneous solutions
from equation~(\ref{eq:bulgingHomogeneousQties}) as
\begin{equation}
  \begin{array}{rll}
    \tmmathbf{E}_{\tmmathbf{h}} & = & \left(\begin{array}{c|c|c}
      \{ h_1 - h_2 \mymult T \}_T & \{ - \nu \mymult (h_1 - \mymult h_2 \mymult T)
      \}_T & \{ 0 \}_T
    \end{array}\right)\\
    W_{\text{hom}} (\tmmathbf{h}) & = & \frac{1}{2} \mymult (Y \mymult a \mymult
    h_1^2 + Y \mymult I \mymult h_2^2) \\
    \tmmathbf{\Sigma}_{\tmmathbf{h}} \cdot \delta \tmmathbf{E} & = & \int_{- a
    + 2}^{+ a / 2} \left(\begin{array}{c|c|c}
      Y \mymult (h_1 - h_2 \mymult T) & 0 & 0
    \end{array}\right) \cdot \delta \tmmathbf{E} (T) \mymult \mathd T\\
    \delta \tmmathbf{E} \cdot \tmmathbf{K}_{\tmmathbf{h}} \cdot \delta
    \tmmathbf{E} & = & \int_{- a + 2}^{+ a / 2} \delta \tmmathbf{E} (T) \cdot
    \left(\begin{array}{ccc}
      \ast & \ast & 0\\
      \ast & 2 \mymult \mu + \lambda & 0\\
      0 & 0 & 4 \mymult \mu
    \end{array}\right) \cdot \delta \tmmathbf{E} (T) \mymult \mathd T,
  \end{array} \label{eq:blockResultsHomogeneous}
\end{equation}
where $I = \int_{- a / 2}^{+ a / 2} T^2 \mymult \mathd T =
\frac{a^3}{12}$ is the geometric moment of inertia of the
cross-section.

\subsection{Change of microscopic variable}

Seeking the microscopic displacement as $\tmmathbf{y} (S)
=\tmmathbf{y}_{\tmmathbf{h}} (S) +\tmmathbf{z} (S)$ from equation~(\ref{eq:guess}), we can calculate the
strain in terms of the new microscopic variable $\tmmathbf{z}$ as
\begin{multline}
     \tmmathbf{e}_{\tmmathbf{h}} (\tmmathbf{h}^{\dag}, \tmmathbf{h}^{\ddag} ;
     \tmmathbf{z}, \tmmathbf{z}^{\dag}, \tmmathbf{z}^{\ddag}) = \Big(
     \begin{array}{l|l|l}
       \{ h_1 + z^{\dag}_1 (T) - h_2 \mymult T \}_T & \{ - \nu \mymult (h_1 -
       h_2 \mymult T) + \partial_T \mymult z_2 \}_T & \ldots
     \end{array} \\
     \begin{array}{l|l}
       & \left\{ \frac{1}{2} \mymult \left( - \nu \mymult h^{\dag}_1 \mymult T +
       \nu \mymult h^{\dag}_2 \left( \frac{T^2}{2} - \frac{a^2}{24} \right) +
       \partial_T z_1 + z^{\dag}_2 \right) \right\}_T
     \end{array} \Big).\nonumber
 \end{multline}
where $\tmmathbf{h}= (h_1, h_2)$, $\tmmathbf{h}^{\dag} =
(h_1^{\dag}, h_2^{\dag})$, $\tmmathbf{z}= (\{ z_1 (T) \}_T, \{ z_2 (T) \}_T)$
and $\tmmathbf{z}^{\dag} = (\{ z_1^{\dag} (T) \}_T, \{ z_2^{\dag} (T) \}_T)$.
For the linear elastic block, the structure coefficients do not depend on
$\tmmathbf{h}^{\ddag}$ or $\tmmathbf{z}^{\ddag}$.

The structure coefficients introduced in equation~(\ref{eq:structureCoefs})
are then obtained as
\[ \begin{array}{c}
     \begin{array}{l}
       \tmmathbf{e}^{1 \nocomma 0}_{0 \nocomma 0 \nocomma 0} (\tmmathbf{h})
       \cdot \tmmathbf{h}^{\dag} = \left(\begin{array}{c|c|c}
         \tmmathbf{0} & \tmmathbf{0} & \left\{ - \frac{\nu}{2} \mymult \left(
         \mymult h^{\dag}_1 \mymult T - \mymult h^{\dag}_2 \left( \frac{T^2}{2} -
         \frac{a^2}{24} \right) \right) \right\}_T
       \end{array}\right)
     \end{array}\\
     \begin{array}{llll}
       \tmmathbf{e}^{2 \nocomma 0}_{0 \nocomma 0 \nocomma 0} (\tmmathbf{h})
       =\tmmathbf{0} & \tmmathbf{e}^{0 \nocomma 1}_{0 \nocomma 0 \nocomma 0}
       (\tmmathbf{h}) =\tmmathbf{0} & \tmmathbf{h}^\dag \cdot \tmmathbf{e}^{1 \nocomma 0}_{1 \nocomma
    	0 \nocomma 0} (\tmmathbf{h}) \cdot \tmmathbf{z}=\tmmathbf{0} &
       \tmmathbf{e}^{0 \nocomma 0}_{1 \nocomma 0 \nocomma 0} (\tmmathbf{h})
       \cdot \tmmathbf{z}= \left(\begin{array}{c|c|c}
         \tmmathbf{0} & \{ \partial_T \mymult z_2 \}_T & \left\{ \frac{1}{2}
         \mymult \partial_T z_1 \right\}_T
       \end{array}\right)
     \end{array}\\
     \begin{array}{ll}
       \tmmathbf{z} \cdot \tmmathbf{e}^{0 \nocomma 0}_{2 \nocomma 0 \nocomma
       0} (\tmmathbf{h}) \cdot \tmmathbf{z}=\tmmathbf{0} & \tmmathbf{e}^{0
       \nocomma 0}_{0 \nocomma 1 \nocomma 0} (\tmmathbf{h}) \cdot
       \tmmathbf{z}^{\dag} = \left(\begin{array}{c|c|c}
         \{ z_1^{\dag} \}_T & \tmmathbf{0} & \left\{ \frac{1}{2} \mymult
         z_2^{\dag} \right\}_T
       \end{array}\right).
     \end{array}
   \end{array} \]
Next, the operators introduced in~(\ref{eq:preliminaryOperatorsDef}) are
calculated as follows,
\begin{equation}
 \begin{array}{rcl}
     \tmmathbf{A}_{\tmmathbf{h}} & = & \tmmathbf{0}\\
     \tmmathbf{C}_{\tmmathbf{h}}^{(0)} & = & \tmmathbf{0}\\
     \tmmathbf{C}_{\tmmathbf{h}}^{(1)} \cdot \tmmathbf{z} & = & \int_{- a +
     2}^{+ a / 2} Y \mymult (h_1 - h_2 \mymult T) \mymult z_1 \mymult \mathd T\\
     \frac{1}{2} \mymult \tmmathbf{h}^{\dag} \cdot
     \tmmathbf{B}_{\tmmathbf{h}}^{(0)} \cdot \tmmathbf{h}^{\dag} & = &
     \frac{1}{2} \mymult \int_{- a + 2}^{+ a / 2} \mu \mymult \nu^2 \mymult
     \left( \mymult h^{\dag}_1 \mymult T - \frac{h^{\dag}_2}{2} \left( T^2 -
     \frac{a^2}{12} \right) \right)^2 \mathd T\\
     & = & \frac{1}{2} \mymult \mu \mymult \nu^2 \mymult \left( \frac{a^3}{12}
     \mymult {h^{\dag}_1}^2 + \frac{a^5}{720} \mymult {h^{\dag}_2}^2 \right)\\
     \tmmathbf{h}^{\dag} \cdot \tmmathbf{B}_{\tmmathbf{h}}^{(1)} \cdot
     \tmmathbf{z} & = & - \nu \mymult \mu \mymult \int_{- a + 2}^{+ a / 2}
     \left( \mymult h^{\dag}_1 \mymult T - \frac{h^{\dag}_2}{2} \mymult \left(
     T^2 - \frac{a^2}{12} \right) \right) \mymult \partial_T z_1 \mymult \mathd
     T - Y \mymult \int_{- a + 2}^{+ a / 2} (h^{\dag}_1 - h^{\dag}_2 \mymult T)
     \mymult z_1 \mymult \mathd T\\
     \frac{1}{2} \mymult \tmmathbf{z} \cdot
     \tmmathbf{B}_{\tmmathbf{h}}^{\text{} (2)} \cdot \tmmathbf{z} & = &
     \frac{1}{2} \mymult \int_{- a + 2}^{+ a / 2} (2 \mymult \mu + \lambda)
     \mymult (\partial_T \mymult z_2)^2 + \mu \mymult (\partial_T z_1)^2 \mymult
     \mathd T.
   \end{array} \label{eq:operatorsACBblock}
   \end{equation}

\subsection{Local optimization problem}

The correction $\tmmathbf{z}_{\text{opt}} = \left( z^{\text{opt}}_1,
z^{\text{opt}}_2 \right)$ to the cross-sectional displacement is found by
writing down the variational problem~(\ref{eq:localOptimizationProblem}),
\begin{multline}
     \forall (\hat{z}_1 (T), \hat{z}_2 (T)) \quad \int_{- a + 2}^{+ a / 2}
     \left[ \mu \left( \partial_T z^{[1]}_1 - \nu \left( - h'_2 \mymult
     \frac{T^2}{2} + h'_1 \mymult T + h'_2 \mymult \mymult \frac{a^2}{24} \right)
     \right) \mymult \partial_T \hat{z}_1 - Y \mymult (h'_1 - h'_2 \mymult T)
     \mymult \hat{z}_1 \right] \mymult \mathd T \ldots\\
     + \int_{- a + 2}^{+ a / 2} \left[ (2 \mymult \mu + \lambda) \mymult
     \partial_T \mymult z^{\text{opt}}_2 \mymult \partial_T \mymult \hat{z}_2
     \right] \mymult \mathd T \nobracket \nobracket - \frac{1}{a} \mymult
     \int_{- a / 2}^{+ a / 2} \left( f_1^{\text{opt}} \mymult \hat{z}_1 +
     f_2^{\text{opt}} \mymult \hat{z}_2 \right) \mymult \mathd T = 0.
	 \nonumber
 \end{multline}
We proceed to solve this variational problem together with the incremental
constraint $\left\langle z_1^{\text{opt}} \right\rangle = \left\langle
z_2^{\text{opt}} \right\rangle = 0$.
As there is no source term in factor of $\hat{z}_2$, the transverse solution
is easily found as $z^{\text{opt}}_2 (T) = 0$ and $f_2^{\text{opt}} = 0$.

The remaining terms in the variational problem above concern the axial
correction, and can be rearranged as
\[ \forall \hat{z}_1 (T) \quad \int_{- a + 2}^{+ a / 2} \left( \left(
   \frac{f_1^{\text{opt}}}{a} + Y \mymult h'_1 \right) - Y \mymult h'_2 \mymult T
   \right) \mymult \hat{z}_1 - \mu \left( \partial_T z^{\text{opt}}_1 - \nu
   \left( - h'_2 \mymult \frac{T^2}{2} + h'_1 \mymult T + h'_2 \mymult \mymult
   \frac{a^2}{24} \right) \right) \mymult \partial_T \hat{z}_1 \mymult \mathd T
   = 0. \]
The solution to this equation satisfying the constraint $\left\langle
z_1^{\text{opt}} \right\rangle = 0$ can be worked out as
\[ \begin{array}{rll}
     f_1^{\text{opt}} & = & - a \mymult h'_1 \mymult Y\\
     z_1^{\text{opt}} (T) & = & \frac{h'_1 \mymult \nu}{2} \mymult \left( T^2 -
     \frac{a^2}{12} \right) - \frac{h'_2}{24} \mymult ((6 + 5 \mymult \nu)
     \mymult a^2 \mymult T - 4 \mymult (2 + \nu) \mymult T^3).
   \end{array} \]
The detailed expression of $z_1^{\text{opt}}$ will not be used, other than to
evaluate the following integral,
\[ \int_{- a / 2}^{a / 2} \left( \partial_T z_1^{\text{opt}} \right)^2 \mymult
   \mathd T = \frac{a^3 \mymult \nu^2}{12} \mymult {h'_1}^2 + \left( \frac{1}{30}
   + \frac{11 \mymult \nu}{180} + \frac{7 \mymult \nu^2}{240} \right) \mymult a^5
   \mymult {h'_2}^2 . \]

To sum up, the displacement correction $\tmmathbf{z}_{\text{opt}} = \left(
z_1^{\text{opt}}, z_2^{\text{opt}} \right)$ can be written in terms of a fixed
basis of functions as $\tmmathbf{z}_{\text{opt}}
=\tmmathbf{Z}^{\tmmathbf{h}}_{\text{opt}} \cdot \tmmathbf{h}'$, where
\[ \tmmathbf{Z}^{\tmmathbf{h}}_{\text{opt}} = \left(\begin{array}{cc}
     \frac{\nu}{2} \mymult \left( T^2 - \frac{a^2}{12} \right) & - \frac{1}{24}
     \mymult ((6 + 5 \mymult \nu) \mymult a^2 \mymult T - 4 \mymult (2 + \nu)
     \mymult T^3)\\
     0 & 0
   \end{array}\right).\]
The entries in the top-left (respectively top-right) slot is a
longitudinal displacement along $\tmmathbf{e}_x$ in response to a
gradient of axial strain $h_1 = U'$ (respectively, to a gradient of
curvature $h_2 = V''$).

The necessary stability condition from section~\ref{ssec:generalStability}
requires $2 \mymult \mu + \lambda \geq 0$ and $\mu \geq 0$, which are standard condition of material stability in 2d, as discussed for example in~\cite{Rivlin97}.

\subsection{Regularized model}\label{ssec:blockRegularizedModel}

Two last operators are defined in~equation(\ref{eq:finalOperatorsDef}). They
can now be evaluated as
\begin{multline} \tmmathbf{h}^{\dag} \cdot \tmmathbf{B}_{\tmmathbf{h}} \cdot
   \tmmathbf{h}^{\dag} = \mu \mymult \nu^2 \mymult \left( \frac{a^3}{12} \mymult
   {h^{\dag}_1}^2 + \frac{a^5}{720} \mymult {h^{\dag}_2}^2 \right) - \mu \mymult
   \int_{- a + 2}^{+ a / 2} (\partial_T z_1)^2 \mymult \mathd T \\
   = - \mu \mymult
   a^5 \mymult \left( \frac{1}{30} + \frac{11 \mymult \nu}{180} +
   \frac{\nu^2}{36} \right) \mymult {h^{\dag}_2}^2 = - Y \mymult a^5 \mymult
   \frac{6 + 5 \mymult \nu}{360} \mymult {h^{\dag}_2}^2
   \label{eq:hBhBlock}
   \end{multline}
and
\[ \tmmathbf{C}_{\tmmathbf{h}} \cdot \tmmathbf{h}^{\dag} = \int_{- a + 2}^{+ a
   / 2} Y \mymult (h_1 - h_2 \mymult T) \mymult \left(\begin{array}{c}
     \frac{\nu}{2} \mymult \left( T^2 - \frac{a^2}{12} \right)\\
     - \frac{1}{24} \mymult ((6 + 5 \mymult \nu) \mymult a^2 \mymult T - 4 \mymult
     (2 + \nu) \mymult T^3)
   \end{array}\right) \cdot \tmmathbf{h}^{\dag} \mymult \mathd T = Y \mymult a^5
   \mymult \frac{12 + 11 \mymult \nu}{720} \mymult h_2 \mymult h_2^{\dag}. \]
In addition, recall that $\tmmathbf{A}_{\tmmathbf{h}} =\tmmathbf{0}$.

Using equation~(\ref{eq:RelaxedEnergy}), we obtain the energy function of the 1d model as
\begin{equation}
  \Phi^{\star} [\tmmathbf{h}] = \int_0^L \frac{1}{2} \mymult (Y \mymult a \mymult
  h_1^2 + Y \mymult I \mymult h_2^2) \mymult \mathd S + Y \mymult a^5 \mymult
  \frac{12 + 11 \mymult \nu}{720} \mymult \mymult [h_2 \mymult h_2']_0^L -
  \frac{1}{2} \mymult Y \mymult a^5 \mymult \frac{6 + 5 \mymult \nu}{360} \mymult
  \int_0^L h_2^{\prime 2} \mymult \mathd S, 
 \label{eq:blockFinalEnergy}
\end{equation}
where $h_1 = U'$ denotes the axial stretch and $h_2 = V''$ denotes the
curvature.

\subsection{Comments}\label{ssec:blockDiscussion}

We have recovered the classical Euler-Bernoulli rod model at the
dominant order, with a stretching modulus $Y \mymult a$ and a bending
modulus $Y \mymult I$.  Indeed, the first step of the reduction method
yields classical structural model (i.e., without the gradient effect),
see the expression of $W_{\text{hom}}$ in
equation~(\ref{eq:blockResultsHomogeneous}).

As the operator $\frac{1}{2} \mymult \tmmathbf{z} \cdot
\tmmathbf{B}_{\tmmathbf{h}}^{\text{} (2)} \cdot \tmmathbf{z}$ in
equation~(\ref{eq:operatorsACBblock}) is non-negative, the necessary
condition of stability with respect to the microscopic degrees of
freedom is satisfied, see~\S\ref{ssec:generalStability}.  However, the
coefficient of the gradient term $h_{2}'^{2}$
in~(\ref{eq:blockFinalEnergy}) is negative, like the second-gradient
modulus $\tmmathbf{B}_{\tmmathbf{h}}$ acting on the macroscopic
degrees of freedom, see~(\ref{eq:hBhBlock}).  As a result,
$\Phi^{\star} [\tmmathbf{h}]$ can be decreased without bound by means 
of small-scale oscillations.  This
behavior can likely be regularized by pushing the expansion to a
higher order.  In its present form, the
functional~(\ref{eq:blockFinalEnergy}) should not be used to set up a
minimization or to analyze stability; it is still useful, as its
stationary points provide a more accurate approximation of the 3d
solution than that of the Euler-Bernoulli model.

To connect with the existing literature, we have worked out this
example in the limited context of linear elasticity but the derivation
can be extended easily to deal, e.g., with a nonlinear constitutive
law, a nonlinear geometry, as demonstrated in the following section.
While previous work has been focused on deriving order by order
solutions to the 3d equilibrium equations, our relaxation method makes
use of the variational structure of these equations by working
directly on the energy.  This, together with the fact that the hard
work underlying the derivation of the general method in
section~\ref{sec:mainResults} and~\ref{sec:detailedProofs} has been done once for all, simplifies the reduction of
any particular structural model considerably.

The only gradient effect present in
equation~(\ref{eq:blockFinalEnergy}) comes from the gradient of
bending.  The absence of a gradient term for stretching is a
peculiarity of the linear elasticity model which we started from: a
gradient effect involving the axial strain $h_{1}$ is restored if we
start instead from a finite-elasticity theory, as the next example
will show.  In fact, it does not make much sense to derive a
{\tmem{higher-order rod model}}, which aims at identifying subdominant
corrections to the elastic energy, starting from a {\tmem{linear
elasticity model}}: the linearization underlying the linear elasticity
theory suppresses subdominant contributions to the strain energy
\emph{a priori}.  This fact has not been well appreciated, as most of
the earlier work on higher-order beam models has been done in the
framework of linear elasticity.

\section{Application to a hyperelastic cylinder in
tension}\label{sec:hyperCylinder}

In our third example, we address the axisymmetric deformation of a
hyperelastic cylinder.  This problem is motivated by the necking of
bars, a situation where deformations become localized.  Necking
typically involves plasticity but it can be analyzed using an
equivalent elastic constitutive law obtained by the $J_{2}$
deformation theory, as long as the loading is proportional and
monotonous: the equivalent hyperelastic law is such that the curve for
homogeneous traction displays a maximum of the force as a function of
the stretch, leading to localization.  This section derives the 1d
model obtained
by~{\citet{Audoly-Hutchinson-Analysis-of-necking-based-2016}} using a
dedicated expansion method, this time using the general method of
section~\ref{sec:mainResults}.  This worked example combines a
continuous cross-section, a nonlinear elastic model, and kinematic
constraints.

\subsection{Finite-strain elasticity model for an axisymmetric bar}

\begin{figure}
  \centerline{\includegraphics{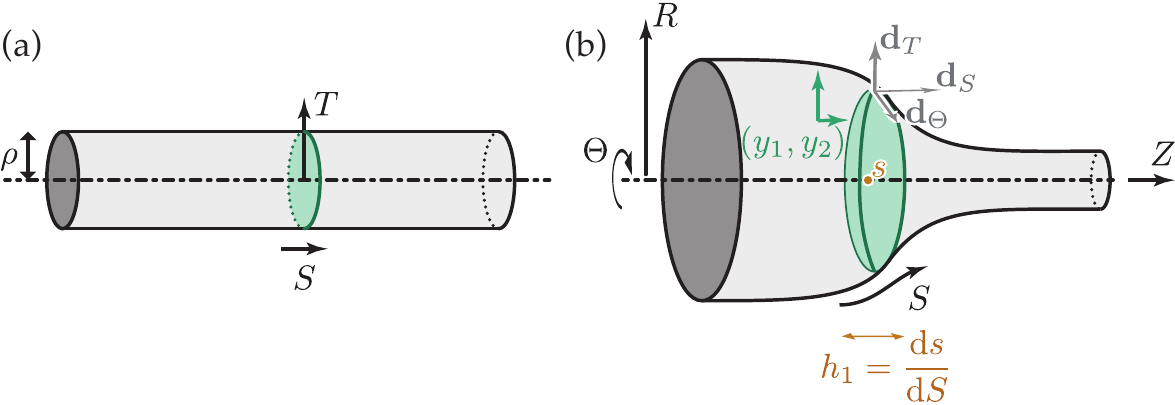}}
  \caption{A nonlinearly elastic cylinder in~(a) reference configuration and
  (b)~current configuration.\label{fig:bar-geom}}
\end{figure}

In its reference configuration, the bar is a cylinder with length $L$ and radius
$\rho$, and we denote by $S$, $T$ and $\Theta$ the axial, radial and azimuthal
coordinates, respectively, with $0 \leq S \leq L$, $0 \leq T
\leq \rho$, $0 \leq \Theta \leq 2 \mymult \pi$. We consider a
transversely isotropic material whose elastic properties are functions of $T$
but not of $\Theta$ or $S$ (isotropy is a particular case of transverse
isotropy, so this includes homogeneous isotropic materials): this warrants
that axisymmetric solutions possessing cylindrical invariance exist when the
bar is subject to traction. In fact, we restrict attention to axisymmetric
solutions, ignoring the possibility of localized modes involving shear
bands~{\citep{Triantafyllidis-Scherzinger-EtAl-Post-bifurcation-equilibria-in-the-plane-strain-2007}}.

The coordinates $(S, T)$ are used as Lagrangian coordinates, and we denote by
$\tmmathbf{x} (S, T) = (Z (S, T), R (S, T))$ the axial and radial cylindrical
coordinates of a material point initially located as $(S, T)$, see
figure~\ref{fig:bar-geom}(b). In 3d space, the final position is $Z (S, T)
\mymult \tmmathbf{d}_S (\Theta) + R (S, T) \mymult \tmmathbf{d}_T$, where
$(\tmmathbf{d}_S (\Theta), \tmmathbf{d}_T, \tmmathbf{d}_{\Theta} (\Theta))$ is
the local cylindrical basis, as sketched in the figure. Denoting partial
derivatives using commas in subscript, the deformation gradient is
$\tmmathbf{F}= Z_{, S} \mymult \tmmathbf{d}_S \otimes \tmmathbf{d}_S + Z_{, T}
\mymult \tmmathbf{d}_S \otimes \tmmathbf{d}_T + R_{, S} \mymult \tmmathbf{d}_T
\otimes \tmmathbf{d}_S + R_{, T} \mymult \tmmathbf{d}_T \otimes \tmmathbf{d}_T
+ \frac{R}{T} \mymult \tmmathbf{d}_{\Theta} \otimes \tmmathbf{d}_{\Theta}$, and
the strain writes $\tmmathbf{E}= \left(\begin{array}{c|c|c|c}
E_1 & E_2& E_3 & E_4
\end{array}\right)$ where
\begin{equation}
  \begin{array}{rll}
    E_1 & = & \frac{1}{2} \mymult (Z_{, S}^2 + R_{, S}^2 - 1)\\
    E_2 & = & \frac{1}{2} \mymult (Z_{, T}^2 + R_{, T}^2 - 1)\\
    E_3 & = & \frac{1}{2} \mymult (Z_{, S} \mymult Z_{, T} + R_{, S} \mymult R_{,
    T})\\
    E_4 & = & \frac{1}{2} \mymult \left( \left( \frac{R}{T} \right)^2 - 1
    \right) .
  \end{array} \label{eq:neckingStrainClassical}
\end{equation}
Here, $\left(\begin{array}{c|c|c|c}
E_1 & E_2& E_3 & E_4
\end{array}\right) = \left(\begin{array}{c|c|c|c}
\overline{E}_{S \nocomma S} & \overline{E}_{T \nocomma T} &
\overline{E}_{S \nocomma T} & \overline{E}_{\Theta \nocomma \Theta}
\end{array}\right)$ are the components of the Green--St-Venant strain
$\overline{\tmmathbf{E}} = \frac{1}{2} \mymult (\tmmathbf{F}^T \cdot
\tmmathbf{F}-\tmmathbf{I})$ from finite-elasticity theory (symbols
bearing a bar on top are relevant to the full model) and
$\tmmathbf{I}$ is the identity matrix.

For a transversely isotropic material, the strain energy of the bar per unit
length can be written in the form
\begin{equation}
  W (\tmmathbf{E}) = \int_0^{\rho} \overline{W} (E_1 (S), E_3^2 (S), E_2 (S) +
  E_4 (S), E_2^2 (S) + E_4^2 (S), E_2 (S) \mymult E_3^2 (S)) \mymult 2 \mymult
  \pi \mymult T \mymult \mathd T,
\label{eq:NLElasticCylinderEnergy}
\end{equation}
where $\overline{W} (\overline{E}_{S \nocomma S}, \overline{E}_{S
\nocomma T}^2, \overline{E}_{T \nocomma T} + \overline{E}_{\Theta
\nocomma \Theta}, \overline{E}_{T \nocomma T}^2 + \overline{E}_{\Theta
\nocomma \Theta}^2, \overline{E}_{T \nocomma T} \mymult
\overline{E}_{S \nocomma T}^2)$ is the elastic potential of the
material model, written in terms of a set of invariants relevant to
the transverse isotropic symmetry.

\subsection{Macroscopic and microscopic variables}

Let us consider the coordinate $s (S)$ of the center of mass of the deformed
cross-section, i.e.,
\begin{equation}
  s (S) = \langle Z \rangle (S),
\label{eq:neckingCenterOfMassS}
\end{equation}
where $\langle f \rangle = \frac{1}{\pi \mymult \rho^2} \mymult \int_0^{\rho} f
(T) \mymult 2 \mymult \pi \mymult T \mymult \mathd T$ denotes the average of a
quantity over the cross-section. This $s (S)$ is denoted by the orange dot in
figure~\ref{fig:bar-geom}(b); it does not correspond to any material point. We
use a single macroscopic strain variable, defined as the apparent axial
stretch
\[ \tmmathbf{h} (S) = (h_1 (S)) = \left( \frac{\mathd s}{\mathd S} (S) \right)
   . \]

We use as microscopic degrees of freedom $\tmmathbf{y}$ the position
of the current relative to the center of mass of the cross-section,
\[ \tmmathbf{y}= (\{ y_1 \}_T, \{ y_2 \}_T), \]
such that the position in deformed configuration can be reconstructed as
\[ \tmmathbf{x} (S, T) = (Z (S, T), R (S, T)) = (s (S) + y_1 (S, T), y_2 (S,
   T)).\]
In view of equation~(\ref{eq:neckingCenterOfMassS}), one must enforce the
kinematic constraint $\tmmathbf{q} (\tmmathbf{y} (S)) = 0$ for all $S$, where
\[ \tmmathbf{q} (\tmmathbf{y}) = \left( \int_0^{\rho} y_1 (T) \mymult 2 \mymult
   \pi \mymult T \mymult \mathd T \right) . \]

In terms of the macroscopic strain and microscopic displacement, the map of
strain over a cross section writes, from
equation~(\ref{eq:neckingStrainClassical}),
\begin{equation}
  \tmmathbf{E} (\tmmathbf{h}, \tmmathbf{h}^{\dag} ; \tmmathbf{y},
  \tmmathbf{y}^{\dag}, \tmmathbf{y}^{\ddag}) = \left(\begin{array}{c}
    \left\{ \frac{1}{2} \mymult ((h_1 + y_1^{\dag})^2 + (y_2^{\dag})^2 - 1)
    \right\}_T\\
    \left\{ \frac{1}{2} \mymult ((\partial_T y_1)^2 + (\partial_T y_2)^2 - 1)
    \right\}_T\\
    \left\{ \frac{1}{2} \mymult ((h_1 + y_1^{\dag}) \mymult \partial_T y_1 +
    y_2^{\dag} \mymult \partial_T y_2) \right\}_T\\
    \left\{ \frac{1}{2} \mymult \left( \left( \frac{y_2}{T} \right)^2 - 1
    \right) \right\}_T
  \end{array}\right) . \label{eq:nlCylinderEh}
\end{equation}
As earlier with the elastic block, and as implied by the $\{ \ldots \}_T$
notation, each component of $\tmmathbf{E} (S)$ is a function defined over the
cross-section.

\subsection{Homogeneous
solutions}\label{ssec:nlCylinderHomogeneousSolsSummary}

A detailed analysis of homogeneous solutions is done in
\ref{app:nlCylinderHomogeneous}.  The main results are summarized as
follows.

At the microscopic level, the cross-sections remains planar and undergo a
uniform dilation with a stretch ratio $\mu_{(h_1)}$ depending on the (uniform)
longitudinal stretch $h_1$, i.e., the microscopic displacement is of the form
\[ y_1^{(h_1)} = \{0\}_T \qquad y_2^{(h_1)} = \{\mu_{(h_1)} \mymult T\}_T. \]

Due to the material symmetry, the microscopic stress is equi-biaxial, with a
longitudinal stress $\overline{\Sigma}_{\parallel} (h_1, \mu_{(h_1)}) =
\overline{\Sigma}_{S \nocomma S}$ and a transverse stress
$\overline{\Sigma}_{\perp} (h_1, \mu_{(h_1)}) = \overline{\Sigma}_{T \nocomma
T} = \overline{\Sigma}_{\Theta \nocomma \Theta}$ given in terms of the elastic
constitutive model by
\begin{equation}
  \begin{array}{rll}
    \overline{\Sigma}_{\parallel} (h_1, \mu) & = & \partial_1 \overline{W}
    \left( \frac{1}{2} \mymult (h_1^2 - 1), 0, \mu^2 - 1, \frac{1}{2} \mymult
    (\mu^2 - 1)^2, 0 \right)\\
    \overline{\Sigma}_{\perp} (h_1, \mu) & = & \partial_3 \overline{W} \left(
    \frac{1}{2} \mymult (h_1^2 - 1), 0, \mu^2 - 1, \frac{1}{2} \mymult (\mu^2 -
    1)^2, 0 \right) + (\mu^2 - 1) \mymult \partial_4 \overline{W} \left(
    \frac{1}{2} \mymult (h_1^2 - 1), 0, \mu^2 - 1, \frac{1}{2} \mymult (\mu^2 -
    1)^2, 0 \right).
  \end{array} \label{eq:nlCylinderSigma}
\end{equation}
Here, $\partial_i \overline{W}$ denotes the partial derivative of the strain
energy with respect to the $i$th argument.

The equilibrium of the lateral boundary yields an implicit equation for the
transverse stretch $\mu_{(h_1)}$ in terms of the longitudinal stretch,
\begin{equation}
  \overline{\Sigma}_{\perp} (h_1, \mu_{(h_1)}) = 0. \label{eq:nlCylinderEquil}
\end{equation}

The homogeneous microscopic strain, strain energy density, microscopic stress
and tangent elastic stiffness are then given by
equation~(\ref{eq:defineHomogeneousProperties}) as
\[ \begin{array}{rll}
     \tmmathbf{E}_{\tmmathbf{h}} & = & \left(\begin{array}{c|c|c|c}
       \left\{ \frac{1}{2} \mymult (h_1^2 - 1) \right\}_T & \left\{ \frac{1}{2}
       \mymult (\mu_{(h_1)}^2 - 1) \right\}_T & \{ 0 \}_T & \left\{ \frac{1}{2}
       \mymult (\mu_{(h_1)}^2 - 1) \right\}_T
     \end{array}\right)\\
     W_{\text{hom}} (\tmmathbf{h}) & = & \int_0^{\rho} \overline{W} \left(
     \frac{1}{2} \mymult (h_1^2 - 1), 0, \mu_{(h_1)}^2 - 1, \frac{1}{2} \mymult
     (\mu_{(h_1)}^2 - 1)^2, 0 \right) \mymult 2 \mymult \pi \mymult T \mymult
     \mathd T\\
     \tmmathbf{\Sigma}_{\tmmathbf{h}} \cdot \delta \tmmathbf{E} & = &
     \int_0^{\rho} \overline{\Sigma}_{\parallel} (h_1, \mu_{(h_1)}) \mymult
     \delta E_1 (T) \mymult 2 \mymult \pi \mymult T \mymult \mathd T\\
     \delta \tmmathbf{E} \cdot \tmmathbf{K}_{\tmmathbf{h}} \cdot \delta
     \tmmathbf{E} & = & \int_0^{\rho} \left[ 4 \mymult \overline{K}_{S \nocomma
     T}^{S \nocomma T} \mymult (\delta E_3 (T))^2 + \left(\begin{array}{c}
       \delta E_1 (T)\\
       \delta E_2 (T)\\
       \delta E_4 (T)
     \end{array}\right) \cdot \left(\begin{array}{ccc}
       \ast & \ast & \ast\\
       \ast & \overline{K}_{T \nocomma T}^{T \nocomma T} & \overline{K}_{T
       \nocomma T}^{\Theta \nocomma \Theta}\\
       \ast & \overline{K}_{T \nocomma T}^{\Theta \nocomma \Theta} &
       \overline{K}_{T \nocomma T}^{T \nocomma T}
     \end{array}\right) \cdot \left(\begin{array}{c}
       \delta E_1 (T)\\
       \delta E_2 (T)\\
       \delta E_4 (T)
     \end{array}\right) \right] \mymult 2 \mymult \pi \mymult T \mymult \mathd T
   \end{array} \]
where the incremental shearing modulus reads
\begin{equation}
  \overline{K}_{S \nocomma T}^{S \nocomma T} = \frac{1}{4} \mymult (2 \mymult
  \partial_2 \overline{W} + (\mu_{(h_1)}^2 - 1) \mymult \partial_5
  \overline{W}). \label{eq:nlCylinderKSTST}
\end{equation}
In the right-hand side, the derivatives of the elastic potential
$\overline{W}$ must be evaluated in the homogeneous solution $(\overline{E}_{S
\nocomma S}, \overline{E}_{S \nocomma T}^2, \overline{E}_{T \nocomma T} +
\overline{E}_{\Theta \nocomma \Theta}, \overline{E}_{T \nocomma T}^2 +
\overline{E}_{\Theta \nocomma \Theta}^2, \overline{E}_{T \nocomma T} \mymult
\overline{E}_{S \nocomma T}^2) = \left( \frac{1}{2} \mymult (h_1^2 - 1), 0,
\mu^2 - 1, \frac{1}{2} \mymult (\mu^2 - 1)^2, 0 \right)$, as in
equation~(\ref{eq:nlCylinderSigma}). The expressions of the other elastic
moduli do not play any role in the 1d model.

The symmetry of the material and of the homogeneous solution accounts for the
particular form of the tangent moduli found above. For instance, the tangent
modulus $\overline{K}_{T \nocomma T}^{T \nocomma T}$ in factor of $(\delta
E_4)^2 = (\delta E_{\Theta \nocomma \Theta})^2$ in the second variation of the
elastic potential $\delta \tmmathbf{E} \cdot \tmmathbf{K}_{\tmmathbf{h}} \cdot
\delta \tmmathbf{E}$ is identical to that in factor of $(\delta E_2)^2 =
(\delta E_{T \nocomma T})^2$ due to the transverse isotropy, i.e.,
$\overline{K}_{\Theta \nocomma \Theta}^{\Theta \nocomma \Theta} =
\overline{K}_{T \nocomma T}^{T \nocomma T}$.

\subsection{Structure coefficients, local optimization
problem}\label{ssec:localOptimizationNecking}

Using the homogeneous solution just obtained and the expression of the strain
$\tmmathbf{E}$ in equation~(\ref{eq:nlCylinderEh}), one can calculate the
structure coefficients relevant to the nonlinear cylinder as (details can be
found in \ref{ssec:nlCylAppendixEh})
\begin{equation}
  \begin{array}{l}
    \begin{array}{l}
      \tmmathbf{e}^{1 \nocomma 0}_{0 \nocomma 0 \nocomma 0} (\tmmathbf{h})
      \cdot \tmmathbf{h}^{\dag} = \left(\begin{array}{c|c|c|c}
        \tmmathbf{0} & \tmmathbf{0} & \left\{ \frac{\mu_{(h_1)} \mymult \nabla
        \mu_{(h_1)}}{2} \mymult T \mymult h^{\dag}_1 \right\}_T & \tmmathbf{0}
      \end{array}\right)
    \end{array}\\
    \begin{array}{l}
      \tmmathbf{h}^{\dag} \cdot \tmmathbf{e}^{2 \nocomma 0}_{0 \nocomma 0
      \nocomma 0} (\tmmathbf{h}) \cdot \tmmathbf{h}^{\dag} =
      \left(\begin{array}{c|c|c|c}
        \{ (\nabla \mu_{(h_1)})^2 \mymult T^2 \mymult (h^{\dag}_1)^2 \}_T &
        \tmmathbf{0} & \tmmathbf{0} & \tmmathbf{0}
      \end{array}\right)
    \end{array}\\
    \begin{array}{ll}
      \tmmathbf{e}^{0 \nocomma 1}_{0 \nocomma 0 \nocomma 0} (\tmmathbf{h})
      =\tmmathbf{0} & \tmmathbf{h}^\dag \cdot \tmmathbf{e}^{1 \nocomma 0}_{1 \nocomma
			0 \nocomma 0} (\tmmathbf{h}) \cdot (z_1, z_2) = \left(\begin{array}{c|c|c|c}
        \tmmathbf{0} & \tmmathbf{0} & \ast & \tmmathbf{0}
      \end{array}\right)
    \end{array}\\
    \begin{array}{ll}
      & \tmmathbf{e}^{0 \nocomma 0}_{1 \nocomma 0 \nocomma 0} (\tmmathbf{h})
      \cdot (z_1, z_2) = \left(\begin{array}{c|c|c|c}
        \tmmathbf{0} & \{ \mu_{(h_1)} \mymult \partial_T z_2 (T) \}_T & \left\{
        \frac{h_1}{2} \mymult \partial_T z_1 (T) \right\}_T & \left\{
        \mu_{(h_1)} \mymult \frac{z_2 (T)}{T} \right\}_T
      \end{array}\right)
    \end{array}\\
    \begin{array}{l}
      (z_1, z_2) \cdot \tmmathbf{e}^{0 \nocomma 0}_{2 \nocomma 0 \nocomma 0}
      (\tmmathbf{h}) \cdot (z_1, z_2) = \left(\begin{array}{c|c|c|c}
        \tmmathbf{0} & \{ (\partial_T z_1 (T))^2 + (\partial_T z_2 (T))^2 \}_T
        & \tmmathbf{0} & \left\{ \left( \frac{z_2 (T)}{T} \right)^2 \right\}_T
      \end{array}\right)
    \end{array}\\
    \begin{array}{l}
      \tmmathbf{e}^{0 \nocomma 0}_{0 \nocomma 1 \nocomma 0} (\tmmathbf{h})
      \cdot (z_1^{\dag}, z_2^{\dag}) = \left(\begin{array}{c|c|c|c}
        \{ h_1 \mymult z_1^{\dag} (T) \}_T & \tmmathbf{0} & \ast & \tmmathbf{0}
      \end{array}\right) .
    \end{array}
  \end{array} \label{eq:nlCylinderStructureCoefficients}
\end{equation}
This yields the first set of operators as
\begin{equation}
  \begin{array}{rcl}
    \tmmathbf{A}_{\tmmathbf{h}} & = & \tmmathbf{0}\\
    \tmmathbf{C}_{\tmmathbf{h}}^{(0)} & = & \tmmathbf{0}\\
    \tmmathbf{C}_{\tmmathbf{h}}^{(1)} \cdot \tmmathbf{z} & = &
    \overline{\Sigma}_{\parallel} (h_1, \mu_{(h_1)}) \mymult h_1 \mymult
    \int_0^{\rho} z_1 (T) \mymult 2 \mymult \pi \mymult T \mymult \mathd T\\
    \frac{1}{2} \mymult \tmmathbf{h}^{\dag} \cdot
    \tmmathbf{B}_{\tmmathbf{h}}^{(0)} \cdot \tmmathbf{h}^{\dag} & = &
    \frac{1}{2} \mymult (h^{\dag}_1)^2 \mymult \frac{2 \mymult \pi \mymult
    \rho^4}{4} \mymult (\nabla \mu_{(h_1)})^2 \mymult (\mu_{(h_1)}^2 \mymult
    \overline{K}_{S \nocomma T}^{S \nocomma T} + \overline{\Sigma}_{\parallel}
    (h_1, \mu_{(h_1)}))\\
    \tmmathbf{h}^{\dag} \cdot \tmmathbf{B}_{\tmmathbf{h}}^{(1)} \cdot
    \tmmathbf{z} & = & h_1^{\dag} \mymult \left( \overline{K}_{S \nocomma T}^{S
    \nocomma T} \mymult \mu_{(h_1)} \mymult \nabla \mu_{(h_1)} \mymult h_1 \mymult
    \int_0^{\rho} T \mymult \partial_T z_1 (T) \mymult 2 \mymult \pi \mymult T
    \mymult \mathd T \right)\\
    &  & \nobracket \nobracket \quad - h_1^{\dag} \mymult \frac{\mathd
    (\overline{\Sigma}_{\parallel} (h_1, \mu_{(h_1)}) \mymult h_1)}{\mathd h_1}
    \mymult \int_0^{\rho} z_1 (T) \mymult 2 \mymult \pi \mymult T \mymult \mathd
    T\\
    \frac{1}{2} \mymult \tmmathbf{z} \cdot \tmmathbf{B}_{\tmmathbf{h}}^{\text{}
    (2)} \cdot \tmmathbf{z} & = & \frac{1}{2} \mymult \int_0^{\rho} h_1^2
    \mymult \overline{K}_{S \nocomma T}^{S \nocomma T} \mymult (\partial_T z_1
    (T))^2 \mymult 2 \mymult \pi \mymult T \mymult \mathd T\\
    &  & \nobracket \nobracket \quad + \frac{1}{2} \mymult \int_0^{\rho}
    \left(\begin{array}{c}
      \partial_T z_2 (T)\\
      \frac{z_2 (T)}{T}
    \end{array}\right) \cdot \tmmathbf{Q} (h_1) \cdot \left(\begin{array}{c}
      \partial_T z_2 (T)\\
      \mymult \frac{z_2 (T)}{T}
    \end{array}\right) \mymult 2 \mymult \pi \mymult T \mymult \mathd T
  \end{array} \label{eq:nlCylOperators}
\end{equation}
where
\[ \tmmathbf{Q} (h_1) = \mu_{(h_1)}^2 \mymult \mymult \left(\begin{array}{cc}
     \overline{K}_{T \nocomma T}^{T \nocomma T} & \overline{K}_{T \nocomma
     T}^{\Theta \nocomma \Theta}\\
     \overline{K}_{T \nocomma T}^{\Theta \nocomma \Theta} & \overline{K}_{T
     \nocomma T}^{T \nocomma T}
   \end{array}\right) \]
and we recall that $\nabla$ is a gradient with the respect to the macroscopic
strain $\tmmathbf{h}$, i.e.,
\[ \nabla \mu_{(h_1)} = \frac{\mathd \mu_{(h)}}{\mathd h} (h_1) . \]
This quantity $\nabla \mu_{(h_1)}$ can be found by differentiating the
implicit equation~(\ref{eq:nlCylinderEquil}).

The local correction $\tmmathbf{z}=\tmmathbf{z}^{\text{opt}}$ is obtained by
making the quantity $\frac{1}{2} \mymult \tmmathbf{z} \cdot
\tmmathbf{B}_{\tmmathbf{h}}^{\text{} (2)} \cdot
\tmmathbf{z}+\tmmathbf{h}^{\dag} \cdot \tmmathbf{B}_{\tmmathbf{h}}^{(1)} \cdot
\tmmathbf{z}$ stationary for given values of $\tmmathbf{h}$ and~$\tmmathbf{h}^{\dag}$,
subject to the constraint $\tmmathbf{q} (\tmmathbf{z}) =\tmmathbf{0}$. This
variational problem is solved in \ref{app:nlCylinderZCorrection}. The
solution can be cast in the form $\tmmathbf{z}_{\text{opt}} (S)
=\tmmathbf{Z}_{\text{opt}}^{\tmmathbf{h} (S)} \cdot \tmmathbf{h}' (S)$
announced in equation~(\ref{eq:correctiveDisplacementFactorOutHPrime}) with
\begin{equation}
	\tmmathbf{Z}^{\tmmathbf{h} (S)}_{\text{opt}} = \left(\begin{array}{c}
     \left\{ - \frac{1}{2} \mymult \frac{\mu_{(h 1)} \mymult \nabla
     \mu_{(h_1)}}{h_1} \mymult \left( T^2 - \frac{\rho^2}{2} \right)
     \right\}_T\\
     \left\{ 0 \vphantom{\frac{1}{2}} \right\}_T
   \end{array}\right) .
   \label{eq:ZoptCylinder}
\end{equation}
Stated differently, the corrective displacement is purely longitudinal
($z_2^{\text{opt}} = 0$), and parabolic: $z_1^{\text{opt}} = - \frac{1}{2}
\mymult \frac{\mu_{(h 1)} \mymult \nabla \mu_{(h_1)}}{h_1} \mymult \left( T^2 -
\frac{\rho^2}{2} \right) \mymult h_1' (S)$.

The following integral, which is required later on, can be calculated based on
the expressions of $z_1^{\text{opt}}$ just found as
\begin{equation}
  \int_0^{\rho} \left( \partial_T z_1^{\text{opt}} (T) \right)^2 \mymult 2
  \mymult \pi \mymult T \mymult \mathd T = \frac{\pi \mymult \rho^4}{2} \mymult
  \left( \frac{\mu_{(h_1)} \mymult \nabla \mu_{(h_1)}}{h_1} \right)^2 \mymult
  h_1^{\prime 2} . \label{eq:capillaryTubeIntegraldTz1}
\end{equation}

A necessary condition for the stability of this microscopic solution
(see~{\textsection}\ref{ssec:generalStability}) is that $\overline{K}_{S
\nocomma T}^{S \nocomma T} \geq 0$ and the submatrix of the tangent
moduli $\left(\begin{array}{cc}
  \overline{K}_{\Theta \nocomma \Theta}^{\Theta \nocomma \Theta} &
  \overline{K}_{T \nocomma T}^{\Theta \nocomma \Theta}\\
  \overline{K}_{T \nocomma T}^{\Theta \nocomma \Theta} & \overline{K}_{T
  \nocomma T}^{T \nocomma T}
\end{array}\right)$ is positive.

\subsection{Regularized model}

We can finally calculate the two operators entering in the 1d
model as
\[ \begin{array}{rcl}
     \frac{1}{2} \mymult \tmmathbf{h}' \cdot \tmmathbf{B}_{\tmmathbf{h}} \cdot
     \tmmathbf{h}^{'} & = & \frac{1}{2} \mymult \tmmathbf{h}' \cdot
     \tmmathbf{B}_{\tmmathbf{h}}^{(0)} \cdot \tmmathbf{h}^{'} - \frac{1}{2}
     \mymult \tmmathbf{z}_{\text{opt}} \cdot
     \tmmathbf{B}_{\tmmathbf{h}}^{\text{} (2)} \cdot
     \tmmathbf{z}_{\text{opt}}\\
     & = & \frac{1}{2} \mymult h_1^{\prime 2} \mymult \frac{2 \mymult \pi \mymult
     \rho^4}{4} \mymult (\nabla \mu_{(h_1)})^2 \mymult
     (\overline{\Sigma}_{\parallel} (h_1, \mu_{(h_1)}) + \mu_{(h_1)}^2 \mymult
     \overline{K}_{S \nocomma T}^{S \nocomma T}) - \frac{1}{2} \mymult h_1^2
     \mymult \overline{K}_{S \nocomma T}^{S \nocomma T} \mymult \int_0^{\rho}
     (\partial_T z_1 (T))^2 \mymult 2 \mymult \pi \mymult T \mymult \mathd T\\
     & = & \frac{h_1^{\prime 2}}{2} \mymult \frac{\pi \mymult \rho^4}{2} \mymult
     (\nabla \mu_{(h_1)})^2 \mymult \overline{\Sigma}_{\parallel} (h_1,
     \mu_{(h_1)})
   \end{array} \]
where we have used the equilibrium condition~(\ref{eq:nlCylinderEquil}) and
the identity~(\ref{eq:capillaryTubeIntegraldTz1}).

The operator $\tmmathbf{C}_{\tmmathbf{h}}$ reads
\[ \tmmathbf{C}_{\tmmathbf{h}} \cdot \tmmathbf{h}^{'}
   =\tmmathbf{C}_{\tmmathbf{h}}^{(0)} \cdot \tmmathbf{h}^{'}
   +\tmmathbf{C}_{\tmmathbf{h}}^{(1)} \cdot
   \tmmathbf{z}_{\text{opt}}^{\tmmathbf{h}} = \overline{\Sigma}_{\parallel}
   (h_1, \mu_{(h_1)}) \mymult h_1 \mymult \int_0^{\rho} z_1^{\text{opt}} (T)
   \mymult 2 \mymult \pi \mymult T \mymult \mathd T = 0. \]

Switching to the more familiar notation $\lambda (S)$ for the apparent
stretch $\lambda = h_1$, we finally obtain the 1d energy
governing the cylinder as
\begin{subequations}
	\label{eq:cylNeckingFinalEnergy}
\begin{equation}
  \Phi^{\star} [\lambda] = \int_0^L W_{\text{hom}} (\lambda (S)) \mymult \mathd
  S + \frac{1}{2} \mymult \int_0^L B (\lambda (S)) \mymult \lambda^{\prime 2}
  (S) \mymult \mathd S,
\end{equation}
where $W_{\text{hom}} (\lambda) = \int_0^{\rho} \overline{W} \left(
\frac{1}{2} \mymult (\lambda^2 - 1), 0, \mu_{(\lambda)}^2 - 1, \frac{1}{2}
\mymult (\mu_{(\lambda)}^2 - 1)^2, 0 \right) \mymult 2 \mymult \pi \mymult T
\mymult \mathd T$ is the energy of the homogeneous solution per unit length,
the strain-gradient modulus is given by
\begin{equation}
 B (\lambda) = \frac{\pi \mymult \rho^4}{2} \mymult \left( \frac{\mathd
   \mu_{(\lambda)}}{\mathd \lambda} \right)^2 \mymult
   \overline{\Sigma}_{\parallel} (\lambda, \mu_{(\lambda)}),
   \end{equation}
\end{subequations}
and the transverse stretch $\mu_{(\lambda)}$ is found by solving the
transverse equilibrium of a homogeneous solution, $\overline{\Sigma}_{\perp}
(\lambda, \mu_{(\lambda)}) = 0$.

\subsection{Comments}

We have recovered in equation~(\ref{eq:cylNeckingFinalEnergy}) the energy
functional derived by
{\citet{Audoly-Hutchinson-Analysis-of-necking-based-2016}}. 
As earlier with the membrane model, see {\textsection}\ref{sec:membranes}, the
strain-gradient modulus $B (\lambda)$ is directly proportional to the
pre-stress $\overline{\Sigma}_{\parallel}$ of the homogeneous solution, and
does not depend on the elastic moduli. This can be explained as follows. The
expression of the operators $\tmmathbf{B}_{\tmmathbf{h}}^{(i)}$ in
equation~(\ref{eq:preliminaryOperatorsDef}), reveal that the contribution to
the strain-gradient modulus coming from the elastic moduli arises fully from
the expansion of $\frac{1}{2} \mymult \tmmathbf{e}_{[1]} \cdot
\tmmathbf{K}_{\tmmathbf{h}} \cdot \tmmathbf{e}_{[1]}$, where
$\tmmathbf{e}_{[1]} =\tmmathbf{e}^{1 \nocomma 0}_{0 \nocomma 0 \nocomma 0}
\cdot \tmmathbf{h}' +\tmmathbf{e}^{0 \nocomma 0}_{1 \nocomma 0 \nocomma 0}
\cdot \tmmathbf{z}$ is the first-order correction to the strain, see also
equation~(\ref{eq:e1+e2}). If the correction $\tmmathbf{z}$ manages to cancel
out entirely the strain $\tmmathbf{e}^{1 \nocomma 0}_{0 \nocomma 0 \nocomma 0}
\cdot \tmmathbf{h}'$ arising from the gradient effect, then
$\tmmathbf{e}_{[1]} =\tmmathbf{0}$ and the strain-gradient modulus arises from
the pre-strain $\tmmathbf{\Sigma}_{\tmmathbf{h}}$ only. This is what happens
with both the axisymmetric membrane, for which $\tmmathbf{e}^{1 \nocomma 0}_{0
\nocomma 0 \nocomma 0} \cdot \tmmathbf{h}' =\tmmathbf{0}$ from
equation~(\ref{eq:membraneStructureCoefficients}), and for the axisymmetric
cylinder, for which $\tmmathbf{e}^{1 \nocomma 0}_{0 \nocomma 0 \nocomma 0}
\cdot \tmmathbf{h}'$ is a shear strain that is canceled out by the out of
plane deformation of the cross-section, as discussed by
{\citet{Audoly-Hutchinson-Analysis-of-necking-based-2016}}. For
the bending of an elastic block, however, the corrective displacement
$\tmmathbf{z}$ does not fully suppresses the first order strain,
$\tmmathbf{e}_{[1]} \neq \tmmathbf{0}$, and the elastic moduli enter into the
expression of the strain-gradient modulus.

\section{Conclusion and discussion}\label{sec:conclusion}

We have presented a systematic reduction method which, given a
structural model representing a prismatic elastic solid, yields a 1d
model that captures the strain gradient effect.  The method implements
a two-scale expansion and is asymptotically exact.  It is based on a
choice of macroscopic strain variables which are retained in the 1d
model, and a choice of microscopic variables which are relaxed during
the reduction process.  It can be applied as a simple recipe, i.e., it
requires one to follow a systematic sequence of steps in order.  The
method retains the nonlinearity of the initial model, and can account
for large and inhomogeneous changes in the shapes of cross-sections.

As illustrated by the worked examples, this method can be used to
recover known 1d models for structures in a systematic and unified
way.  In future work, it will be used to derive original 1d models,
e.g., for structures possessing highly deformable cross-sections, such
as tape springs, or having large and inhomogeneous
pre-stress~{\citep{Liu-Huang-EtAl-Structural-Transition-from-2014,Lestringant-Audoly-Elastic-rods-with-2016}}.
The method can also be applied to the analysis of localization, which
is ubiquitous in slender structures.  In the absence of 1d models
capturing the gradient effect, the various examples of localization
have been addressed using equations that are specific to each
particular structure, such as the axisymmetric membrane theory for
bulges in
balloons~{\citep{Fu-Pearce-EtAl-Post-bifurcation-analysis-of-a-thin-walled-2008,Pearce-Fu-Characterization-and-stability-of-localized-2010}};
the authors have suggested recently that, by using dimension
reduction, it is possible to analyze the various examples of
localization in a unified mathematical
framework~{\citep{Lestringant-Audoly-A-diffuse-interface-model-2018}}.

We close this paper with a few general remarks.

In the reduction method, the choice of the macroscopic strain
$\tmmathbf{h}$ has been left to the user.  This choice actually
reflects how the applied load scales with the slenderness parameter
(note that the external load has not appeared in our reduction
reduction method).  The scaling assumptions for the load are a key
ingredient in dimension reduction, and different assumptions can lead
to different 1d
model~{\citep{Marigo-Hierarchy-of-One-Dimensional-Models-2006}}.
Similarly, different choices of $\tmmathbf{h}$ ultimately lead to
different 1d models using our method.  Consider for instance what
happens if we add a `strong' external shearing force $\tmmathbf{g}$ on
the linear elastic block, $\tmmathbf{g} (S, T) = G (S) \mymult
\frac{T}{a^3 / 12} \mymult \tmmathbf{e}_x$.  This external load
induces a moment $G (S) \mymult \tmmathbf{e}_z$ perpendicular to the
plane of the block, in each cross-section.  The kinematic quantity
conjugate to $\tmmathbf{g}$ is the average rotation of the
cross-section, which in the linear setting reads $\frac{1}{a^2 / 12}
\langle \tmmathbf{y} \cdot (T \mymult \tmmathbf{e}_x) \rangle =
\frac{\langle T \mymult u \rangle}{a^2 / 12} + V'$.  This external
load can therefore be handled by including an additional kinematic
constraint $\langle T \mymult u \rangle (S) = \varphi (S)$ in the
reduction method, and by augmenting macroscopic strain $\tmmathbf{h}$
with this new internal variable, $h_3 = \varphi$.  The reduction
method has to be redone, yielding this time a Timoshenko beam model.
The external load is then taken into account simply by coupling the
load intensity $G (S)$ with $\frac{\varphi (S)}{a^2 / 12} + V' (S)$.
Note that this modification of the reduction procedure is required if,
and only if, the external load is strong and inhomogeneous enough that
it modifies significantly the natural microscopic displacement.  For
mild applied force, the external moment $G (S)$ can be coupled
directly to the macroscopic rotation $V'$, saving one from the need to
amend the dimension reduction.

In their analysis of a 1d model for stretched bars,
{\citet{Coleman-Newman-On-the-rheology-of-cold-drawing.-1988}} have
proposed a derivation of the gradient effect starting from the
assumption that the kinematics of the classical bar model without
gradient effect remains valid, i.e., that the cross-sections remain
planar.  A similar approximation has been used by several authors in
various derivations of strain gradient models.  In our notation, this
amounts to neglecting the microscopic correction $\tmmathbf{z}$
arising from the gradient effect, i.e., to set
$\tmmathbf{Z}_{\text{opt}}^{\tmmathbf{h} (S)} =\tmmathbf{0}$, see
equation~(\ref{eq:correctiveDisplacementFactorOutHPrime}).
Equation~(\ref{eq:finalOperatorsDef}) then yields
$\tmmathbf{B}_{\tmmathbf{h}} =\tmmathbf{B}_{\tmmathbf{h}}^{(0)}$.  The
inequality~(\ref{eq:ColemanNewmanInequality}) shows that the strain
gradient modulus derived from this kinematic assumption has a larger
strain-gradient elastic modulus $\tmmathbf{B}_{\tmmathbf{h}}^{(0)}$
than the true modulus predicted by our method with due account for the
microscopic relaxation $\tmmathbf{z}$, as noted
by~{\citet{Audoly-Hutchinson-Analysis-of-necking-based-2016}}.  This
is not surprising, as our approach relaxes
the strain energy optimally by design, while the former does not.

For the sake of simplicity, we have assumed that the full model is
invariant along the longitudinal direction.  It is quite simple to
extend the method to the case where the geometry and/or elastic
properties of the full model vary slowly in the longitudinal
direction, i.e., depend on the stretched variable $\tilde{S}$ used in
\ref{sec:detailedProofs}.  An additional explicit dependence on
$\tilde{S}$ must then be added to the various quantities entering into
the analysis, such as $W$, $\tmmathbf{E}_{\tmmathbf{h}}$,
$W_{\text{hom}}$, $\tmmathbf{\Sigma}_{\tmmathbf{h}}$, etc.  This
brings in a single significant change: in the integrand of
$\tilde{\Phi}_{[2]}$ in equation~(\ref{eq:phiTilde2}), one needs to
include two extra terms $- \left[ \frac{\mathd
\tmmathbf{C}_{\tmmathbf{h}}^{(0)}}{\mathd \tilde{S}} (\tilde{S})
\right]_{\tmmathbf{h}=\tmmathbf{h} (\tilde{S})} \cdot
\dot{\tmmathbf{h}}$ and $- \left[ \frac{\mathd
\tmmathbf{C}_{\tmmathbf{h}}^{(1)}}{\mathd \tilde{S}} (\tilde{S})
\right]_{\tmmathbf{h}=\tmmathbf{h} (\tilde{S})} \cdot
\tmmathbf{z}_{[1]}$ in order to cancel out the extra terms coming from
the expansion of the {\tmem{total}} derivatives $\frac{\mathd
(\tmmathbf{C}_{\tmmathbf{h} (\tilde{S})}^{(0)} (\tilde{S}))}{\mathd
\tilde{S}} \cdot \dot{\tmmathbf{h}} + \frac{\mathd
(\tmmathbf{C}_{\tmmathbf{h} (\tilde{S})}^{(1)} (\tilde{S}))}{\mathd
\tilde{S}} \cdot \tmmathbf{z}_{[1]}$, when the
$\tmmathbf{C}_{\tmmathbf{h}}^{(j)} (\tilde{S})$'s depend explicitly on
$\tilde{S}$.  These two additional terms make their way into the final
expression of the 1d energy.

The illustration examples in
sections~\ref{sec:membranes}--\ref{sec:hyperCylinder} were simple
enough that they could be solved analytically.  When this is not
possible, the proposed reduction method lends itself naturally to a
numerical implementation: by solving numerically a series of
elasticity problem over the cross-section, it is possible to build a
numerical representation of the various operators
$\tmmathbf{A}_{\tmmathbf{h}}$, $\tmmathbf{B}_{\tmmathbf{h}}$ and
$\tmmathbf{C}_{\tmmathbf{h}}$, i.e., to evaluate numerically the
coefficients of the 1d energy $\Phi^{\star}$.

This paper was prepared using
TeXmacs~\citep{Hoeven-Grozin-EtAl-GNU-TEXmacs:-a-scientific-editing-2013},
an outstanding and freely available scientific text editor.

\appendix\section{Proof of the main results}\label{sec:detailedProofs}

In this appendix, we offer a detailed justification of the results
announced in~{\textsection}\ref{sec:mainResults}.

\subsection{Change of microscopic variable}

We return to the relaxation problem~(\ref{eq:relaxedEnergy}) for a prescribed,
non-homogeneous distribution of macroscopic strain $\{ \tmmathbf{h} (S) \}_S$,
and seek the microscopic displacement achieving the optimum of the functional
$\Phi [\tmmathbf{h}, \tmmathbf{y}]$ in equation~(\ref{eq:generalStrainEnergy})
subject to the constraint $\tmmathbf{q} (\tmmathbf{y} (S))$ in
equation~(\ref{eq:qConstraint}). We seek $\tmmathbf{y} (S)$ in the form~(\ref{eq:guess}), $\tmmathbf{y} (S)
=\tmmathbf{y}_{\tmmathbf{h} (S)} +\tmmathbf{z} (S)$. Since
$\tmmathbf{y}_{\tmmathbf{h} (S)}$ satisfies the linear constraint by
construction, the new unknown $\tmmathbf{z} (S)$ must satisfy the constraint
as well, $\forall S \quad \tmmathbf{q} (\tmmathbf{z} (S)) =\tmmathbf{0}.$

In terms of $\tmmathbf{z} (S)$, the gradients of the original microscopic
displacement write
\[ \begin{array}{lll}
     \tmmathbf{y}' (S) & = & \tmmathbf{h}' (S) \cdot \nabla
     \tmmathbf{y}_{\tmmathbf{h} (S)} +\tmmathbf{z}' (S)\\
     \tmmathbf{y}'' (S) & = & \tmmathbf{h}'' (S) \cdot \nabla
     \tmmathbf{y}_{\tmmathbf{h} (S)} +\tmmathbf{h}' (S) \cdot \nabla
     \tmmathbf{y}_{\tmmathbf{h} (S)} \cdot \tmmathbf{h}' (S) +\tmmathbf{z}''
     (S)
   \end{array} \]
where $\nabla$ denotes the gradient with respect to the macroscopic parameter
$\tmmathbf{h}$, see equation~(\ref{eq:nablaDefinition}). In view of this, the
strain appearing in the strain energy in
equation~(\ref{eq:generalStrainEnergy}) can be expressed in terms of the new
unknown as
\[ \tmmathbf{E}=\tmmathbf{e}_{\tmmathbf{h} (S)} (\tmmathbf{h}' (S),
   \tmmathbf{h}'' (S) ; \tmmathbf{z} (S), \tmmathbf{z}' (S), \tmmathbf{z}''
   (S)), \]
where $\tmmathbf{e}_{\tmmathbf{h}} (\tmmathbf{h}^{\dag}, \tmmathbf{h}^{\ddag}
; \tmmathbf{z}, \tmmathbf{z}^{\dag}, \tmmathbf{z}^{\ddag})$ is the strain
function introduced in equation~(\ref{eq:strainInTermsOfZ}).

Note that the principle of virtual work for the homogeneous
solution~(\ref{eq:outlineHomPVW}) can be rewritten in hindsight using the
structure coefficient $\tmmathbf{e}^{0 \nocomma 0}_{1 \nocomma 0 \nocomma 0}$
as
\begin{equation}
  \begin{array}{l}
    \forall \hat{\tmmathbf{y}} \quad -\tmmathbf{\Sigma}_{\tmmathbf{h}} \cdot
    (\tmmathbf{e}^{0 \nocomma 0}_{1 \nocomma 0 \nocomma 0} (\tmmathbf{h})
    \cdot \hat{\tmmathbf{y}}) +\tmmathbf{f}_{\tmmathbf{h}} \cdot \tmmathbf{q}
    (\hat{\tmmathbf{y}}) = 0\\
    \tmmathbf{q} (\tmmathbf{y}_{\tmmathbf{h}}) =\tmmathbf{0}.
  \end{array} \label{eq:homogeneousPVWcompact}
\end{equation}

\subsection{Expansion method}

Given the distribution of macroscopic strain $\tmmathbf{h} (S)$, the
strain energy of the full model is expressed in terms of the corrective
displacement $\tmmathbf{z} (S)$ as
\[ \Phi [\tmmathbf{h}, \tmmathbf{y}_{\tmmathbf{h}} +\tmmathbf{z}] = \int_0^L W
   (\tmmathbf{e}_{\tmmathbf{h} (S)} (\tmmathbf{h}' (S), \tmmathbf{h}'' (S) ;
   \tmmathbf{z} (S), \tmmathbf{z}' (S), \tmmathbf{z}'' (S))) \mymult \mathd S.
\]
This energy must be relaxed with respect to $\tmmathbf{z}$, subject to the
constraint
\[ \forall S \quad \tmmathbf{q} (\tmmathbf{z} (S)) =\tmmathbf{0}. \]

This relaxation problem is treated by an asymptotic method, which assumes that
the prescribed strain $\tmmathbf{h} (S)$ is a slowly varying function of $S$.
Accordingly, we introduce a stretched variable $\tilde{S} = \gamma \mymult S$
where $\gamma \ll 1$ is a small parameter. We denote by a dot the derivation
with respect to the new variable,
\[ \dot{f} = \frac{\mathd f}{\mathd \tilde{S}} . \]
The assumption of slow axial variations is implemented by requiring that the
dependence of the various functions on $S$ encountered so far is replaced with
a dependence on the slow variable $\tilde{S}$. This amounts to assuming
$\dot{f} =\mathcal{O} (1)$, $\ddot{f} =\mathcal{O} (1)$, etc.~while the
original derivatives scale as
\[ f' = \gamma \mymult \dot{f} =\mathcal{O} (\gamma) \quad f'' = 
\gamma^{2} \mymult
   \ddot{f} =\mathcal{O} (\gamma^2) \quad \text{etc.} \]

We seek the microscopic correction as an expansion
\begin{equation}
  \tmmathbf{z} (\tilde{S}) = \gamma \mymult \tmmathbf{z}_{[1]} (\tilde{S}) +
  \gamma^2 \mymult \tmmathbf{z}_{[2]} (\tilde{S}) + \cdots
  \label{eq:zExpansion}
\end{equation}
where the absence of any term of order $\gamma^0 = 1$ follows from the
observation that in the homogeneous case, corresponding to the formal limit
$\gamma \rightarrow 0$, we have $\tmmathbf{z} (S) = 0$.

The strain energy to be relaxed can be rescaled as $\tilde{\Phi} = \Phi /
\gamma$, where
\begin{equation}
  \tilde{\Phi} [\tmmathbf{h}, \tmmathbf{z}] = \int_0^{\tilde{L}} W
  (\tmmathbf{e}_{\tmmathbf{h}} (\gamma \mymult \dot{\tmmathbf{h}} (\tilde{S}),
  \gamma^2 \mymult \ddot{\tmmathbf{h}} (\tilde{S}) ; \gamma \mymult
  \tmmathbf{z}_{[1]} (\tilde{S}) + \gamma^2 \mymult \tmmathbf{z}_{[2]}
  (\tilde{S}) + \cdots, \gamma^2 \mymult \dot{\tmmathbf{z}}_{[1]} (\tilde{S}) +
  \cdots, \tmmathbf{0})) \mymult \mathd \tilde{S} +\mathcal{O} (\gamma^3),
  \label{eq:ObjectiveFunctionAllOrders}
\end{equation}
and $\tilde{L} = \gamma \mymult L$, subject to the constraint
\begin{equation}
  \forall i \geq 1 \quad \forall \tilde{S} \quad \tmmathbf{q}
  (\tmmathbf{z}_{[i]} (\tilde{S})) =\tmmathbf{0}. \label{eq:constraintForZis}
\end{equation}
We now proceed to derive an expansion of the energy $\tilde{\Phi}$ in powers
of $\gamma$, and to solve this problem order by order.

\subsection{Strain expansion}

The argument $\tmmathbf{e}_{\tmmathbf{h}}$ of $W$
in~(\ref{eq:ObjectiveFunctionAllOrders}) is the strain. It can be expanded as
\begin{equation}
    \tmmathbf{e}_{\tmmathbf{h}} (\gamma \mymult \dot{\tmmathbf{h}} (\tilde{S}),
    \gamma^2 \mymult \ddot{\tmmathbf{h}} (\tilde{S}) ; \gamma \mymult
    \tmmathbf{z}_{[1]} (\tilde{S}) + \gamma^2 \mymult \tmmathbf{z}_{[2]}
    (\tilde{S}), \gamma^2 \mymult \dot{\tmmathbf{z}}_{[1]} (\tilde{S}),
    \tmmathbf{0}) 
     =\tmmathbf{E}_{\tmmathbf{h} (\tilde{S})} + \gamma
    \mymult \tmmathbf{e}_{[1]} (\tilde{S}) + \gamma^2 \mymult \tmmathbf{e}_{[2]}
    (\tilde{S}) +\mathcal{O} (\gamma^3) \label{eq:strainExpansion}
\end{equation}
where term of order $\gamma^0$ is the homogeneous strain
$\tmmathbf{E}_{\tmmathbf{h} (\tilde{S})} =\tmmathbf{E} (\tmmathbf{h}
(\tilde{S}), \tmmathbf{0}; \tmmathbf{y}_{\tmmathbf{h} (\tilde{S})},
\tmmathbf{0}, \tmmathbf{0})$ defined in equation~(\ref{eq:defineHomogeneousProperties}), and the
linear and quadratic corrections can be written in terms of the structure
coefficients as
\begin{equation}
  \begin{array}{lll}
    \tmmathbf{e}_{[1]} (\tilde{S}) & = & \tmmathbf{e}^{1 \nocomma 0}_{0
    \nocomma 0 \nocomma 0} (\tmmathbf{h} (\tilde{S})) \cdot \dot{\tmmathbf{h}}
    (\tilde{S}) +\tmmathbf{e}^{0 \nocomma 0}_{1 \nocomma 0 \nocomma 0}
    (\tmmathbf{h} (\tilde{S})) \cdot \tmmathbf{z}_{[1]} (\tilde{S})\\
    \tmmathbf{e}_{[2]} (\tilde{S}) & = & \frac{1}{2} \mymult
    (\dot{\tmmathbf{h}} \cdot \tmmathbf{e}^{2 \nocomma 0}_{0 \nocomma 0
    \nocomma 0} (\tmmathbf{h}) \cdot \dot{\tmmathbf{h}} + 2 \mymult
    \dot{\tmmathbf{h}} \cdot \tmmathbf{e}^{1 \nocomma 0}_{1 \nocomma 0
    \nocomma 0} (\tmmathbf{h}) \cdot \tmmathbf{z}_{[1]} +\tmmathbf{z}_{[1]}
    \cdot \tmmathbf{e}^{0 \nocomma 0}_{2 \nocomma 0 \nocomma 0} (\tmmathbf{h})
    \cdot \tmmathbf{z}_{[1]}) \cdots\\
    &  & \hspace{2.5cm} +\tmmathbf{e}^{0 \nocomma 1}_{0 \nocomma 0 \nocomma
    0} (\tmmathbf{h}) \cdot \ddot{\tmmathbf{h}} +\tmmathbf{e}^{0 \nocomma
    0}_{0 \nocomma 1 \nocomma 0} (\tmmathbf{h}) \cdot \dot{\tmmathbf{z}}_{[1]}
    +\tmmathbf{e}^{0 \nocomma 0}_{1 \nocomma 0 \nocomma 0} (\tmmathbf{h})
    \cdot \tmmathbf{z}_{[2]}.
  \end{array} \label{eq:e1+e2}
\end{equation}
All the quantities appearing in the right-hand side of $\tmmathbf{e}_{[2]}
(\tilde{S})$ must be evaluated at $\tilde{S}$, like those in the right-hand
side of $\tmmathbf{e}_{[2]} (\tilde{S})$.

\subsection{Energy expansion}

The strain expansion~(\ref{eq:strainExpansion}) can be inserted into the
energy in~(\ref{eq:ObjectiveFunctionAllOrders})
\[ \tilde{\Phi} [\tmmathbf{h}, \tmmathbf{z}] = \int_0^{\tilde{L}} (W
   (\tmmathbf{e}_{\tmmathbf{h} (\tilde{S})} + \gamma \mymult \tmmathbf{e}_{[1]}
   (\tilde{S}) + \gamma^2 \mymult \tmmathbf{e}_{[2]} (\tilde{S})) +\mathcal{O}
   (\gamma^3)) \mymult \mathd \tilde{S} \]
This yields an expansion of the energy as
\begin{equation}
  \tilde{\Phi} [\tmmathbf{h}, \tmmathbf{z}] = \tilde{\Phi}_{[0]}
  [\tmmathbf{h}] + \gamma \mymult \tilde{\Phi}_{[1]} [\tmmathbf{h}] + \gamma^2
  \mymult \tilde{\Phi}_{[2]} [\tmmathbf{h}, \tmmathbf{z}_{[1]}] + \cdots
  \label{eq:PhiTildaExpansion}
\end{equation}
where the first orders in the expansion read
\[ \begin{array}{rll}
     \tilde{\Phi}_{[0]} [\tmmathbf{h}] & = & \int_0^{\tilde{L}} W_{\text{hom}}
     (\tmmathbf{h} (\tilde{S})) \mymult \mathd \tilde{S}\\
     \tilde{\Phi}_{[1]} [\tmmathbf{h}] & = & \int_0^{\tilde{L}}
     \tmmathbf{\Sigma}_{\tmmathbf{h} (\tilde{S})} \cdot \tmmathbf{e}_{[1]}
     (\tilde{S}) \mymult \mathd \tilde{S}\\
     \tilde{\Phi}_{[2]} [\tmmathbf{h}, \tmmathbf{z}_{[1]}] & = &
     \int_0^{\tilde{L}} \left( \frac{1}{2} \mymult \tmmathbf{e}_{[1]}
     (\tilde{S}) \cdot \tmmathbf{K}_{\tmmathbf{h} (\tilde{S})} \cdot
     \tmmathbf{e}_{[1]} (\tilde{S}) +\tmmathbf{\Sigma}_{\tmmathbf{h}
     (\tilde{S})} \cdummy \tmmathbf{e}_{[2]} (\tilde{S}) \right) \mymult \mathd
     \tilde{S}.
   \end{array} \]
Our notation anticipates on the fact that $\tilde{\Phi}_{[1]}$ does not depend
on $\tmmathbf{z}$, and that $\tilde{\Phi}_{[2]}$ depends on $\tmmathbf{z}$
through its dominant contribution $\tmmathbf{z}_{[1]}$ only, as we prove in
the next section.

\subsection{Rearranging the energy contributions}\label{ssec:eliminateZDot}

Using the operator $\tmmathbf{A}_{\tmmathbf{h}}$ introduced in
equation~(\ref{eq:preliminaryOperatorsDef}), the energy contribution
$\tilde{\Phi}_{[1]}$ can be written as
\[ \tilde{\Phi}_{[1]} [\tmmathbf{h}] = \int_0^{\tilde{L}}
   (\tmmathbf{A}_{\tmmathbf{h} (\tilde{S})} \cdot \dot{\tmmathbf{h}}
   (\tilde{S}) +\tmmathbf{\Sigma}_{\tmmathbf{h} (\tilde{S})} \cdot
   (\tmmathbf{e}^{0 \nocomma 0}_{1 \nocomma 0 \nocomma 0} (\tmmathbf{h}
   (\tilde{S})) \cdot \tmmathbf{z}_{[1]} (\tilde{S}))) \mymult \mathd \tilde{S}.
\]
The second term in the integrand can be simplified by using the principle of
virtual work for homogeneous solutions~(\ref{eq:homogeneousPVWcompact}):
setting the virtual motion as $\hat{\tmmathbf{y}} =\tmmathbf{y}_{\tmmathbf{h}
(\tilde{S})}$, one has $\tmmathbf{\Sigma}_{\tmmathbf{h} (\tilde{S})} \cdot
(\tmmathbf{e}^{0 \nocomma 0}_{1 \nocomma 0 \nocomma 0} (\tmmathbf{h}
(\tilde{S})) \cdot \tmmathbf{y}_{\tmmathbf{h} (\tilde{S})})
=\tmmathbf{f}_{\tmmathbf{h} (\tilde{S})} \cdot \tmmathbf{q}
(\tmmathbf{y}_{\tmmathbf{h} (\tilde{S})}) = 0$ since
$\tmmathbf{y}_{\tmmathbf{h} (\tilde{S})}$ satisfies the constraint
$\tmmathbf{q}$. This shows that the second term in the integrand above
vanishes, and that $\tilde{\Phi}_{[1]} [\tmmathbf{h}]$ is actually independent
of the microscopic displacement $\tmmathbf{z}$, as anticipated in our notation,
\[ \tilde{\Phi}_{[1]} [\tmmathbf{h}] = \int_0^{\tilde{L}}
   \tmmathbf{A}_{\tmmathbf{h} (\tilde{S})} \cdot \dot{\tmmathbf{h}}
   (\tilde{S}) \mymult \mathd \tilde{S} . \]

Most structures are invariant by the reflection $S \leftarrow (- S)$, which
implies that the operator $\tmmathbf{A}_{\tmmathbf{h}}$, as well as the
first-order correction to the energy $\tilde{\Phi}_{[1]} [\tmmathbf{h}]$,
vanish. This makes it important to determine the expansion of the energy to
second order.

At order $\gamma^2$, we have
\[ \tilde{\Phi}_{[2]} [\tmmathbf{h}, \tmmathbf{z}_{[1]}] = \int_0^{\tilde{L}}
   \left( \frac{1}{2} \mymult \tmmathbf{e}_{[1]} (\tilde{S}) \cdot
   \tmmathbf{K}_{\tmmathbf{h} (\tilde{S})} \cdot \tmmathbf{e}_{[1]}
   (\tilde{S}) +\tmmathbf{\Sigma}_{\tmmathbf{h} (\tilde{S})} \cdummy
   \tmmathbf{e}_{[2]} (\tilde{S}) \right) \mymult \mathd \tilde{S} \]
Inserting the expression of $\tmmathbf{e}_{[1]}$ and $\tmmathbf{e}_{[2]}$ in
the integrand, and using the operators defined
in~(\ref{eq:preliminaryOperatorsDef}), we can rewrite this as
\begin{equation}
  \tilde{\Phi}_{[2]} [\tmmathbf{h}, \tmmathbf{z}_{[1]}] = \int_0^{\tilde{L}}
  \left( \begin{array}{l}
    \tmmathbf{C}_{\tmmathbf{h}}^{(0)} \cdot \ddot{\tmmathbf{h}} + \frac{\mathd
    (\tmmathbf{C}_{\tmmathbf{h} (\tilde{S})}^{(0)})}{\mathd \tilde{S}} \cdot
    \dot{\tmmathbf{h}} +\tmmathbf{C}_{\tmmathbf{h}}^{(1)} \cdot
    \dot{\tmmathbf{z}}_{[1]} + \frac{\mathd (\tmmathbf{C}_{\tmmathbf{h}
    (\tilde{S})}^{(1)})}{\mathd \tilde{S}} \cdot \tmmathbf{z}_{[1]} \ldots\\
    \nobracket \nobracket + \frac{1}{2} \mymult \dot{\tmmathbf{h}} \cdot
    \tmmathbf{B}_{\tmmathbf{h}}^{(0)} \cdot \dot{\tmmathbf{h}} +
    \dot{\tmmathbf{h}} \cdot \tmmathbf{B}_{\tmmathbf{h}}^{(1)} \cdot
    \tmmathbf{z}_{[1]} + \frac{1}{2} \mymult \tmmathbf{z}_{[1]} \cdot
    \tmmathbf{B}_{\tmmathbf{h}}^{\text{} (2)} \cdot \tmmathbf{z}_{[1]}
    \cdots\\
    \nobracket \nobracket +\tmmathbf{\Sigma}_{\tmmathbf{h}} \cdot
    (\tmmathbf{e}^{0 \nocomma 0}_{1 \nocomma 0 \nocomma 0} (\tmmathbf{h})
    \cdot \tmmathbf{z}_{[2]})
  \end{array} \right) \mymult \mathd \tilde{S} \label{eq:phiTilde2}
\end{equation}
where all quantities in the integrand must evaluated at $\tilde{S}$,
namely $\tmmathbf{h}=\tmmathbf{h} (\tilde{S})$, $\dot{\tmmathbf{h}} =
\dot{\tmmathbf{h}} (\tilde{S})$, $\tmmathbf{z}_{[i]}
=\tmmathbf{z}_{[i]} (\tilde{S})$ and $\dot{\tmmathbf{z}}_{[1]} =
\dot{\tmmathbf{z}}_{[1]} (\tilde{S})$.  In this expression, we have
made appear the term $\frac{\mathd (\tmmathbf{C}_{\tmmathbf{h}
(\tilde{S})}^{(0)})}{\mathd \tilde{S}} \cdot \dot{\tmmathbf{h}}
(\tilde{S}) = \dot{\tmmathbf{h}} (\tilde{S}) \cdot \nabla
\tmmathbf{C}_{\tmmathbf{h} (\tilde{S})}^{(0)} \cdot \dot{\tmmathbf{h}}
(\tilde{S})$, which cancels out with the last term introduced in the
definition of $\frac{1}{2} \mymult \dot{\tmmathbf{h}} \cdot
\tmmathbf{B}_{\tmmathbf{h}}^{(0)} \cdot \dot{\tmmathbf{h}}$, see
equation~(\ref{eq:preliminaryOperatorsDef}).  The same holds for
$\frac{\mathd (\tmmathbf{C}_{\tmmathbf{h} (\tilde{S})}^{(1)})}{\mathd
\tilde{S}} \cdot \tmmathbf{z}_{[1]}$ and the last term in
$\dot{\tmmathbf{h}} \cdot \tmmathbf{B}_{\tmmathbf{h}}^{(1)} \cdot
\tmmathbf{z}_{[1]}$.

In addition, the last term in the integrand of $\tilde{\Phi}_{[2]}
[\tmmathbf{h}, \tmmathbf{z}_{[1]}]$ cancels by the same argument as earlier,
i.e., using the homogeneous principle of virtual
work~(\ref{eq:homogeneousPVWcompact}) and the
constraint~(\ref{eq:constraintForZis}) at second order. As a result,
$\tilde{\Phi}_{[2]}$ depends on $\tmmathbf{z}_{[1]}$ but not on
$\tmmathbf{z}_{[2]}$, as anticipated in our notation. Noting that the terms on the first line in the integrand form an exact
derivative, and can be integrated, we have
\begin{equation}
  \tilde{\Phi}_{[2]} [\tmmathbf{h}, \tmmathbf{z}_{[1]}] =
  [\tmmathbf{C}_{\tmmathbf{h}}^0 \cdot \dot{\tmmathbf{h}}
  +\tmmathbf{C}_{\tmmathbf{h}}^1 \cdot \tmmathbf{z}_{[1]}]_0^{\tilde{L}} +
  \int_0^{\tilde{L}} \left( \frac{1}{2} \mymult \dot{\tmmathbf{h}} \cdot
  \tmmathbf{B}_{\tmmathbf{h}}^{(0)} \cdot \dot{\tmmathbf{h}} +
  \dot{\tmmathbf{h}} \cdot \tmmathbf{B}_{\tmmathbf{h}}^{(1)} \cdot
  \tmmathbf{z}_{[1]} + \frac{1}{2} \mymult \tmmathbf{z}_{[1]} \cdot
  \tmmathbf{B}_{\tmmathbf{h}}^{\text{} (2)} \cdot \tmmathbf{z}_{[1]} \right)
  \mymult \mathd \tilde{S} \label{eq:phi2TildeTmp}
\end{equation}
where all the quantities in the integrand side are implicitly evaluated at
$\tilde{S}$ in our notation.

The expansion of the energy $\Phi [\tmmathbf{h}, \tmmathbf{y}_{\tmmathbf{h}}
+\tmmathbf{z}]$ in non-scaled form, announced earlier in
equation~(\ref{eq:PhiExpansionAnnounce}), is readily obtained by inserting
into equation~(\ref{eq:PhiTildaExpansion}) the expressions of
$\tilde{\Phi}_{[0]}$, $\tilde{\Phi}_{[1]}$ and $\tilde{\Phi}_{[2]}$ just
derived, and restoring the original (scaled) gradients $\tmmathbf{h}'$,
$\tmmathbf{h}''$, $\tmmathbf{z}'$ and $\tmmathbf{z}''$, as well as the
unscaled longitudinal coordinate $S$.

\subsection{Dominant order correction found by a local problem}

We now proceed to minimize the energy~(\ref{eq:ObjectiveFunctionAllOrders}) under the constraint~(\ref{eq:constraintForZis})
order by order, using the expansion of $\tilde{\Phi} [\tmmathbf{h},
\tmmathbf{z}]$ just found.
Thanks to the integration by parts, the dependence of the strain energy
$\tilde{\Phi}_{[2]} [\tmmathbf{h},\tmmathbf{z}_{[1]}]$ on the axial gradient
$\tmmathbf{z}_{[1]}'$ of the corrective displacement has been removed.
Therefore, in every cross-section $\tilde{S}$, $\tmmathbf{z}_{[1]}
(\tilde{S})$ is the solution of a {\tmem{local}} optimization problem:
$\tmmathbf{z}_{[1]} (\tilde{S})$ makes stationary the quantity
$\tmmathbf{B}_{\tmmathbf{h}}^{(1)} \cdot \tmmathbf{z}_{[1]} + \frac{1}{2}
\mymult \tmmathbf{z}_{[1]} \cdot \tmmathbf{B}_{\tmmathbf{h}}^{\text{} (2)}
\cdot \tmmathbf{z}_{[1]}$ among all $\tmmathbf{z}$'s satisfying the constraint
$\tmmathbf{q} (\tmmathbf{z}) =\tmmathbf{0}$. This leads to the variational
problem stated in equation~(\ref{eq:localOptimizationProblem}), where we use
the notation $\tmmathbf{z}_{\text{opt}} = \gamma \mymult \tmmathbf{z}_{[1]}$
for the dominant contribution to the correction $\tmmathbf{z}$, see
equation~(\ref{eq:zExpansion}), and $\tmmathbf{f}_{\text{opt}} = \gamma \mymult
\tmmathbf{f}_{[1]}$ for the scaled Lagrange multiplier. In terms of
$\tmmathbf{z}_{[1]}$, the variational problem can be written equivalently as
$\forall \hat{\tmmathbf{z}} \quad \dot{\tmmathbf{h}} (\tilde{S}) \cdot
\tmmathbf{B}_{\tmmathbf{h} (\tilde{S})}^{(1)} \cdot \hat{\tmmathbf{z}}
+\tmmathbf{z}_{[1]} (\tilde{S}) \cdot \tmmathbf{B}_{\tmmathbf{h}
(\tilde{S})}^{\text{} (2)} \cdot \hat{\tmmathbf{z}} -\tmmathbf{f}_{[1]}
(\tilde{S}) \cdot \tmmathbf{q} (\hat{\tmmathbf{z}}) = 0$ and $\tmmathbf{q}
(\tmmathbf{z}_{[1]} (\tilde{S})) =\tmmathbf{0}$.

\subsection{Relaxed energy}

By using the particular virtual motion $\hat{\tmmathbf{z}} =\tmmathbf{z}_{[1]}
(\tilde{S})$ in the above variational problem, we obtain the identity
\[ \dot{\tmmathbf{h}} (\tilde{S}) \cdot \tmmathbf{B}_{\tmmathbf{h}
   (\tilde{S})}^{(1)} \cdot \tmmathbf{z}_{[1]} (\tilde{S}) =
   -\tmmathbf{z}_{[1]} (\tilde{S}) \cdot \tmmathbf{B}_{\tmmathbf{h}
   (\tilde{S})}^{\text{} (2)} \cdot \tmmathbf{z}_{[1]} (\tilde{S}) . \]
When the optimal displacement $\tmmathbf{z}_{[1]}^{\text{opt}}$ is inserted into
$\tilde{\Phi}_{[2]} [\tmmathbf{h}, \tmmathbf{z}_{[1]}]$ in
equation~(\ref{eq:phi2TildeTmp}), and this identity is used, we obtain the
second-order contribution to the relaxed energy as
\[ \tilde{\Phi}_{[2]}^{\star} [\tmmathbf{h}] = \tilde{\Phi}_{[2]}
   [\tmmathbf{h}, \tmmathbf{z}_{[1]}^{\text{opt}}] = [\tmmathbf{C}_{\tmmathbf{h}}^0 \cdot
   \dot{\tmmathbf{h}} +\tmmathbf{C}_{\tmmathbf{h}}^1 \cdot
   \tmmathbf{z}_{[1]}^{\text{opt}}]_0^{\tilde{L}} + \int_0^{\tilde{L}} \left( \frac{1}{2}
   \mymult \dot{\tmmathbf{h}} \cdot \tmmathbf{B}_{\tmmathbf{h}}^{(0)} \cdot
   \dot{\tmmathbf{h}} - \frac{1}{2} \mymult \tmmathbf{z}_{[1]} \cdot
   \tmmathbf{B}_{\tmmathbf{h}}^{\text{} (2)} \cdot \tmmathbf{z}_{[1]}^{\text{opt}} \right)
   \mymult \mathd \tilde{S}.\]
This yields the final
expressions~(\ref{eq:RelaxedEnergy}--\ref{eq:finalOperatorsDef}) of the
relaxed model.

\section{Detailed calculations for the nonlinear
cylinder}\label{app:cylinder}

\subsection{Analysis of homogeneous
solutions}\label{app:nlCylinderHomogeneous}

In this section, we provide a detailed analysis of the homogeneous solutions
for the hyperelastic cylinder, with the aim to justify the main results
announced in section~\ref{ssec:nlCylinderHomogeneousSolsSummary}.

The map of strain for homogeneous solutions writes, from
equation~(\ref{eq:ETildeH}) and~(\ref{eq:neckingStrainClassical}),
\[ \tilde{\tmmathbf{E}}(\tmmathbf{h},\tmmathbf{y}) = \left(\begin{array}{c}
     \left\{ \frac{1}{2} \mymult (h_1^2 - 1) \right\}_T\\
     \left\{ \frac{1}{2} \mymult ((\partial_T y_1)^2 + (\partial_T y_2)^2 - 1)
     \right\}_T\\
     \left\{ \frac{1}{2} \mymult h_1 \mymult \partial_T y_1 \right\}_T\\
     \left\{ \frac{1}{2} \mymult \left( \left( \frac{y_2}{T} \right)^2 - 1
     \right) \right\}_T
   \end{array}\right) . \]

In view of the material symmetries, we seek equi-biaxial solutions in the
particular form $y_1 = \{0\}_T$ and $y_2 =\{ \mu \mymult T\}_T$. Then,
\[  \tilde{\tmmathbf{E}}(\tmmathbf{h},\tmmathbf{y})=
   \tilde{\tmmathbf{E}} (\tmmathbf{h},(\{ 0 \}_T, \{ \mu \mymult T \}_T)) =
   \left(\begin{array}{c}
     \left\{ \frac{1}{2} \mymult (h_1^2 - 1) \right\}_T\\
     \left\{ \frac{1}{2} \mymult (\mu^2 - 1) \right\}_T\\
     \{ 0 \}_T\\
     \left\{ \frac{1}{2} \mymult (\mu^2 - 1) \right\}_T
   \end{array}\right) \quad \frac{\partial
   \tilde{\tmmathbf{E}}}{\partial \tmmathbf{y}} (\tmmathbf{h},(\{ 0 \}_T, \{
   \mu \mymult T \}_T)) \cdot \hat{\tmmathbf{y}} = \left(\begin{array}{c}
     \{ 0 \}_T\\
     \{ \mu \mymult \partial_T \hat{y}_2 \}_T\\
     \left\{ \frac{1}{2} \mymult h_1 \mymult \partial_T \hat{y}_1 \right\}_T\\
     \left\{ \frac{\mu \mymult \hat{y}_2}{T} \right\}_T
   \end{array}\right). \]
For this type of strain, the material symmetry warrants $\overline{\Sigma}_{S
\nocomma T} = 0$ and $\overline{\Sigma}_{T \nocomma T} =
\overline{\Sigma}_{\Theta \nocomma \Theta}$. We denote the longitudinal stress
as $\overline{\Sigma}_{\parallel} (h_1, \mu) = \overline{\Sigma}_{S \nocomma
S}$ and the isotropic transverse stress as $\overline{\Sigma}_{\perp} (h_1,
\mu) = \overline{\Sigma}_{T \nocomma T} = \overline{\Sigma}_{\Theta \nocomma
\Theta}$. Then, the first variation of the density of strain energy reads
\[ \begin{array}{lll}
     \frac{\mathd W}{\mathd \tmmathbf{E}} (\tilde{\tmmathbf{E}}(\tmmathbf{h},\tmmathbf{y})) \cdot \delta \tmmathbf{E} & = & \int_0^{\rho}
     \left(\begin{array}{c}
       \overline{\Sigma}_{S \nocomma S}\\
       \overline{\Sigma}_{T \nocomma T}\\
       2 \mymult \overline{\Sigma}_{S \nocomma T}\\
       \overline{\Sigma}_{\Theta \nocomma \Theta}
     \end{array}\right) \cdot \left(\begin{array}{c}
       \delta E_1 (T)\\
       \delta E_2 (T)\\
       \delta E_3 (T)\\
       \delta E_4 (T)
     \end{array}\right) \mymult 2 \mymult \pi \mymult T \mymult \mathd T\\
     & = & \int_0^{\rho} (\overline{\Sigma}_{\parallel} (h_1, \mu) \mymult
     \delta E_1 (T) + \overline{\Sigma}_{\perp} (h_1, \mu) \mymult (\delta E_2
     (T) + \delta E_4 (T))) \mymult 2 \mymult \pi \mymult T \mymult \mathd T.
   \end{array} \]
By combining these equations with equation~(\ref{eq:outlineHomPVW}), we obtain
a principle of virtual work for the homogeneous radial displacement
$y_2^{\tmmathbf{h}} = \{\mu_{\tmmathbf{h}} \mymult T\}_T$ as
\[ \forall \hat{y}_1, \hat{y}_2 \quad \int_0^{\rho} \left[ f_1^{\tmmathbf{h}}
   \mymult \hat{y}_1 (T) - \mu \mymult \overline{\Sigma}_{\perp} (h_1,
   \mu_{\tmmathbf{h}}) \mymult \left( \frac{\hat{y}_2 (T)}{T} + \partial_T
   \hat{y}_2 (T) \right) \right] \mymult 2 \mymult \pi \mymult T \mymult \mathd T
   = 0 \]
It can be seen that the Lagrange multiplier is zero, $f_1^{(h_1)} = 0$, which
is a consequence of the fact that our trial function satisfies the constraint
already. The other term in the integrand above can be rewritten as
$\int_0^{\rho} [- 2 \mymult \pi \mymult \mu \mymult \overline{\Sigma}_{\perp}
(h_1, \mu_{\tmmathbf{h}}) \mymult \partial_T (T \mymult \hat{y}_2)] \mymult
\mathd T = - 2 \mymult \pi \mymult \mu \mymult \overline{\Sigma}_{\perp} (h_1,
\mu_{\tmmathbf{h}}) \rho \mymult \hat{y}_2 (\rho)$ and so the principle of
virtual work yields the equation $\overline{\Sigma}_{\perp} (h_1, \mu_{(h_1)})
= 0$ as announced in equation~(\ref{eq:nlCylinderEquil}).

We proceed to present a derivation of the stress and the tangent moduli in the
homogeneous solution.

The stress is found by identifying the first variation of the elastic
potential $\overline{W} (E_1 (S), E_3^2 (S), E_2 (S) + E_4 (S), E_2^2 (S) +
E_4^2 (S), E_2 (S) \mymult E_3^2 (S))$, namely
\[ \delta \overline{W} = \partial_1 \overline{W} \mymult \delta E_1 + 2 \mymult
   E_3 \mymult \partial_2 \overline{W} \mymult \delta E_3 + (\delta E_2 + \delta
   E_4) \mymult \partial_3 \overline{W} + 2 \mymult (E_2 \mymult \delta E_2 + E_4
   \mymult \delta E_4) \mymult \partial_4 \overline{W} + (E_3^2 \mymult \delta
   E_2 + 2 \mymult E_2 \mymult E_3 \mymult \delta E_3) \mymult \partial_5
   \overline{W}. \]
Identifying the stress components with the definition of Piola-Kirchhoff stress from the general theory of
elasticity, $\delta \overline{W} = \sum_{I \nocomma J} \overline{\Sigma}_{I
\nocomma J} \mymult \delta \overline{E}_{I \nocomma J}$, using of the definition of the $\overline{E}_{I \nocomma J}$'s in terms of
$(E_1, \ldots, E_4)$ (see below equation~(\ref{eq:neckingStrainClassical})) and evaluating these stress components  in the
homogeneous solution yields the equation~(\ref{eq:nlCylinderSigma}).

The tangent moduli are found similarly, by identifying the second variation of
the elastic potential
\[ \delta^2 \overline{W} = \partial_{1 \nocomma 1} \overline{W} \times (\delta
   E_1)^2 + \ldots \]
with the definition of the tangent moduli from the general theory of
elasticity, $\delta^2 \overline{W} = \sum_{I \nocomma J \nocomma L \nocomma M}
\overline{K}_{I \nocomma J}^{L \nocomma M} \mymult \delta \overline{E}_{I
\nocomma J} \mymult \delta \overline{E}_{L \nocomma M}$. This yields the
expression of $\overline{K}_{S \nocomma T}^{S \nocomma T}$ given in
equation~(\ref{eq:nlCylinderKSTST}), together with the additional tangent
elastic moduli
\[ \begin{array}{rcl}
     \overline{K}_{T \nocomma T}^{T \nocomma T} & = & 2 \mymult \partial_4
     \overline{W} + \overline{K}_{T \nocomma T}^{\Theta \nocomma \Theta}\\
     \overline{K}_{T \nocomma T}^{\Theta \nocomma \Theta} & = & \partial_{3
     \nocomma 3} \overline{W} + 2 \mymult (\mu_{(h_1)}^2 - 1) \mymult
     \partial_{3 \nocomma 4} \overline{W} + (\mu_{(h_1)}^2 - 1)^2 \mymult
     \partial_{4 \nocomma 4} \overline{W} .
   \end{array} \]
The factor $1 / 4$ in equation~(\ref{eq:nlCylinderKSTST}) comes from the fact
that there are four possible sets of indices $(I, J, L, M)$ such that $\delta
\overline{E}_{I \nocomma J} \mymult \delta \overline{E}_{L \nocomma M} =
(\delta \overline{E}_{S \nocomma T})^2$.

\subsection{Strain function in terms of the microscopic
correction}\label{ssec:nlCylAppendixEh}

The change of microscopic unknown is carried out by setting $\tmmathbf{y} (S)
=\tmmathbf{y}_{\tmmathbf{h}} (S) +\tmmathbf{z} (S)$, that is $y_1 = z_1$ and
$y_2 = y_2^{(h_1)} + z_2 = \mu_{(h_1)} \mymult T + z_2$. A strain function
$\tmmathbf{e}_{\tmmathbf{h}}$ defined in terms of the new unknown
$\tmmathbf{z}$ has been introduced in equation~(\ref{eq:strainInTermsOfZ}).
Inserting the expression of $\tmmathbf{E}$ from
equation~(\ref{eq:nlCylinderEh}) relevant to nonlinear elastic cylinder
yields
\[ \tmmathbf{e}_{\tmmathbf{h}} (\tmmathbf{h}^{\dag}, \tmmathbf{h}^{\ddag} ;
   \tmmathbf{z}, \tmmathbf{z}^{\dag}, \tmmathbf{z}^{\ddag}) =
   \left(\begin{array}{c}
     \left\{ \frac{1}{2} \mymult ((h_1 + z^{\dag}_1 (T))^2 + (h^{\dag}_1 \mymult
     \nabla \mu_{(h_1)} \mymult T + z^{\dag}_2 (T))^2 - 1) \right\}_T\\
     \left\{ \frac{1}{2} \mymult ((\partial_T z_1 (T))^2 + (\mu_{(h_1)} +
     \partial_T z_2 (T))^2 - 1) \right\}_T\\
     \left\{ \frac{1}{2} \mymult ((h_1 + z^{\dag}_1 (T)) \mymult \partial_T z_1
     (T) + (h^{\dag}_1 \mymult \nabla \mu_{(h_1)} \mymult T + z^{\dag}_2 (T))
     \mymult (\mu_{(h_1)} + \partial_T z_2 (T))) \right\}_T\\
     \left\{ \frac{1}{2} \mymult \left( \left( \mu_{(h_1)} + \frac{z_2 (T)}{T}
     \right)^2 - 1 \right) \right\}_T
   \end{array}\right) . \]
Calculating the Taylor expansion of $\tmmathbf{e}_{\tmmathbf{h}}$ about a
homogeneous solution, i.e., for small $\tmmathbf{h}^{\dag}$,
$\tmmathbf{h}^{\ddag}$, $\tmmathbf{z}$, $\tmmathbf{z}^{\dag}$ and
$\tmmathbf{z}^{\ddag}$, we can then identify the coefficients in this
expansion with the structure coefficients introduced in
equation~(\ref{eq:structureCoefs}).

\subsection{Corrective displacement}\label{app:nlCylinderZCorrection}

In this section, we solve the local optimization problem for the nonlinear
cylinder in traction. We start from the expressions of the operators
$\tmmathbf{B}_{\tmmathbf{h}}^{(1)}$ and $\tmmathbf{B}_{\tmmathbf{h}}^{(2)}$
obtained in equation~(\ref{eq:nlCylOperators}). As the operator
$\tmmathbf{B}_{\tmmathbf{h}}^{\text{} (2)}$ does not couple the unknowns $z_1
(T)$ and $z_2 (T)$, the optimization problem for $\tmmathbf{z}$ obtained in
equation~(\ref{eq:localOptimizationProblem}) splits into a problem for the
axial corrective displacement $z_1$
\begin{equation}
  \left\{\begin{array}{l}
    \forall \hat{z}_1 \quad \tmmathbf{h}' (S) \cdot \tmmathbf{B}_{\tmmathbf{h}
    (S)}^{(1)} \cdot (\hat{z}_1, \tmmathbf{0}) + \left( z^{\text{opt}}_1,
    \tmmathbf{0} \right) \cdot \tmmathbf{B}_{\tmmathbf{h} (S)}^{\text{} (2)}
    \cdot (\hat{z}_1, \tmmathbf{0}) -\tmmathbf{f}_{\text{opt}} (S) \cdot
    \tmmathbf{q} ((\hat{z}_1, \tmmathbf{0})) = 0\\
    \tmmathbf{q} \left( z^{\text{opt}}_1, \tmmathbf{0} \right) = 0,
  \end{array}\right. \label{eq:capillaryTubeZ1Pb}
\end{equation}
and a problem for the radial corrective displacement $z_2$,
\[ \left\{\begin{array}{l}
     \forall \hat{z}_2 \quad \tmmathbf{h}' (S) \cdot
     \tmmathbf{B}_{\tmmathbf{h} (S)}^{(1)} \cdot (\tmmathbf{0}, \hat{z}_2) +
     \left( \tmmathbf{0}, z^{\text{opt}}_2 \right) \cdot
     \tmmathbf{B}_{\tmmathbf{h} (S)}^{\text{} (2)} \cdot (\tmmathbf{0},
     \hat{z}_2) = 0.
   \end{array}\right. \]

In the latter, the first term $\tmmathbf{h}' (S) \cdot
\tmmathbf{B}_{\tmmathbf{h} (S)}^{(1)} \cdot (\tmmathbf{0}, \hat{z}_2)$
vanishes in view of the definition of $\tmmathbf{B}_{\tmmathbf{h}}^{(1)}$ in
equation~(\ref{eq:nlCylOperators}). The solution for the corrective radial
displacement is therefore zero,
\[ z^{\text{opt}}_2 = \{ 0 \}_T . \]

Observe that the second term in the right-hand side of the definition
of $\tmmathbf{B}_{\tmmathbf{h}}^{(1)}$ in
equation~(\ref{eq:nlCylOperators}) is zero, as it is proportional to
the constraint $\tmmathbf{q} \left( z^{\text{opt}}_1, \tmmathbf{0}
\right) = \int_{0}^{\rho}z^{\text{opt}}_1(T)\,2\,\pi\,T\,\mathrm{d}T$.
Inserting the expressions of the operators
$\tmmathbf{B}_{\tmmathbf{h}}^{(1)}$ and
$\tmmathbf{B}_{\tmmathbf{h}}^{(2)}$ from
equation~(\ref{eq:nlCylOperators}), we can rewrite the variational
problem~(\ref{eq:capillaryTubeZ1Pb}) as
\begin{multline}
     \forall \hat{z}_1 \quad \int_0^{\rho} \left( [h_1 \mymult h_1^{'} \mymult
     \overline{K}_{S \nocomma T}^{S \nocomma T} \mymult \mu_{(h_1)} \mymult
     \nabla \mu_{(h_1)}] \mymult T + [\nobracket h_1 \nobracket^2 \mymult
     \overline{K}_{S \nocomma T}^{S \nocomma T}] \mymult \partial_T
     z_1^{\text{opt}} (T) \right) \mymult \partial_T \hat{z}_1 (T) \mymult 2
     \mymult \pi \mymult T \mymult \mathd T \cdots\\
     + \left[ - \tilde{f}_1^{\text{opt}} \right] \mymult \int_0^{\rho}
     \hat{z}_1 \mymult 2 \mymult \pi \mymult T \mymult \mathd T = 0
	 \nonumber
 \end{multline}
where we have made appear a new Lagrange multiplier
$\tilde{f}_1^{\text{opt}} = f_1^{\text{opt}} - h_1^{\dag} \mymult
   \frac{\mathd (\overline{\Sigma}_{\parallel} (h_1, \mu_{(h_1)}) \mymult
   h_1)}{\mathd h_1}$.

To solve this equation, we rewrite it in compact form as
\[ \forall \hat{z}_1 \quad \int_0^{\rho} \left( a \mymult T + b \mymult
   \partial_T z_1^{\text{opt}} \right) \mymult \partial_T \hat{z}_1 \mymult 2
   \mymult \pi \mymult T \mymult \mathd T + d \mymult \int_0^{\rho} \hat{z}_1
   \mymult 2 \mymult \pi \mymult T \mymult \mathd T = 0, \]
where the coefficients $a$, $b$, $d$ are identified with the square brackets
in the equation above. An integration by parts yields
\[ \forall \hat{z}_1 \quad \left[ 2 \mymult \pi \mymult T \mymult \left( a \mymult
   T + b \mymult \partial_T z_1^{\text{opt}} \right) \mymult \hat{z}_1
   \right]_0^{\rho} - \int_0^{\rho} \partial_T \left[ 2 \mymult \pi \mymult T
   \mymult \left( a \mymult T + b \mymult \partial_T z_1^{\text{opt}} - d \mymult
   \frac{T}{2} \right) \right] \mymult \hat{z}_1 \mymult \mathd T = 0 \]
The solution is that $T \mymult \left( a \mymult T + b \mymult \partial_T
z_1^{\text{opt}} - d \mymult \frac{T}{2} \right)$ is a constant, which is found
using of the boundary conditions as
\[ T \mymult \left( a \mymult T + b \mymult \partial_T z_1^{\text{opt}} - d
   \mymult \frac{T}{2} \right) = - \frac{d \mymult \rho^2}{2} . \]
We conclude $\partial_T z_1^{\text{opt}} = \frac{T}{b} \mymult \left(
\frac{d}{2} - a \right) - \frac{d \mymult \rho^2}{2 \mymult b} \mymult
\frac{1}{T}$. The Lagrange multiplier $d = \left[ - \tilde{f}_1^{\text{opt}}
\right]$ must be chosen so as to remove the logarithmic divergence for $T
\rightarrow 0$: this yields $d = 0$. We can integrate the remaining terms and
set the constant of integration such that the average constraint $q_1 \left(
z_1^{\text{opt}} \right) = 0$ is satisfied: the result is $z_1^{\text{opt}}
(T) = - \frac{a}{2 \mymult b} \mymult \left( T^2 - \frac{\rho^2}{2} \right)$. In
terms of the original quantities, the solution for the axial correction to the
displacement writes, after simplification, $z_1^{\text{opt}} (T) = -
\frac{a}{2 \mymult b} \mymult \left( T^2 - \frac{\rho^2}{2} \right) = -
\frac{1}{2} \mymult \frac{\mu_{(h 1)} \mymult \mymult \nabla \mu_{(h_1)}}{h_1}
\mymult h_1' \mymult \left( T^2 - \frac{\rho^2}{2} \right)$. This can be
rewritten as
\[ z_1^{\text{opt}} = \left\{ - \frac{1}{2} \mymult \frac{\mu_{(h 1)} \mymult
   \mymult \nabla \mu_{(h_1)}}{h_1} \mymult \left( T^2 - \frac{\rho^2}{2}
   \right) \right\}_T \mymult h_1' ,\]
   as captured in equation~(\ref{eq:ZoptCylinder}).

 \bigskip
\bibliographystyle{elsarticle-harv}

\begin{thebibliography}{36}
	\expandafter\ifx\csname natexlab\endcsname\relax\def\natexlab#1{#1}\fi
	\expandafter\ifx\csname url\endcsname\relax
	\def\url#1{\texttt{#1}}\fi
	\expandafter\ifx\csname urlprefix\endcsname\relax\def\urlprefix{URL }\fi
	
	\bibitem[{Acerbi et~al.(1991)Acerbi, Buttazzo, and
		Percivale}]{Acerbi-Buttazzo-EtAl-A-variational-definition-of-the-strain-1991}
	Acerbi, E., Buttazzo, G., Percivale, D., 1991. A variational definition of the
	strain energy for an elastic string. Journal of Elasticity 25, 137--148.
	
	\bibitem[{Agostiniani et~al.(2016)Agostiniani, DeSimone, and
		Koumatos}]{Agostiniani-DeSimone-EtAl-Shape-programming-for-narrow-2016}
	Agostiniani, V., DeSimone, A., Koumatos, K., 2016. Shape programming for narrow
	ribbons of nematic elastomers, arxiv 1603.02088v1.
	
	\bibitem[{Audoly and
		Hutchinson(2016)}]{Audoly-Hutchinson-Analysis-of-necking-based-2016}
	Audoly, B., Hutchinson, J.~W., 2016. Analysis of necking based on a
	one-dimensional model. Journal of the Mechanics and Physics of Solids 97,
	68--91.
	
	\bibitem[{Bardenhagen and
		Triantafyllidis(1994)}]{Bardenhagen-Triantafyllidis-Derivation-of-higher-order-1994}
	Bardenhagen, S., Triantafyllidis, N., 1994. Derivation of higher order gradient
	continuum theories in 2,3-d non-linear elasticity from periodic lattice
	models. Journal of the Mechanics and Physics of Solids 42~(1), 111--139.
	
	\bibitem[{Barenblatt and Joseph(1997)}]{Rivlin97}
	Barenblatt, G., Joseph, D. (Eds.), 1997. Collected Papers of R.S. Rivlin.
	Springer, New York, Ch. Stability of an elastic material.
	
	\bibitem[{Bermudez and
		Via{\~n}o(1984)}]{Bermudez-Viano-Une-justification-des-equations-de-la-thermoelasticite-1984}
	Bermudez, A., Via{\~n}o, J.~M., 1984. Une justification des {\'e}quations de la
	thermo{\'e}lasticit{\'e} des poutres {\`a} section variable par des
	m{\'e}thodes asymptotiques. Mod{\'e}lisation Math{\'e}matqiue et Analyse
	Num{\'e}rique 18, 347--376.
	
	\bibitem[{Buannic and Cartaud(2000)}]{Cartaud1}
	Buannic, N., Cartaud, P., 2000. Higher-order effective modeling of periodic
	heterogeneous beams. {I}. {A}symptotic expansion method. International
	Journal of Solids and Structures 38, 7139--7161.
	
	\bibitem[{Calladine(1983)}]{Calladine-Theory-of-shell-structures-1983}
	Calladine, C.~R., 1983. Theory of shell structures. Cambridge University Press.
	
	\bibitem[{Cimeti{\`e}re et~al.(1988)Cimeti{\`e}re, Geymonat, Le~Dret, Raoult,
		and Tutek}]{Cimetiere-Geymonat-EtAl-Asymptotic-theory-and-analysis-1988}
	Cimeti{\`e}re, A., Geymonat, G., Le~Dret, H., Raoult, A., Tutek, Z., 1988.
	Asymptotic theory and analysis for displacements and stress distribution in
	nonlinear elastic straight slender rods. Journal of Elasticity 19, 111--161.
	
	\bibitem[{Coleman and
		Newman(1988)}]{Coleman-Newman-On-the-rheology-of-cold-drawing.-1988}
	Coleman, B.~D., Newman, D.~C., 1988. On the rheology of cold drawing. {I.}
	elastic materials. Journal of Polymer Science: Part B: Polymer Physics 26,
	1801--1822.
	
	\bibitem[{Freddi et~al.(2016)Freddi, Hornung, Mora, and
		Paroni}]{Freddi-Hornung-EtAl-A-variational-model-for-anisotropic-2016}
	Freddi, L., Hornung, P., Mora, M.-G., Paroni, R., 2016. A variational model for
	anisotropic and naturally twisted ribbons. SIAM Journal on Mathematical
	Analysis 48~(6), 3883--3906.
	
	\bibitem[{Freddi et~al.(2004)Freddi, Morassi, and
		Paroni}]{FreddiMorassiParoni-Thin-Walled-Beams-the-Case-of-the-Rectangular-Cross-Section-2004}
	Freddi, L., Morassi, A., Paroni, R., 2004. Thin-walled beams: the case of the
	rectangular cross-section. Journal of Elasticity 76~(1), 45--66.
	
	\bibitem[{Fu et~al.(2008)Fu, Pearce, and
		Liu}]{Fu-Pearce-EtAl-Post-bifurcation-analysis-of-a-thin-walled-2008}
	Fu, Y.~B., Pearce, S.~P., Liu, K.~K., 2008. Post-bifurcation analysis of a
	thin-walled hyperelastic tube under inflation. International Journal of
	Non-Linear Mechanics 43~(8), 697--706.
	
	\bibitem[{Geymonat et~al.(2018)Geymonat, Krasucki, and
		Serpilli}]{Geymonat-Krasucki-EtAl-Asymptotic-derivation-of-linear-2018}
	Geymonat, G., Krasucki, F., Serpilli, M., 2018. Asymptotic derivation of linear
	plate model for soft ferromagnetic material. Chinese Annals of Mathematics,
	Series B 39~(3), 451--460.
	
	\bibitem[{G'Sell et~al.(1983)G'Sell, Aly-Helal, and
		Jonas}]{GSell-Aly-Helal-EtAl-Effect-of-stress-triaxiality-1983}
	G'Sell, C., Aly-Helal, N.~A., Jonas, J.~J., 1983. Effect of stress triaxiality
	on neck propagation during the tensile stretching of solid polymers. Journal
	of Materials Science 18, 1731--1742.
	
	\bibitem[{Hamdouni and
		Millet(2006)}]{Hamdouni-Millet-An-asymptotic-non-linear-model-2006}
	Hamdouni, A., Millet, O., 2006. An asymptotic non-linear model for thin-walled
	rods with strongly curved open cross-section. International Journal of
	Non-Linear Mechanics 41~(3), 396--416.
	
	\bibitem[{Kyriakides and
		Chang(1990)}]{Kyriakides-Chang-On-the-inflation-of-a-long-elastic-1990}
	Kyriakides, S., Chang, Y.-C., 1990. On the inflation of a long elastic tube in
	the presence of axial load. International Journal of Solids and Structures
	26~(9--10), 975--991.
	
	\bibitem[{Kyriakides and
		Chang(1991)}]{Kyriakides-Chang-The-initiation-and-propagation-of-a-localized-1991}
	Kyriakides, S., Chang, Y.-C., 1991. The initiation and propagation of a
	localized instability in an inflated elastic tube. International Journal of
	Solids and Structures 27~(9), 1085--1111.
	
	\bibitem[{Lestringant and
		Audoly(2017)}]{Lestringant-Audoly-Elastic-rods-with-2016}
	Lestringant, C., Audoly, B., 2017. Elastic rods with incompatible strain:
	macroscopic versus microscopic buckling. Journal of the Mechanics and Physics
	of Solids 103, 40--71.
	
	\bibitem[{Lestringant and
		Audoly(2018)}]{Lestringant-Audoly-A-diffuse-interface-model-2018}
	Lestringant, C., Audoly, B., 2018. A diffuse interface model for the analysis
	of propagating bulges in cylindrical balloons. Proceedings of the Royal
	Society A: Mathematical, Physical and Engineering Sciences 474, 20180333.
	
	\bibitem[{Liu et~al.(2014)Liu, Huang, Su, Bertoldi, and
		Clarke}]{Liu-Huang-EtAl-Structural-Transition-from-2014}
	Liu, J., Huang, J., Su, T., Bertoldi, K., Clarke, D., 2014. Structural
	transition from helices to hemihelices. PLoS ONE 9~(4), e93183.
	
	\bibitem[{Mahadevan et~al.(2007)Mahadevan, Vaziri, and
		Das}]{MahadevanVaziriDas-Persistence-of-a-pinch-in-a-pipe-2007}
	Mahadevan, L., Vaziri, A., Das, M., 2007. Persistence of a pinch in a pipe.
	Europhysics Letters 77~(4), 40003.
	
	\bibitem[{Marigo and
		Meunier(2006)}]{Marigo-Hierarchy-of-One-Dimensional-Models-2006}
	Marigo, J.-J., Meunier, N., 2006. Hierarchy of one-dimensional models in
	nonlinear elasticity. Journal of Elasticity 83, 1--28.
	
	\bibitem[{Matsuo and
		Tanaka(1992)}]{Matsuo-Tanaka-Patterns-in-shrinking-gels-1992}
	Matsuo, E.~S., Tanaka, T., 1992. Patterns in shrinking gels. Nature 368, 1735.
	
	\bibitem[{Mora et~al.(2010)Mora, Phou, Fromental, Pismen, and
		Pomeau}]{Mora-Phou-EtAl-Capillarity-Driven-Instability-2010}
	Mora, S., Phou, T., Fromental, J.-M., Pismen, L.~M., Pomeau, Y., Nov 2010.
	Capillarity driven instability of a soft solid. Physical Review Letters 105,
	214301.
	
	\bibitem[{Ogden(1972)}]{Ogden-Large-deformation-isotropic-1972}
	Ogden, R.~W., 1972. Large deformation isotropic elasticity-on the correlation
	of theory and experiment for incompressible rubber-like solids. Proceedings
	of the Royal Society A: Mathematical, Physical and Engineering Science 326,
	565--584.
	
	\bibitem[{Pearce and
		Fu(2010)}]{Pearce-Fu-Characterization-and-stability-of-localized-2010}
	Pearce, S.~P., Fu, Y.~B., 2010. Characterization and stability of localized
	bulging/necking in inflated membrane tubes. IMA Journal of Applied
	Mathematics 75, 581--602.
	
	\bibitem[{Picault et~al.(2016)Picault, Bourgeois, Cochelin, and
		Guinot}]{Picault-Bourgeois-EtAl-A-rod-model-with-thin-walled-2016}
	Picault, E., Bourgeois, S., Cochelin, B., Guinot, F., 2016. A rod model with
	thin-walled flexible cross-section: Extension to {3D} motions and application
	to {3D} foldings of tape springs. International Journal of Solids and
	Structures 84, 64--81.
	
	\bibitem[{Sadowsky(1930)}]{Sadowsky-Ein-elementarer-Beweis-fur-die-Existenz-1930}
	Sadowsky, M., 1930. {Ein elementarer Beweis f{\"u}r die Existenz eines
		abwickelbaren M{\"o}biusschen Bandes und die Zur{\"u}ckf{\"u}hrung des
		geometrischen Problems auf ein Variationsproblem}. In: Sitzungsberichte der
	Preussischen Akademie der Wissenschaften, physikalisch-mathematische Klasse,
	17. Juli 1930, Mitteilung vom 26. Juni. pp. 412--415.
	
	\bibitem[{Sanchez-Hubert and
		Sanchez~Palencia(1999)}]{Sanchez-Hubert-Sanchez-Palencia-Statics-of-curved-rods-1999}
	Sanchez-Hubert, J., Sanchez~Palencia, E., 1999. Statics of curved rods on
	account of torsion and flexion. European Journal of Mechanics. A. Solids 18,
	365--390.
	
	\bibitem[{Seffen and
		Pellegrino(1999)}]{Seffen-Pellegrino-Deployment-dynamics-of-tape-1999}
	Seffen, K.~A., Pellegrino, S., 1999. Deployment dynamics of tape springs.
	Proceedings of the Royal Society of London. Series A: Mathematical, Physical
	and Engineering Sciences 455~(1983), 1003--1048.
	
	\bibitem[{Trabucho and
		Via{\~n}o(1996)}]{Trabucho-Viano-Mathematical-modelling-of-rods-1996}
	Trabucho, L., Via{\~n}o, J.~M., 1996. Mathematical modelling of rods. Handbook
	of numerical analysis 4, 487--974.
	
	\bibitem[{Triantafyllidis et~al.(2007)Triantafyllidis, Scherzinger, and
		Huang}]{Triantafyllidis-Scherzinger-EtAl-Post-bifurcation-equilibria-in-the-plane-strain-2007}
	Triantafyllidis, N., Scherzinger, W.~M., Huang, H.-J., 2007. Post-bifurcation
	equilibria in the plane-strain test of a hyperelastic rectangular block.
	International Journal of Solids and Structures 44~(11--12), 3700--3719.
	
	\bibitem[{van~der Hoeven et~al.(2013)van~der Hoeven, Grozin, Gubinelli, Lecerf,
		Poulain, and
		Raux}]{Hoeven-Grozin-EtAl-GNU-TEXmacs:-a-scientific-editing-2013}
	van~der Hoeven, J., Grozin, A., Gubinelli, M., Lecerf, G., Poulain, F., Raux,
	D., 2013. {GNU TEXmacs}: a scientific editing platform. ACM Communications in
	Computer Algebra 47~(1--2), 59--61.
	
	\bibitem[{Wunderlich(1962)}]{Wunderlich-Uber-ein-abwickelbares-Mobiusband-1962}
	Wunderlich, W., 1962. {\"Uber ein abwickelbares M\"obiusband}. Monatshefte
	f\"ur Mathematik 66~(3), 276--289.
	
	\bibitem[{Xuan and
		Biggins(2017)}]{Xuan-Biggins-Plateau-Rayleigh-instability-in-solids-2017}
	Xuan, C., Biggins, J., May 2017. {Plateau-Rayleigh} instability in solids is a
	simple phase separation. Physical Reivew E 95, 053106.
	
 \end{thebibliography}

\end{document}